\documentclass[master,       
               twoside,        
               BCOR10mm,       
               ngerman,english  
               ]{GAUBM}

\usepackage{setspace}  
\usepackage{babel}     
\usepackage{calc}      
\usepackage{libertine}
\usepackage[T1]{fontenc}       
\usepackage[utf8]{inputenc}    

\usepackage{amsmath,amssymb} 
\usepackage{bm}
\usepackage{bbm}

\usepackage{lmodern} 
\usepackage[sorting=none,natbib,citestyle=numeric-comp,backend=biber,backref=true,firstinits=true,alldates=iso8601]{biblatex}
\DeclareSourcemap{ 
	\maps[datatype=bibtex]{
		\map[overwrite=true]{
			\step[fieldset=abstract, null]
			\step[fieldset=url, null]
			\step[fieldset=doi, null]
			\step[fieldset=eprint, null]
			\step[fieldset=issn, null]
			\step[fieldset=isbn, null]
		}
	}
}
\newcommand{\tabheadfont}[1]{\textbf{#1}} 
\usepackage{booktabs}                      
\usepackage{longtable}                     
\usepackage{array}                         

\usepackage[pdfstartview=FitH,      
            breaklinks=true,        
            bookmarksopen=true,     
            bookmarksnumbered=true  
            ]{hyperref}
            
\usepackage{subcaption}
\usepackage{xcolor}
\usepackage{placeins}
\usepackage{rotating}
\usepackage{adjustbox}

\begin{document}
\ThesisAuthor{Nick}{Scholand}
\PlaceOfBirth{Arnsberg}
\ThesisTitle{Model-Based Reconstruction for Quantitative MRI using the Bloch Equations}{Modellbasierte Rekonstruktion in der Quantitativen MRT unter Verwendung der Bloch-Gleichungen}
\FirstReferee{Prof. Dr. Martin Uecker}
\Institute{Institute for Diagnostic and Interventional Radiology\\ at the University Medical Center in  Göttingen}
\SecondReferee{Prof. Dr. Ulrich Parlitz}
\ThesisBegin{20}{08}{2018}
\ThesisEnd{12}{02}{2019}
\frontmatter
\maketitle
\cleardoublepage
\onehalfspacing
\begin{abstract}
  In dieser Arbeit wird eine Methode der modellbasierten Rekonstruktion für die Magnet-Resonanz-Tomographie entwickelt. Während bisherige Verfahren vereinfachte Signalmodelle benutzen, werden hier die Bloch-Gleichungen direkt in die Rekonstruktion eingebunden, um die Relaxationsparameter $T_1$ und $T_2$ einer selbst erstellten Messprobe zu bestimmen. Dazu wird eine kalibrationsfreie parallele Bildgebung mit einem Runge-Kutta 5(4) basierten Vorwärtsoperator kombiniert, der die Spindynamik beliebiger Sequenzen simulieren kann. Zusammen mit einem IRGNM-FISTA Algorithmus können damit die Parameterkarten für $T_1$, $T_2$ und $M_0$, sowie die Spulen-Sensitivitäten $c_N$ direkt aus den Rohdaten ohne vorherige Rekonstruktion bestimmt werden. Die Gradientenberechnung für den Optimierungsalgorithmus wird mittels einer Sensitivitätsanalyse realisiert, deren gewöhnliche Differenzialgleichungen parallel zur Simulation gelöst werden. Die Datenaufnahme erfolgt radial mit vollabgetasteten Multiinversions- und stark unterabgetastete Einzelinversionsmessungen.
  \bigskip\par
  \textbf{Stichwörter:} Model-Basierte Rekonstruktion, Sensitivitätsanalyse, Bloch-Simulationen, Quantitative MRT
\end{abstract}
\begin{otherlanguage}{english}
\begin{abstract}
  In this work a generic model-based reconstruction for the quantification of relaxation parameters is developed. In contrast to previous approaches that rely on simplified models derived from the Bloch equations, this work includes the Bloch equations directly into the reconstruction. Therefore, non-linear calibrationless parallel imaging is combined with a generic Runge-Kutta 5(4) based forward operator to simulate spin dynamics described by arbitrary sequences. Gradients are determined by using a sensitivity analysis and solving the resulting ordinary differential equations parallel to the signal simulation. Based on this formulation an IRGNM-FISTA algorithm is used to estimate quantitative maps for $T_1$, $T_2$, $M_0$, and the coil profiles $c_N$ from fully-sampled multi-inversion and golden-angle single-shot inversion-recovery radial bSSFP measurement with a custom-built $T_1$-$T_2$ phantom. 
\end{abstract}
\end{otherlanguage}

\cleardoublepage
\onehalfspacing
\tableofcontents

\begin{nomenclature}
\section*{Abbreviations}
\begin{longtable}[l]{p{0.2\textwidth}p{0.8\textwidth-4\tabcolsep}}
  \tabheadfont{Shortcut}&\tabheadfont{Meaning}\\\midrule\endhead
  BART			& Berkeley Advanced Reconstruction Toolbox	\\
  bSSFP			& Balanced Steady-State Free Precession	\\
  BWTP			& Bandwidth-Time-Product		\\
  Dopri			& Dormand and Prince Algorithm	\\
  FA			& Flip-Angle					\\
  FFT			& Fast-Fourier Transformation	\\ 
  FID			& Free Induction Decay			\\
  FISTA			& Fast Iterative Shrinkage/Thresholding Algorithm	\\
  FLASH			& Fast Low-Angle Shot			\\
  FoV			& Field of View					\\
  FSM			& Forward Sensitivity Method	\\
  IDEA			& Integrated Development Environment for (MR) Applications \\
  IRGNM			& Iteratively Regularized Gauss Newton Method	\\
  MR			& Magnetic Resonance			\\
  MRI			& Magnetic Resonance Imaging 	\\
  ODE			& Ordinary Differential Equation	\\
  PI			& Parallel Imaging				\\
  RF			& Radio-Frequency				\\
  RK			& Runge-Kutta					\\
  ROI			& Region of Interest			\\
  SSFP			& Steady-State Free Precession	\\
  TE			& Echo Time						\\
  TI			& Inversion Time				\\
  TR			& Repetition Time				\\
\end{longtable}
\end{nomenclature}

\mainmatter   

\chapter{Introduction}

Magnetic resonance imaging (MRI) is one of the most important imaging techniques in clinical practice. In comparison to other tomographic methods like computed tomography is it based on non-ionizing electro-magnetic radiation and has a good soft tissue contrast, which can be adjusted by using different sequences or tuning their parameters to visualize abnormal tissue like edema or scars. Often contrast agents are used to characterize the tissue by its metabolism.\\
Besides the structural imaging possibilities, MRI offers a wide range of different applications. They go from temperature measurements over flow characterization up to measurements of oxygen levels in the brain. An other growing field is the quantitative MRI. It works on determining the relaxation parameters of the visualized tissue, which allows to actually identify the tissue, for example to find edemas of scars in the heart \cite{Okur_Diagn.Interv.Radiol._2014}\cite{Taylor_JACC:CardiovascularImaging_2016}, but also to simplify image segmentation and to speed up the acquisition time based on synthetic MRI \cite{Warntjes_Magn.Reson.Med._2008}\cite{Tanenbaum_AJNR_2017}.\\
While the synonym qMRI for quantitative MRI was first used in 1997 during a meeting of the UK Institute
of Physics and Engineering in Medicine \cite[p.12]{Tofts__2003}
today various methods exist to quantify the relaxation parameter. Typically they exploit simplified models of the Bloch equations and combine them with highly-specialized sequences. The latter are designed to be sensitive to the desired tissue characteristic, but insensitive to others. After acquiring the data, the images are reconstructed and pixel-wisely fitted with the derived model. This results in robust methods for estimating single-parameters \cite{Crawley_Magn.Reson.Med._1988}\cite{Messroghli_Magn.Reson.Med._2004}
, but also in rather sensitive ones for multiple estimates \cite{Schmitt_Magn.Reson.Med._2004}\cite{Deoni_Magn.Reson.Med._2005}. \\
Therefore techniques are developed which are based on dictionaries and which exploits the strong increase of computational power and memory \cite{Ma_Nature_2013}. Others have chosen model-based approaches, where the model is directly added to the reconstruction and not afterwards applied. They exploit iterative optimization algorithms to allow a more efficient handling of the acquired data \cite{Block_IEEETrans.Med.Imaging_2009}\cite{Sumpf_J.Magn.Reson.Imaging_2011}\cite{Wang_Magn.Reson.Med._2018}\cite{Maier_Magn.Reson.Med._2018}\footnote{And there are mixed developments \cite{Doneva_Magn.Reson.Med._2010}.}.
They accelerate quantitative MRI procedures by directly using k-space data without any previous reconstructions or databases and do not have discretization errors, which occur through the limited sampling density in dictionaries. On the other hand, model-based reconstructions are also based on previously derived and approximated model. That means they rely on specialized sequences and can often acquire a single relaxation parameter only. \\
In this work a new model-based reconstruction algorithm is proposed, which directly uses the Bloch equations as signal model. Therefore, in principle no approximations are needed to derive a simplified signal behavior and the method is not limited to a special sequence. Additionally, non-linear calibrationless parallel imaging is included to directly estimate the coil profiles and therefore to avoid preparation scans. For this thesis, the developed method is tested with the well known inversion-prepared bSSFP sequence. In the past it was shown to be sensitive to $T_1$ and $T_2$ \cite{Schmitt_Magn.Reson.Med._2004}\cite{Ma_Nature_2013}\cite{Scheffler_Magn.Reson.Med._2001}. To improve the accuracy of the IRGNM-FISTA based optimization, the calculation of the scaling factor $M_0$ is included.\\
This thesis starts with an introduction to MRI in general, followed by a chapter about the creation of a $T_1$-$T_2$ phantom to test quantitative MRI methods. Afterwards the used sequences are discussed as well as the development of the simulation tool, which is used in the later designed forward operator. In the end the final steps for creating the model-based reconstruction and the used IRGNM-FISTA optimization are introduced and the results are presented as well as summarized and discussed.

\chapter{Theory}

The following pages give an introduction to the basic principles of magnetic resonance imaging. This contains physical, mathematical as well as technical basics, which are important for a better understanding of the methods used in this work.

\section{Fundamental Principles of MRI}

	\subsection[Spin]{Spin}
\begin{figure}[!h]
			\centering
			\includegraphics[width=0.5\linewidth]{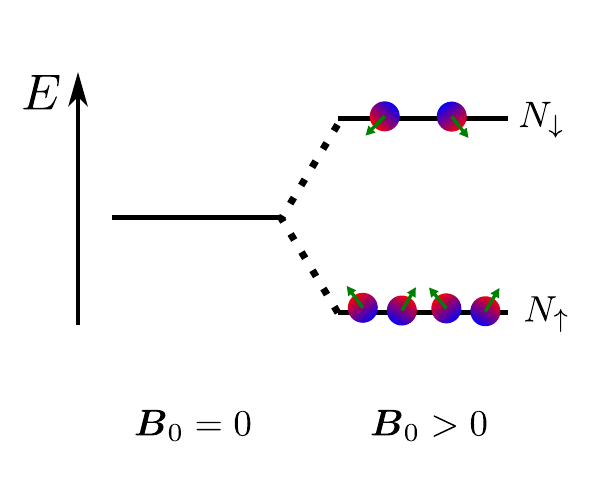}
			\caption{Visualization of the Zeeman-splitting in an external magnetic field $\bm{B}_0$. The two energy levels  $E_{\uparrow}$ and  $E_{\downarrow}$ are illustrated together with the number of spins in both levels $N_{\uparrow}$ and  $N_{\downarrow}$. The small green arrows represent the magnetic moment $\bm{\mu}$ of the spins, which is tilted by a angle $\cos\theta=\frac{\mu_z}{\mu}$ \cite[p.61]{Liang__1999}.}
			\label{fig::spin_distribution}
		\end{figure}

		\textit{This section follows \cite[p. 57ff]{Liang__1999}.}\\
		 The typically used particles in MRI are $^1$H, $^{13}$C, $^{19}$F and $^{31}$P. Due to its high occurrence in human tissue the proton is the most common one. It has a spin quantum number of $I$=1/2, which leads to an intrinsic magnetic moment magnitude of 
		\begin{align}
			\mu = \gamma\hbar\sqrt{I(I+1)},
		\end{align} 
		with the gyromagnetic ratio $\gamma$ and the Planck-constant $\hbar$. While $\mu$ is always the same, in the classical model its orientation can change. Without an external magnetic field $\bm{B_0} = 0$, thermal motion leads to a randomly distributed orientation of the microscopic $\bm{\mu}$ in a cubic volume element of tissue, called voxel. Therefore no macroscopic net magnetization occurs.\\
		If the protons are placed in a strong magnetic field $\bm{B_0} = B_0 \bm{\hat{e}}_z > 0$, the z-component of $\bm{\mu}$ is forced to have discrete values
		\begin{align}
			\mu_z=\gamma m_I \hbar,
		\end{align}
		with the magnetic quantum number $m_I = -I,-I+1,...,I$. This property leads to the Zeeman-splitting and allows $^1$H atoms to have two different spin-orientations in an external magnetic field, like sketched in Figure \ref{fig::spin_distribution}. Both differ in energy $E$:
		\begin{align}
			E = -\bm{\mu}\cdot\bm{B}_0 = -\gamma\hbar m_I B_0.
		\end{align}		
		While spin-up spins with $m_I$=1/2 have a low energy of $E_{\uparrow}=-\frac{1}{2}\gamma\hbar B_0$, the spin-down ones with $m_I$=-1/2 have a higher energy of $E_{\downarrow}=\frac{1}{2}\gamma\hbar B_0$. Their difference follows with 
		\begin{align}
			\Delta E = \gamma\hbar B_0
		\end{align}
		and can be used to determine the distribution of spins in both states following the Boltzmann-distribution
		\begin{align}
			\frac{N_{\uparrow}}{N_{\downarrow}}=\exp\left(-\frac{\Delta {E}}{{k}_BT}\right),
		\end{align}
		with the temperature $T$, the Boltzmann constant $k_B$ and the number of spins in both energy states $N_{\uparrow}$ as well as $N_{\downarrow}$.

		Because the macroscopic magnetization $\bm{M}$ of a voxel is defined as
		\begin{align}
			\bm{M} = \sum\limits_{n=1}^{N_s}\bm{\mu}_s,
		\end{align}
		with the number of microscopic spins $N_s$ and their magnetic moment $\bm{\mu}_s$, the asymmetry of the Boltzmann distribution for non-zero temperatures and an external magnetic field leads to a net magnetization $M>0$ pointing in the same direction as $\bm{B}_0$.

	\subsection[Rotating and Laboratory Frame]{Rotating and Laboratory Frame}
\begin{figure}[!h]
		\centering
		\includegraphics[width=\linewidth]{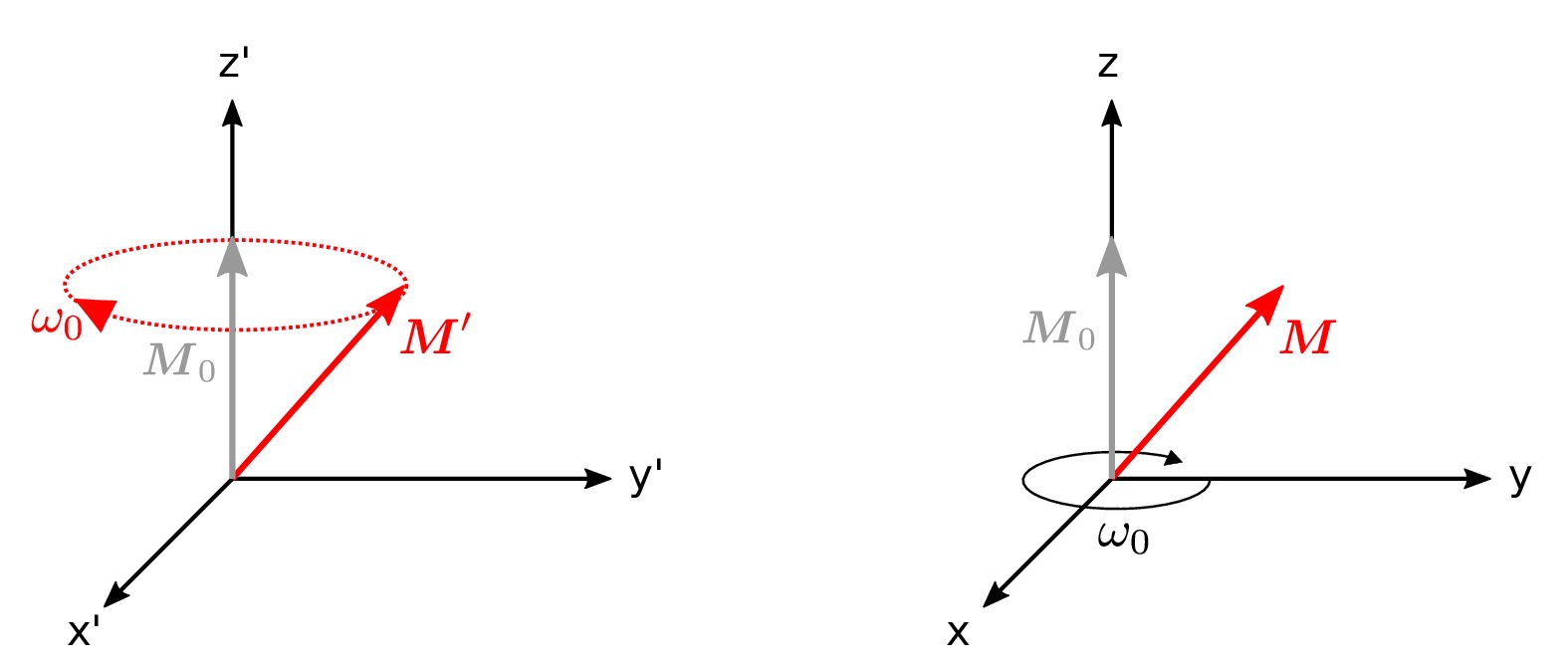}
		\caption{(\textbf{Left}) Visualization of the laboratory frame. The magnetization $\bm{M'}$ rotates with the Larmor frequency $\omega_0$, while in the rotating frame (\textbf{right}) the coordinate system is rotated with $\omega_0$ an $\bm{M}$ is static.}
		\label{fig::theory_rot_lab_frame}
	\end{figure}

	\textit{This section follows \cite[p.23, 26f, 69f]{Bernstein__2004}.}\\
	To understand the time-development of the magnetization $\bm{M}$ the concepts of a rotating and a laboratory frame, presented in Figure \ref{fig::theory_rot_lab_frame}, are introduced. They are based on a geometrical description of the spin-ensemble based on $\bm{M}$ of one voxel.\\
	For notational reasons the time-dependent magnetization vector in the rotating frame is defined as $\bm{M}$ and in the stationary laboratory frame as $\bm{M'}$. The three Cartesian axis are noted as ($x'$, $y'$, $z'$) for the laboratory and ($x$, $y$, $z$) for the rotating frame with the angular velocity $\omega$ equal to the Larmor frequency 
	\begin{align}
	\omega_0 = \gamma B_0
	\label{eq::Larmor-frequency}
	\end{align}	
	around the z-axis.
	
	Therefore the magnetization vector can be expressed in both systems
	\begin{align}
	\textrm{rot. frame:}~~ &\bm{M} = M_x\bm{\hat{e}}_x + M_y\bm{\hat{e}}_y + M_z\bm{\hat{e}}_z~~~\textrm{and}\\
	\textrm{lab. frame:}~~ &\bm{M}' = M_{x'}\bm{\hat{e}}_{x'} + M_{y'}\bm{\hat{e}}_{y'} + M_{z'}\bm{\hat{e}}_{z'},
	\end{align}
	using the basis vectors $\bm{\hat{e}}_i$ for $i\in$ ($x$, $y$, $z$, $x'$, $y'$, $z'$) and assuming $\bm{\hat{e}}_{z} = \bm{\hat{e}}_{z'}$, because the rotation is around the $z'$-axis. The relation between both notations is given by the rotation matrix $\mathfrak{R}$
	\begin{align}
	\begin{pmatrix}
	M_x\\ M_y\\	M_z
	\end{pmatrix}=
	\begin{pmatrix}
	\cos(\omega_0 t) & -\sin(\omega_0 t) & 0\\
	\sin(\omega_0 t) & \cos(\omega_0 t) & 0 \\
	0 & 0 & 1
	\end{pmatrix}
	\begin{pmatrix}
	M_{x'}\\ M_{y'}\\ M_{z'}
	\end{pmatrix} = \mathfrak{R}\bm{M}'.
	\label{eq::labToRotFrame_Matrix}
	\end{align}
	Following \cite{Schlichter__1990} the temporal changes between both frames for $\bm{M}$ are defined as
	\begin{align}
	\frac{\textrm{d}\bm{M}}{\textrm{d}t} = \frac{\textrm{d}\bm{M'}}{\textrm{d}t} - \bm{\Omega}\times\bm{M}\label{eq:labToRot_derivatives}\\
	\textrm{with} ~~~~ \bm{\Omega} = -\omega_0\bm{\hat{e}}_z,
	\label{eq:labToRot_Omega}
	\end{align}
	which will be needed in the following chapter.

	\subsection[Radio-Frequency Excitation]{Radio-Frequency Excitation}
	
\begin{figure}[!h]
		\centering
		\includegraphics[width=0.4\linewidth]{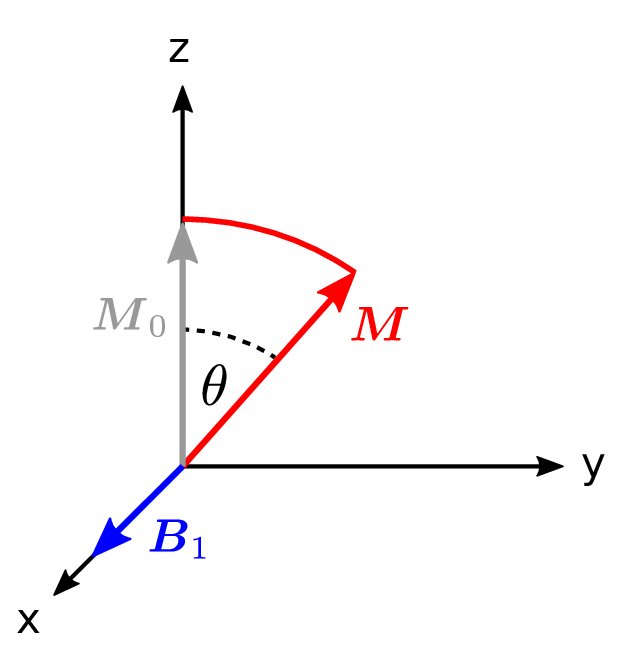}
		\caption{Visualization of the spin-excitation in the rotating frame. For on-resonant excitation $\bm{B}_1$ is constant in time, while $\bm{M}$ precesses around it. The flip-angle $\theta$ follows from equation \ref{eq::theory_flipangle}.}
		\label{fig::spin_precession}
	\end{figure}

	\textit{This section follows \cite[p.23, 26f, 69f]{Bernstein__2004}.}\\
	For the excitation of a spin system we use a radio-frequency pulse (RF-pulse) $\bm{B'_1}$ with the angular frequency $\omega_{\textrm{rf}}$ in transverse (xy-) plane initialized in x-direction without loss of generality. For the laboratory frame we get
	\begin{align}
	\bm{B'_1}(t) = B_1(t)\cos(\omega_{\textrm{rf}} t)~\bm{\hat{e}}_{x'}-B_1(t)\sin(\omega_{\textrm{rf}} t)~\bm{\hat{e}}_{y'}.
	\label{eq:labFrame_B1}
	\end{align}
	For simplification we now assume a spin system without relaxation and diffusion, which will be explained in the next sections. The temporal change of $\bm{M}$ follows with the precession around the total magnetization vector $\bm{B'}(t) = \bm{B_0} + \bm{B'_1}(t)$ through its angular momentum and becomes:
	\begin{align}
	\frac{\textrm{d}\bm{M'}}{\textrm{d}t} &= \gamma\bm{M}\times\bm{B'}(t)\nonumber\\
	&= \gamma \bm{M}\times \left[B_1(t)\cos(\omega_{\textrm{rf}} t)~\bm{\hat{e}}_{x'}-B_1(t)\sin(\omega_{\textrm{rf}} t)~\bm{\hat{e}}_{y'} + B_0\bm{\hat{e}}_z\right].
	\label{eq:labFrame_M}
	\end{align}
	This is still in the laboratory frame and has to be transformed into the rotating one for a more intuitive representation. In a first step equation \ref{eq:labFrame_M} is inserted into \ref{eq:labToRot_derivatives} which leads to
	\begin{align}
	\frac{\textrm{d}\bm{M}}{\textrm{d}t} = \gamma \bm{M}\times\left(B_1(t)\cos(\omega_{\textrm{rf}} t)~\bm{\hat{e}}_{x}-B_1(t)\sin(\omega_{\textrm{rf}} t)~\bm{\hat{e}}_{y} + B_0\bm{\hat{e}}_z + \frac{\bm{\Omega}}{\gamma}\right).
	\end{align}
	In a second one $\bm{B'_1}(t)$ is transformed into the rotating frame. Because this is done the same way as for $\bm{M}$, equation \ref{eq:labFrame_B1} is practically inserted into \ref{eq::labToRotFrame_Matrix}. The result combined with equation \ref{eq:labToRot_Omega} gives
	\begin{align}
	\frac{\textrm{d}\bm{M}}{\textrm{d}t} = \gamma \bm{M}\times
	\underbrace{\left(B_1(t)\cos([\omega_{\textrm{rf}}-\omega_0] t)~\bm{\hat{e}}_{x}-B_1(t)\sin([\omega_{\textrm{rf}}-\omega_0] t)~\bm{\hat{e}}_{y} + \bm{\hat{e}}_z \left(B_0 -\frac{\omega_{\textrm{rf}}}{\gamma}\right)
		\right)}_{ \textstyle B_{\textrm{eff}} },
	\label{eq:rotFrame_Beff}
	\end{align}
	where the part inside the large brackets is typically called effective field $B_{\textrm{eff}}$.\\
	For the on-resonant excitation the carrier frequency of the radio-frequency pulse is equal to the one of the rotating frame/spin-system: $\omega_{\textrm{rf}} = \omega_0$. Equation \ref{eq:rotFrame_Beff} therefore simplifies to
	\begin{align}
	\frac{\textrm{d}\bm{M}}{\textrm{d}t} = \gamma \bm{M}\times
	\left[\underbrace{B_1(t)~\bm{\hat{e}}_{x}}_{\textrm{rf-field}} + \underbrace{\bm{\hat{e}}_z \left(B_0 -\frac{\omega_{\textrm{rf}}}{\gamma}\right)}_{\textrm{freq. offset}} \right].
	\end{align}
	This shows the rotation of the magnetization vector around the axis of the RF-pulse like sketched in Figure \ref{fig::spin_precession}, but also indicates that an off-resonant excitation leads to an additional rotation around the $z$-axis, which results in a more complex temporal behavior of $\bm{M}$. A typical application for those pulses is MR-tagging \cite[p. 407ff]{Brown__2014}\cite{Nasiraei-Moghaddam_Magn.Reson.Med._2014}. 
	
	The flip-angle of the excitation is
	\begin{align}
	\theta(t) = \gamma\int\limits_{t'=0}^{t}B_1(t') \textrm{d}t'.
	\label{eq::theory_flipangle}
	\end{align}

	\subsection[Radio-Frequency Pulse-Shapes]{Radio-Frequency Pulse-Shapes}
	
\begin{figure}[!h]
		\centering
		\includegraphics[width=0.8\linewidth]{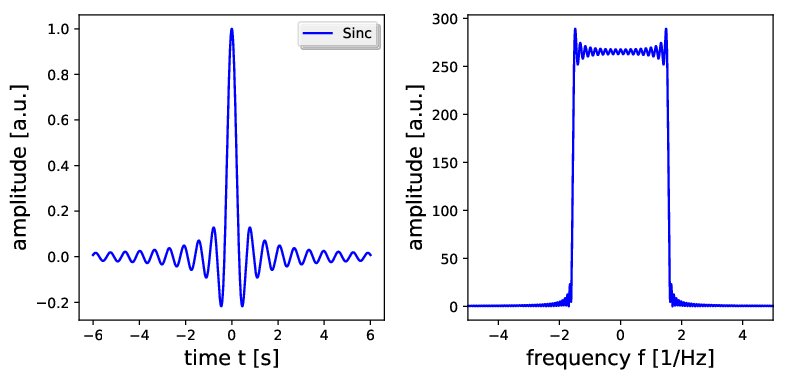}
		\caption{(\textbf{Left}) Figure illustrates a truncated sinc funtion (\textbf{left}) with its Fourier transform on the \textbf{right}. Following to equation \ref{eq::sinc} the defining parameters are: \mbox{$N_L=N_R=N$ = 6, $A$ = 1, $t_0$ = 1} and $\alpha$ = 0.}
		\label{fig::sinc_and_ft}
	\end{figure}

	\textit{This section follows \cite[p.38f]{Bernstein__2004}.}\\
	The most common radio-frequency pulse-shape is the sinc pulse shown in Figure \ref{fig::sinc_and_ft} left.
	Considering that the Fourier transform of the envelop of the RF-pulse defines its produced frequency profile, the infinite sinc-pulse in time domain would be ideal to reach a rectangular frequency response, which will become important for locating signals in MRI. The definition of the pulse follows \cite{Bernstein__2004}:
	\begin{align}
	B_1(t) = 
	\begin{cases}
	A~\textrm{sinc}\left(\frac{\pi t}{t_0}\right),~~~~~&-N_Lt_0 \leq t \leq N_Rt_0\\
	0,&\textrm{elsewhere}
	\end{cases},
	\label{eq::sinc}
	\end{align}
	with the peak amplitude $A$ at $t$=0, $N_L$ and $N_R$ as number of zero-crossings to the left and right of the main lope and $t_0$ as half width of the main lope and full width of the side lopes.
	
\begin{figure}[!h]
		\centering
		\includegraphics[width=0.8\linewidth]{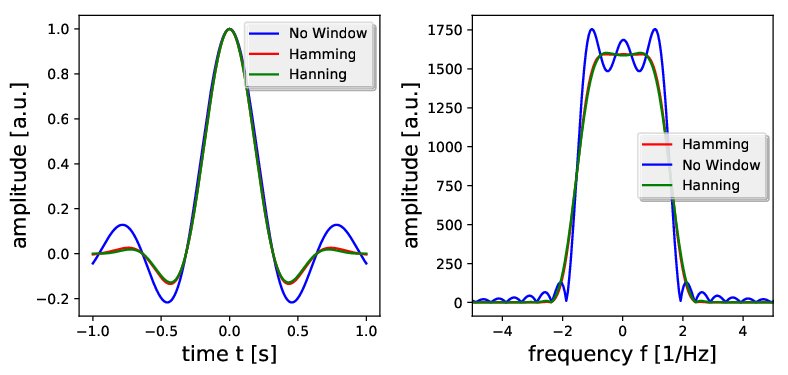}
		\caption{(\textbf{Left}) Figure illustrates a truncated sinc-funtion with its Fourier transformation for various windowing functions. (\textbf{Right}) The Fourier transformations are added. The simulation is based on equation \ref{eq::sinc} and the defining parameters are: \mbox{$N_L=N_R=N$ = 1, $A$ = 1, $t_0$ = 1} and $\alpha$ is adjusted related to the used window.}
		\label{fig::sinc_window_and_ft}
	\end{figure}

	In practice an infinite sinc-pulse is not realizable. Some outer lopes need to be cut out (see Figure \ref{fig::sinc_and_ft} left), which may lead to discontinuities in the derivative of the used pulse-shape. The resulting ringing at the edges of the Fourier transform (\ref{fig::sinc_and_ft} right) disturbs its later application in slice selection. \\
	One possible solution is the apodization of the pulse. This includes a window, which smooths the amplitude especially in the areas where the derivative shows discontinuities. This changes equation \ref{eq::sinc} to \cite{Bernstein__2004}
	\begin{align}
	B_1(t) = 
	\begin{cases}
	A\left[(1-\alpha)+\alpha\cos\left(\frac{\pi t}{Nt_0}\right)\right]~\textrm{sinc}\left(\frac{\pi t}{t_0}\right),~~~~~&-N_Lt_0 \leq t \leq N_Rt_0\\
	0,&\textrm{elsewhere.}
	\end{cases}
	\label{eq::rf_pulse_shape_sinc}
	\end{align}
	In addition to the previous described parameters equation \ref{eq::rf_pulse_shape_sinc} contains $N = \textrm{max}(N_L,N_R)$ and $\alpha$, which controls the windowing type. Possible values are $\alpha$ = 0/0.46/0.5 for no/Hamming/Hanning window. The different pulse-shapes and their Fourier transforms are shown in Figure \ref{fig::sinc_window_and_ft}.
	
	\subsection{Relaxation Parameters}
\begin{figure}[!h]
		\centering
		\includegraphics[width=0.6\linewidth]{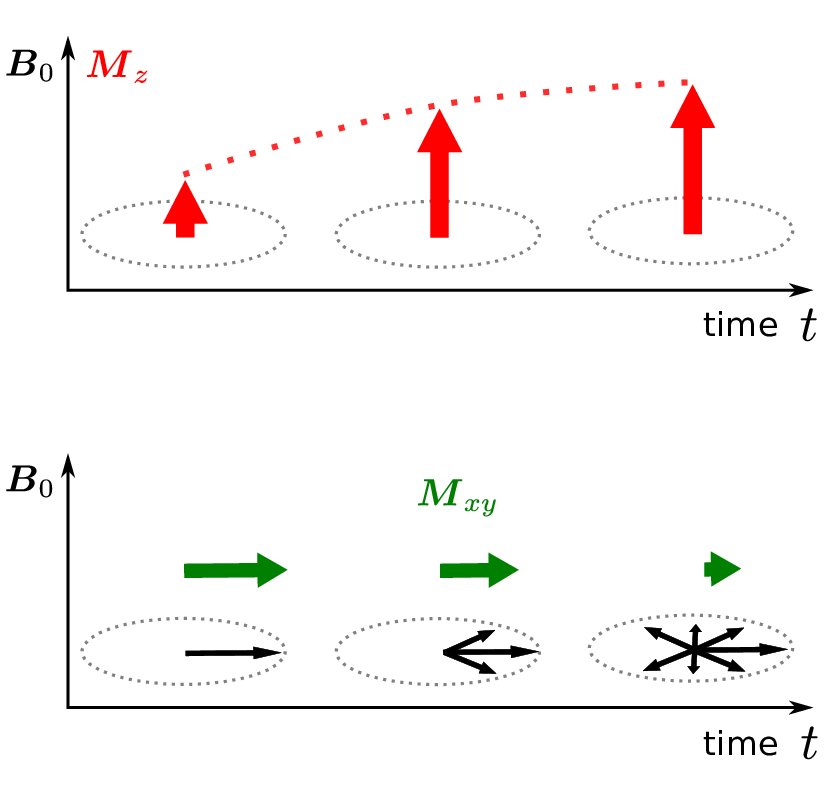}
		\caption{ Visualization of $T_1$ (\textbf{top}) and $T_2$ relaxation (\textbf{bottom}). The direction of the main magnetic field $\bm{B}_0$ and the projection of the spin-ensemble's magnetization on the z-axis ($\bm{M}_z$) and xy-plane ($\bm{M}_{xy}$) is added. The additionally illustrated temporal evolution of both projections enable a better understanding of the underlying effects of $T_1$ and $T_2$. }
		\label{fig::spin_relaxation}
	\end{figure}

	In the previous section the excitation of the net-magnetization $\bm{M}$ and its flipping into an excited state is described. Because this is not temporarily stable a relaxation needs to occur. This section focus on the process, which takes $\bm{M}$ back to its original equilibrium state. The effect can be split into two main subprocesses: the $T_1$- and the $T_2$-relaxation. A visualization of both is shown in Figure \ref{fig::spin_relaxation}.
		
	\subsubsection*{$T_1$-Relaxation}
	The $T_1$-relaxation describes the process of the transition of the longitudinal projection of $\bm{M}$ back to its original state before the excitation. It mostly relies on thermal effect where energy is transferred from the spins into their environment. Therefore it is also called spin-lattice relaxation.\\
	Figure \ref{fig::spin_relaxation} shows the increasing of the $M_z$ component in time. The underlying model is given by
	\begin{align}
		M_z = M_0\left(1-e^{-\frac{t}{\textrm{T}_1}}\right),
		\label{eq:signal_T1_relaxation}
	\end{align}
	with the original magnetization $M_0$ and the relaxation constant $T_1$.
	
	\subsubsection*{$T_2$-Relaxation}
	The $T_2$-relaxation, also called spin-spin-relaxation, relies on the interaction between the spins itself. This results in dephasing effects, which decrease the net $M_{xy}$ of the ensemble in the xy-plane over time. A visualization is shown in Figure \ref{fig::spin_relaxation}.\\
	The whole effect is described by
	\begin{align}
		M_{xy} = M_0 e^{-\frac{t}{\textrm{T}_2}},
		\label{eq:signal_T2_relaxation}
	\end{align}
	with the relaxation constant $T_2$.

	\subsubsection*{Other Relaxations}
	Besides the transitions described by $T_1$ and $T_2$, other relaxation processes are observable, too.\\
	One example is the $T_2^*$-relaxation which combines $T_2$ with an additional dephasing effect on the transversal magnetization through external field inhomogeneities. Other relaxations can results from special sequences like $T_1^*$ from Look-Locker methods \cite{Look_Rev.Sci.Instrum._1970}.

	\subsection{Bloch Equations}
	\label{ssec::Bloch_equations}
	The previously described excitation as well as the relaxation can be combined in one model which was introduced by Felix Bloch in 1946: the so called Bloch equations \cite{Bloch_Phys.Rev._1946}.
	A quantum mechanical derivation of this non-linear system of differential equations can be found in \cite{Block__2004}, but here the focus relies on the classical description.\\
	The equations are given by \cite{Bloch_Phys.Rev._1946}
	\begin{align}
		\frac{\textrm{d}\bm{M}}{\textrm{d}t} = \gamma\bm{M}\times\bm{B} - \frac{1}{{T}_2} \left(\bm{\hat{e}}_x \bm{M} + \bm{\hat{e}}_y \bm{M}\right) -\frac{\bm{\hat{e}}_z\bm{M}-M_0}{{T}_1},
		\label{eq::Bloch-Equation}
	\end{align}
	with
	\begin{align}
		\bm{M} = \begin{pmatrix}{M}_x \\ {M}_y \\ {M}_z \end{pmatrix}
		~~~\textrm{and}~~~
		\bm{B} = \begin{pmatrix}{B}_x \\ {B}_y \\ {B}_z \end{pmatrix}.
	\end{align}
	Here $\bm{M}$ describes the temporal evolution of the magnetization, $\bm{B}$ the external magnetic field as well as the gradients, $M_0$ the original equilibrium longitudinal magnetization and $\gamma$ the gyro-magnetic ratio of the excited particles.\\
	Equation \ref{eq::Bloch-Equation} assumes static spins. In cases of diffusion an additional term occurs following Torrey 1956 \cite{Torrey_Phys.Rev._1956}.

\section{Image Acquisition}
	After the introduction into the physical principles of magnetic resonance in the previous chapter, the following should lead to a better understanding of the imaging part in MRI. Therefore the required technical components are introduced and the mathematical methods of the spatial encoding are described.

	\subsection{Setup MRI Scanner}
\begin{figure}[!h]
		\centering
		\includegraphics[width=0.7\linewidth]{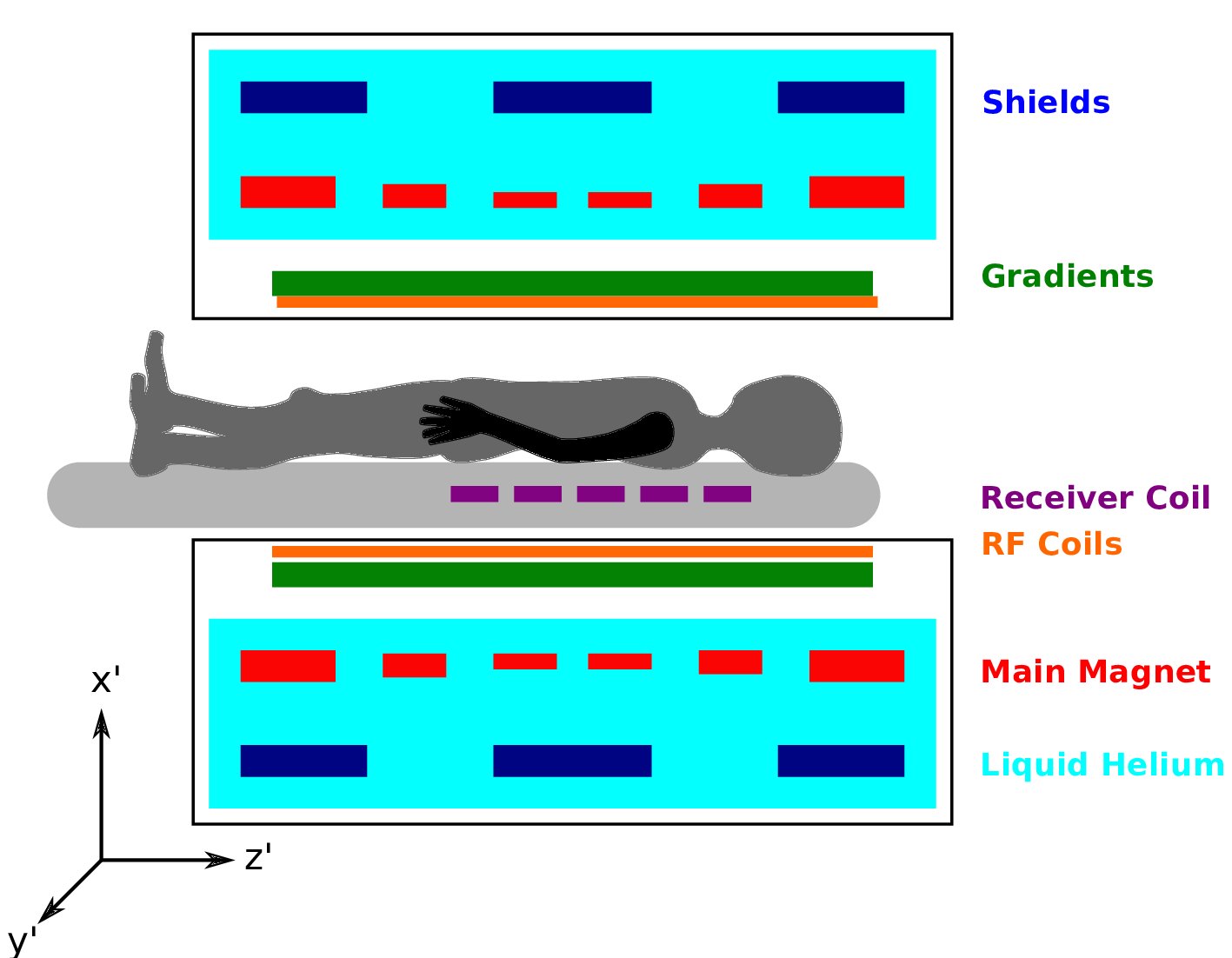}
		\caption{Visualization of the main technical components of a MRI-scanner. Additionally, the coordinates of the laboratory frame are added.}
		\label{fig::mriscanner}
	\end{figure}

	A MRI scanner consists technically of some main parts, like shown in Figure \ref{fig::mriscanner}:  Here the magnet, which is responsible for the main magnetic field, three gradients, one per spatial directions ($x'$, $y'$, $z'$), a transmitting and receiving radio-frequency coil are the most important ones for the spatial encoding.\\
	In the following chapters we define the main magnetic field to be orientated in z-direction. 
		
	Additionally, there are systems for shimming the main magnetic field, which leads to a more homogeneous $B_0$ inside the bore during the measurement as well as shields for reducing the impact of switching gradients on the electronic parts. One other essential part is the cooling. Most MR-systems are large tanks filled with liquid helium, where the magnet is placed in to reach superconductivity.\\

	\subsection{Spatial Encoding}
	The spatial encoding in MRI was first introduced by P.C. Lauterbur in 1973 \cite{Lauterbur_Nature_1973}. It is based on changing the precession-frequency of the magnetized tissue locally by applying additional magnetic field gradients in all spatial dimensions. It is divided into three main techniques: the slice selection, the frequency- and the phase-encoding. Each of them encodes one dimension of the signal.

	\subsubsection[Slice Selection]{Slice Selection}
\begin{figure}[!h]
			\centering
			\includegraphics[width=0.8\linewidth]{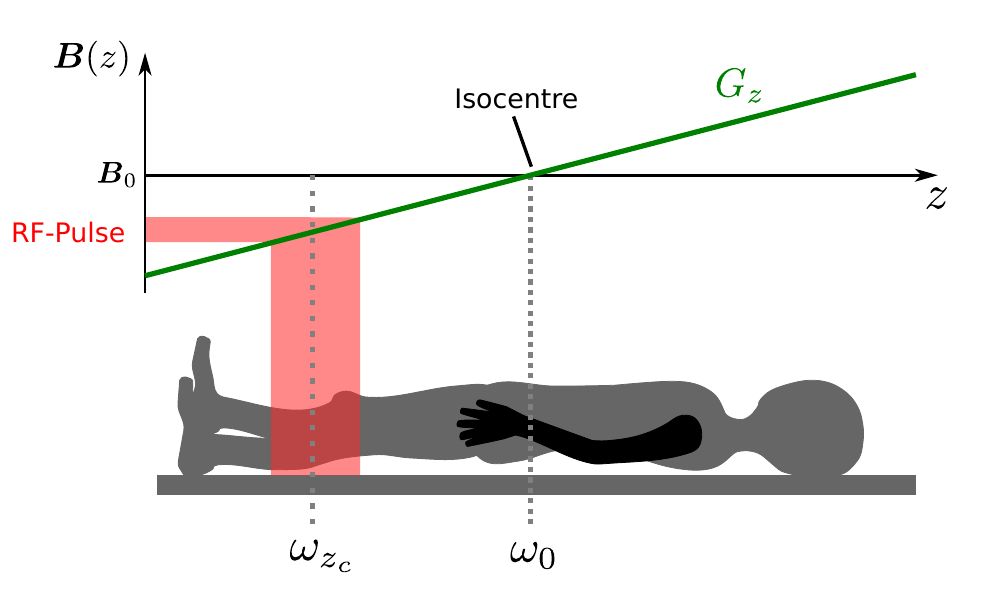}
			\caption{Visualization of the slice selection process in MRI. A patient along the $z$-axis centered around the isocenter is presented. While the spins are precessing with the Larmor frequency $\omega_0$ in the isocenter, the magnetic field $\bm{B}$ farther outside is modified by the gradient $G_z$ depending on the position of the voxel along $z$. Therefore, the resonance-frequency band centered around $\omega_{z_c}$ can be excited by an RF-pulse.  }
			\label{fig::theory_sliceselection}
		\end{figure}

	\textit{This section follows \cite[p.145f]{Liang__1999} and \cite{Rosenzweig__2016_2}.}\\
		During slice selection the origin of a signal contributor is localized within a slice. Assuming linear gradient systems, the realization is based on a gradient in the direction of the slices normal vector $\bm{\hat{e}}_s$: $\bm{G_s} = G_s \bm{\hat{e}}_s$. At the center of its gradient $\bm{r_c}$ the total magnetic field is still equal to $B_0$, while further regions are increased or reduced by fractions of $G_s$ depending on their position $\bm{r}$. This results in an additional magnetic field $\bm{B_s}(\bm{r})$
		\begin{align}
			\bm{B_s}(\bm{r}) = \bm{G_s}~\bm{r}~\bm{\hat{e}}_z,
		\end{align}
		which then leads to a total magnetic field at the location $\bm{r}$ assuming a high magnetic field strength ($\approx$ 3 T) to neglect concomitant fields \cite[p.292f]{Bernstein__2004} :
		\begin{align}
			\bm{B}(\bm{r}) = \bm{B_0} + \bm{B_s}(\bm{r}) = (B_0 + \bm{G_s}\bm{r})\bm{\hat{e}}_z.
			\label{eq:local_Bfield}
		\end{align}
		This directly influences the precession frequency at $\bm{r}$: $\omega_{\bm{r}} = \gamma |\bm{B}(\bm{r})|$ and leads together with the definition of the Larmor frequency \ref{eq::Larmor-frequency} to
		\begin{align}
			\omega_{\bm{r}} = \omega_0 + \gamma \bm{G_s}\bm{r}.
			\label{eq::slice_selection_larmor_freq}
		\end{align}
		\\
		Without loss of generality, we assume slice selection in $z$-direction: $\bm{G_s} = G_z~\bm{\hat{e}}_z$. Following equation \ref{eq:local_Bfield} the local magnetic field then depends only on $z$
		\begin{align}
			\bm{B}(z) = (B_0 + G_z z)~\bm{\hat{e}}_z,
		\end{align}
		and the Larmor frequency of the central position of the slice $\bm{z_c}$ becomes
		\begin{align}
			\omega_{z_c} = \gamma (B_0+G_z z_c),
		\end{align}
		visualized in Figure \ref{fig::theory_sliceselection}. The dependency of the precession-frequency on the location along the $z$-axis is then often used to excite one single slice, but multiple excited slices are also possible \cite{Maudsley_J.Magn.Reson._1980}\cite{Mueller_Magn.Reson.Med._1988}\cite{Larkman_J.Magn.Reson.Imaging_2001}.
		To reach an excitation, the envelope frequency of the RF-pulse needs to be adjusted to match the Larmor frequency of the desired slice. The bandwidth together with the gradient $\bm{G}_z$ define how thick it will be. Afterwards, only the slice is excited and the spatial location in z-direction of the received signal is known.		
			
	\subsubsection[Frequency-Encoding]{Frequency-Encoding}
\begin{figure}[!h]
			\centering
			\includegraphics[width=0.6\linewidth]{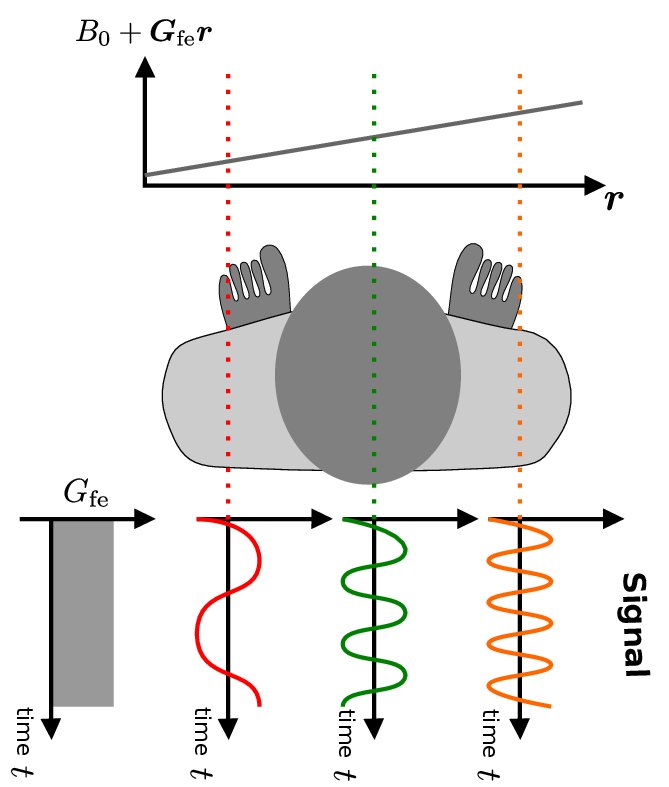}
			\caption{Visualization of the frequency encoding in MRI. The additionally applied gradient $\bm{G}_{\textrm{fe}}$ encodes the tissue along $\bm{r}$ to send out signals with different frequencies depending on $\bm{r}$. Afterwards, they can be separated using a Fourier transformation.}
			\label{fig::frequency_encoding}
		\end{figure}

		\textit{This section follows \cite[p.153f]{Liang__1999}.}\\
		The frequency encoding is used after the slice selection. The detected signal becomes a superposition of all signals from every voxel within the excited volume $\rho$. To disentangle the different signals along one direction $\bm{\hat{e}}_{\textrm{fe}}$ and map them to their real positions the frequency encoding is used. It is based on a gradient $\bm{G_{\textrm{fe}}}$ which modifies the frequency of the voxel-signals with their spatial position inside of the measured object $\rho$. The local Larmor frequency becomes similar to equation \ref{eq::slice_selection_larmor_freq}:
		\begin{align}
			\omega_r = \omega_0 + \gamma\bm{G_{\textrm{fe}}} \bm{r}.
		\end{align}
		This assignment of frequencies $\omega_r$ to spatial positions $\bm{r}$, is the reason why the encoding is called \textit{frequency encoding}.\\
		The resulting local signal $\textrm{d}S$ of the volume $\textrm{d}\bm{r}$ becomes \cite{Liang__1999}
		\begin{align}
			\textrm{d}S(\bm{r},t)=\rho(\bm{r})\textrm{d}\bm{r}~\textrm{e}^{-i\gamma \left(B_0+\bm{G_{\textrm{fe}}} \bm{r}\right)t}.
		\end{align}
		Effects of flip-angles and field strength of $B_0$ on this parameter are neglected to simplify the notation.\\
		For the whole object $\rho$ we get the signal
		\begin{align}
			S(t) = \int_{\textrm{object}}\textrm{d}S(\bm{r},t) = \int\limits_{-\infty}^{\infty}\rho(\bm{r})\textrm{e}^{-i\gamma(B_0+\bm{G_{\textrm{fe}}} \bm{r})t}\textrm{d}\bm{r},
		\end{align}
		where the carrier signal can be excluded and later demodulated:
		\begin{align}
			S(t) =\left[ \int\limits_{-\infty}^{\infty}\rho(\bm{r})\textrm{e}^{-i\gamma\bm{G_{\textrm{fe}}} \bm{r}t}\textrm{d}\bm{r}\right]\textrm{e}^{-i\omega_0 t}
			\underbrace{\Rightarrow}_{\textrm{demodulation}}
			\tilde{S}(t) = \int_{\textrm{object}}\rho(\bm{r})\textrm{e}^{-i\gamma\bm{G_{\textrm{fe}}} \bm{r}t}\textrm{d}\bm{r}.
			\label{eq::Freq_encode_signal_equation}
		\end{align}
		The whole process of frequency encoding is visualized in Figure \ref{fig::frequency_encoding}.
			
	\subsubsection[Phase-Encoding]{Phase-Encoding}
\begin{figure}[!h]
			\centering
			\includegraphics[width=0.6\linewidth]{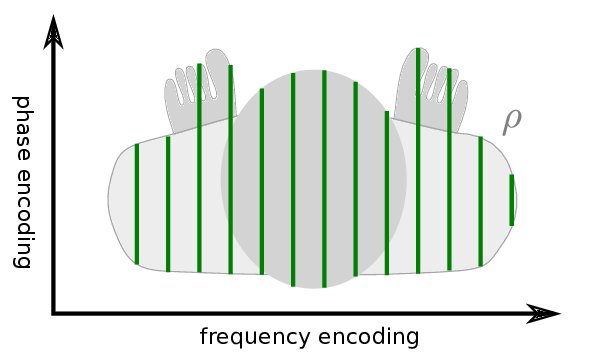}
			\caption{Visualization of the isolines created through frequency encoding within an object $\rho$. Therefore, the signal of the green marked lines is superposed to one output. For a more accurate localization, the direction along the isolines needs to be encoded.}
			\label{fig::frequency_encoding_iso_lines}
		\end{figure}

	\textit{This section follows \cite[p.155ff]{Liang__1999}.}\\
		The slice and frequency encoding combined will lead to signal superpositions from different isofrequency lines of one slice $\rho$ of an object, like visualized in Figure \ref{fig::frequency_encoding_iso_lines}.
		
		To locate individual signals from voxels along these lines phase encoding is used. Therefore a gradient $\bm{G_p}$ is turned on for a time interval $T_p$. This changes the local signal to
		\begin{align}
			\textrm{d}{S}(\bm{r},t) = 
			\begin{cases}
			\rho(\bm{r})~\textrm{e}^{-i\gamma(B_0+\bm{G_p}\bm{r})t},~~~&0\leq t\leq T_p\\
			\rho(\bm{r})~\textrm{e}^{-i\gamma\bm{G_p}\bm{r}T_p}~\textrm{e}^{-i\gamma B_0 t},~~~&T_p\leq t
			\end{cases}
		\end{align}
		and shows how the gradient adds an additional phase $\phi(\bm{r})$ to the individual voxel at location $\bm{r}$:
		\begin{align}
			\phi(\bm{r}) = -\gamma\bm{G_p}\bm{r}T_p.
		\end{align}
		This linear behavior of the voxels additional phase to their position is the reason why it is called \textit{phase encoding}.\\
		In the view of the whole object $\rho$ the signal becomes similar to the case of frequency encoding and demodulation:
		\begin{align}
			S(t) = \left[\int_{\textrm{object}}\rho(\bm{r})\textrm{e}^{-i\gamma\bm{G_p} \bm{r}T_p}\textrm{d}\bm{r}\right]\textrm{e}^{-i\omega_0 t}
			\underbrace{\Rightarrow}_{\textrm{demodulation}}
			\tilde{S}(t) = \int_{\textrm{object}}\rho(\bm{r})\textrm{e}^{-i\gamma\bm{G_p} \bm{r}T_p}\textrm{d}\bm{r}.
			\label{eq::phase_encode_signal_equation}
		\end{align}
			
		\begin{figure}[!h]
			\centering
			\includegraphics[width=0.9\linewidth]{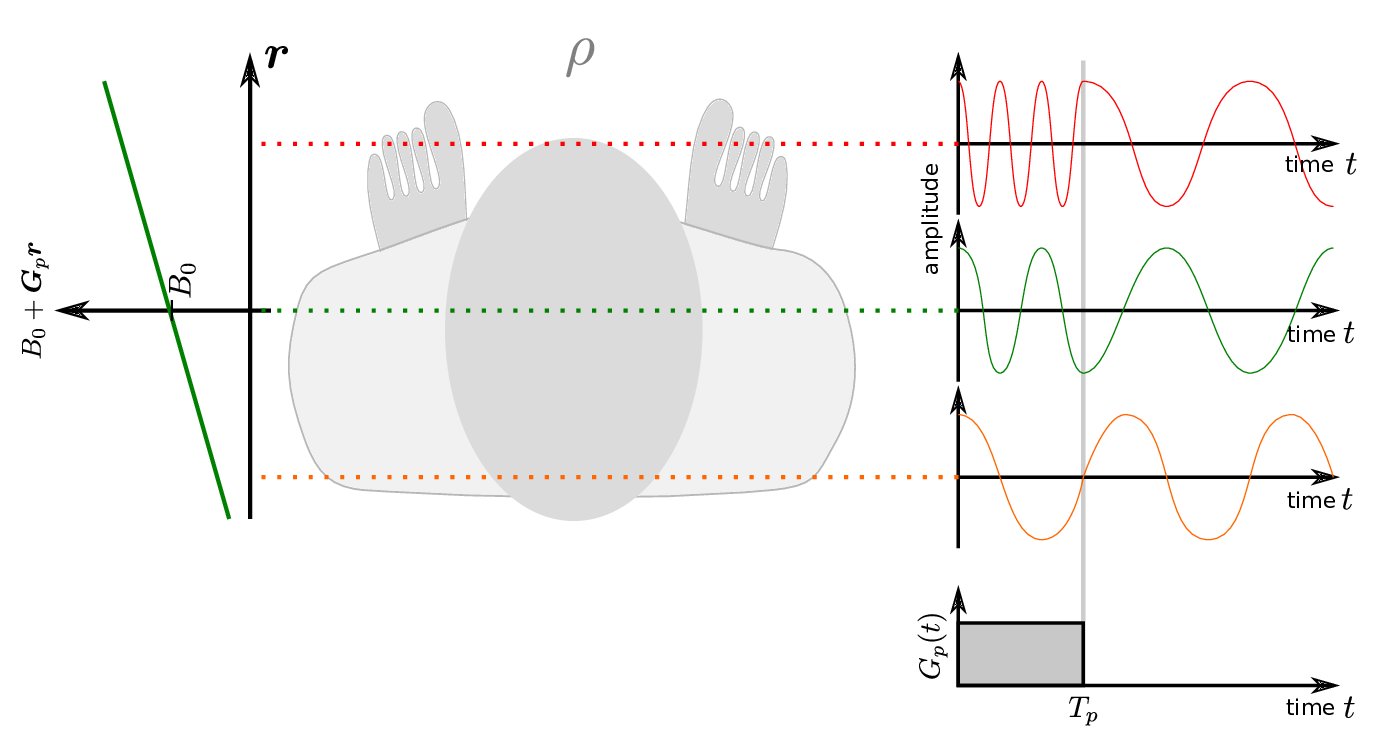}
			\caption{Visualization of the phase encoding in MRI. The gradient $\bm{G}_p$ adds an additional phase to the signal acquired after $T_p$ depending on its position in $\rho$ along $\bm{r}$. }
			\label{fig::theory_phase_encoding}
		\end{figure}
	
		To explain how this additional phase, shown in Figure \ref{fig::theory_phase_encoding}, can help to differentiate between signals from one iso-line, the k-space perspective is explained in the following.

	\subsubsection[Spatial Encoding in k-Space Perspective]{Spatial Encoding in k-Space Perspective}
		\label{sssec: spatial_encoding_kspace_perspective}
		
\begin{figure}[!h]
			\centering
			\includegraphics[width=0.8\linewidth]{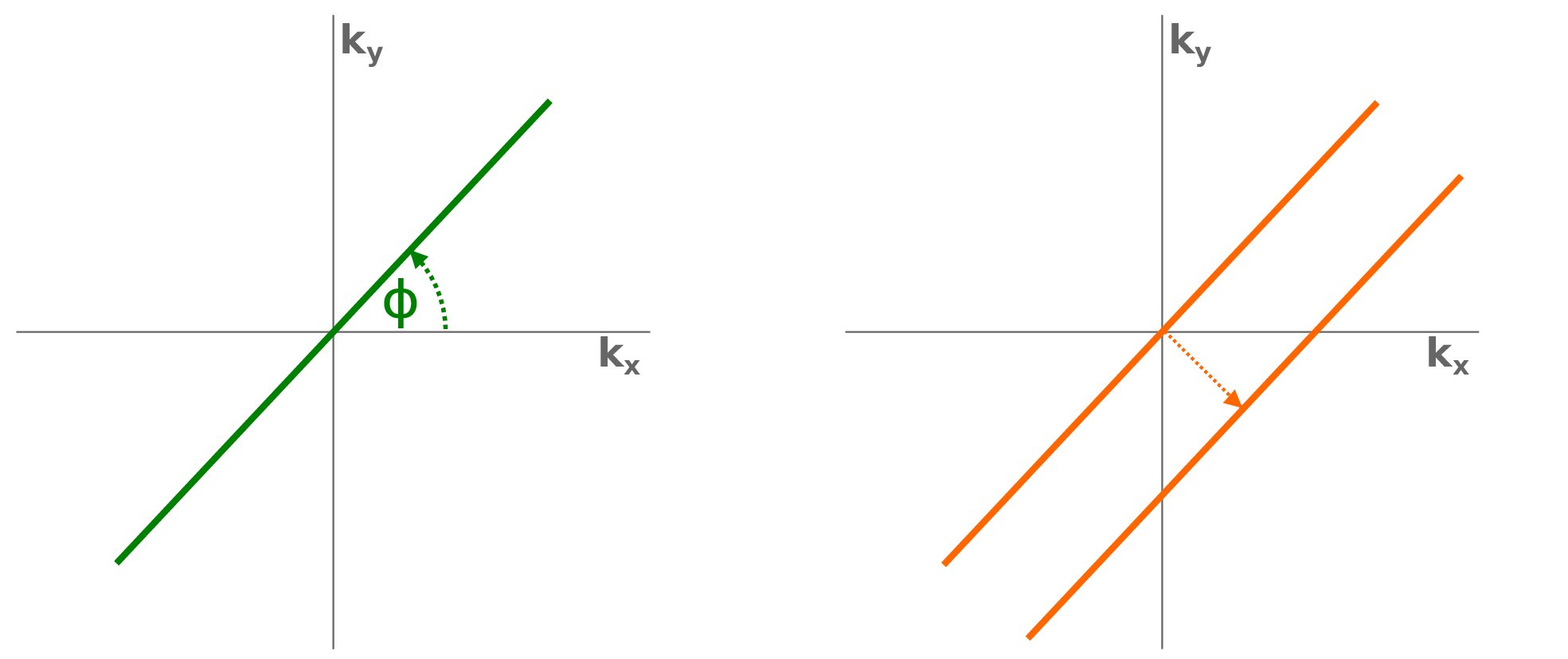}
			\caption{ Figure visualizes the effect of the frequency encoding on the sampling trajectory on the \textbf{left}. On the \textbf{right} the effect of the phase-encoding is illustrated. }
			\label{fig::sampling_trajectory}
		\end{figure}

		\textit{This section follows \cite[p.158ff]{Liang__1999}.}\\
		After discussing the basic concepts of slice-, frequency- and phase-encoding, we are going to interpret the two latter ones in a k-space perspective. This improves the understanding of how the k-space is sampled and how various pattern are created.\\
		The k-space is a collection of different spatial frequencies in the later reconstructed image. In a two-dimensional case their positions are encoded with $k_x$ and $k_y$ where low frequencies are encoded in the center, and higher ones in the outer regions like shown in Figure \ref{fig::sampling_trajectory}. \\
		To determine the effect of \textbf{frequency encoding}, equation \ref{eq::Freq_encode_signal_equation} is used and substituted with
		\begin{align}
			\bm{k}=
			\begin{cases}
			\frac{\gamma}{2\pi}~\bm{G_{\textrm{fe}}}t~~~~~~&\textrm{FID signals}\\
			\frac{\gamma}{2\pi}~\bm{G_{\textrm{fe}}}(t-\textrm{TE})~~~~~~&\textrm{echo signals}.
			\end{cases}
			\label{eq::spatial_encoding_freq_gradient_transform}
		\end{align}
		In equation \ref{eq::spatial_encoding_freq_gradient_transform} the concepts of a free induction decay FID and echo signal are introduced to differentiate between the two possible trajectories $\bm{k}$. While the FID signal occurs directly after excitation, the echo describes the same after refocusing and measured at the echo time TE. So the difference between both in equation \ref{eq::spatial_encoding_freq_gradient_transform} is the temporal shift by TE. The central position at TE corresponds to the initial one of the FID signal.\\
		All in all, substituting equation \ref{eq::spatial_encoding_freq_gradient_transform} in \ref{eq::Freq_encode_signal_equation} results in 
		\begin{align}
			S(\bm{k}) = \int_{\textrm{object}}\rho(\bm{r})\textrm{e}^{-2\pi i\bm{k} \bm{r}}\textrm{d}\bm{r}.
			\label{eq::kspace_traj_equation_frequency_encoding}
		\end{align}
		The gradient $\bm{G_{\textrm{fe}}}$ maps the time to the k-space coordinates $\bm{k}$, which are then called \textit{sampling trajectories}. Additionally, equation \ref{eq::kspace_traj_equation_frequency_encoding} shows that the signal in k-space coordinates corresponds to a Fourier transform of the object $\rho$. \\
		To give an example for the echo signal we assume a two dimensional frequency-encoding:
		\begin{align}
			~\begin{cases}
			k_x = \frac{\gamma}{2\pi}~G_x (t-\textrm{TE})\\
			k_y = \frac{\gamma}{2\pi}~G_y (t-\textrm{TE}),\\
			\end{cases}
		\end{align}
		which can be transformed with $k = \frac{\gamma}{2\pi}~(t-\textrm{TE})\sqrt{G_x^2+G_y^2}$ and $\phi = \tan{-1}\left(\frac{G_y}{G_x}\right)$ to reach
		\begin{align}
			~\begin{cases}
			k_x = k\cos\phi\\
			k_y = k\sin\phi.\\
			\end{cases}
		\end{align}
		For constant gradients the encoding follows a straight line with an angle of $\phi$ in k-space. Figure \ref{fig::sampling_trajectory} illustrates the position in k-space of this sampling trajectory. For the echo signal used in this work the line goes straight through the center, while for a FID one would start at $(k_x,k_y)$=(0,0). 
			
		To create curved lines in k-space, like for example for spiral trajectories, varying gradients are necessary. With non-linear ones even curved slices are possible.\\
		\\
		After the effect of the frequency encoding is discussed in a k-space perspective, now the same is done with the \textbf{phase encoding}.
		Therefore equation \ref{eq::phase_encode_signal_equation} is used and substituted by 
		\begin{align}
			\bm{k}=\frac{\gamma}{2\pi}~\bm{G_p}T_p,
			\label{eq::spatial_encoding_phase_gradient_transform}
		\end{align}
		which leads to
		\begin{align}
			S(\bm{k})=\int_{\textrm{object}}\rho(\bm{r})\textrm{e}^{-2\pi i\bm{k}\bm{r}}\textrm{d}\bm{r}.
		\end{align}
		Again the gradient $\bm{G_p}$, but also the time $T_p$ for how long it is turned on, influences the final k-space position. If this relation is even further generalized to arbitrary gradient shapes, we get 
		\begin{align}
			\bm{k}=\frac{\gamma}{2\pi}~\int\limits_0^{T_p}\bm{G_p}(\tau)\textrm{d}\tau.
		\end{align}
		Both equations show that $\bm{k}$ is no longer a function of time like for frequency encoding (see equation \ref{eq::spatial_encoding_freq_gradient_transform}). This means that the phase-encoding only influences the start position from which the sampling trajectory starts, but not its temporal behavior.

	\subsection{Sequences}
	
		Sequences in magnetic resonance imaging define the protocol the scanner is using. They describe how gradients are switched and RF-pulses are applied, where breaks are done, and if triggers like electrocardiograms are used. Basically, they define which kind of signal is acquired.\\
		By varying sequences, a lot of different physical effects can be examined by exploiting the freedom of modifying the emitted signal.\\
		This section introduces some important sequences which are used in this work.

		\subsubsection[Spin-Echo Sequence]{Spin-Echo Sequence}
		
\begin{figure}[!h]
			\centering
			\includegraphics[width=\linewidth]{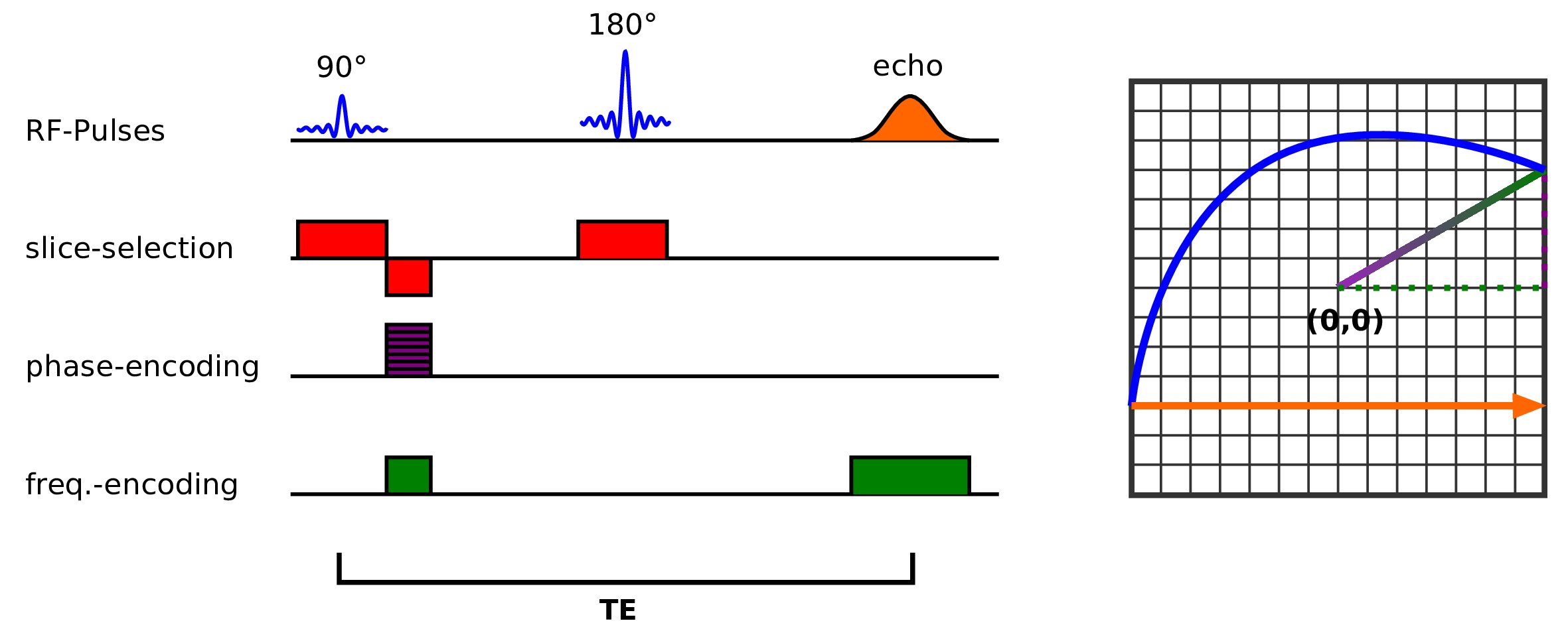}
			\caption{ The \textbf{right} side visualizes the gradient switching and RF-pulse timing of a typical Cartesian spin-echo sequence. The \textbf{left} side illustrates the evolution in k-space during a Cartesian spin-echo sequence. While the starting point is set to the center directly after the RF-pulse, the phase- and frequency-encoding gradients move its position to the upper left side of the k-space. Afterwards, a blue added inversion pulse reflects the current position on the center and finally, the frequency encoding gradient during the echo allows the sampling of the orange marked horizontal line. }
			\label{fig::sequence_spin_echo}
		\end{figure}

		The notation and structure of this section follows \cite[p.630ff]{Bernstein__2004}.\\
		The spin-echo-sequence visualized in Figure \ref{fig::sequence_spin_echo} is one of the most robust sequences against blurring, ghosting, and off-resonance artifacts in MRI. It was originally described by Hahn in 1950 \cite{Hahn_Phys.Rev._1950}. 
			
		Its typical structure contains spin excitations with 90° as well as refocusing pulses which have 180° flip-angles.\\
		Compared to other sequences the required long repetition time (TR) is a disadvantage because it leads to long acquisition times. Each repetition of the sequence needs to start from the same original state and the by 90° flipped magnetization needs a lot of time to recover. Thus the measurements need to be long. There are improvements by using interleaved schemes in 3D imaging where other slices are acquired, while waiting for recovering, but this is not working for single-slices. On the other hand, this sequence provides the clinically useful feature to adjust the image contrast by changing TR and the echo time (TE). The origin of this flexibility can be derived from the Bloch equations introduced in section \ref{ssec::Bloch_equations} and leads to a signal intensity $I_{SE}$ of \cite{Perman_Magn.Reson.Imaging_1984}
		\begin{align}
			I_{SE} = M_0\cdot \textrm{e}^{-\frac{\textrm{TE}}{T_2}}\left(1-2\textrm{e}^{-\frac{\textrm{TR}-\textrm{TE}/2}{T_1}}+\textrm{e}^{-\frac{\textrm{TR}}{T_1}}\right),
			\label{eq::intensity_spin_echo}
		\end{align}
		where $M_0$ describes the proton density.\\
		In practice, we can simplify equation \ref{eq::intensity_spin_echo} even further because a typical experiment fulfills TE $\ll$ TR, which leads to
		\begin{align}
			I_{SE} = M_0\cdot \textrm{e}^{-\frac{\textrm{TE}}{T_2}}\left(1-\textrm{e}^{-\frac{\textrm{TR}}{T_1}}\right).
			\label{eq::spin_echo_signal_equation}
		\end{align}
		The different contrast options based on equation \ref{eq::spin_echo_signal_equation} are shown in Table \ref{tab::spin_echo_contrast_values}.
			
		\begin{table}[!h]
			\centering
			\caption{Table with the different contrast types for a typical spin-echo-sequence in relation to TE and TR following \cite[p.631]{Bernstein__2004}.}
			\begin{tabular}{c|c|c}\addlinespace[2ex]
				& Short TE ($\leq$ 20 ms) & Long TE ($\geq$ 80 ms)\\\hline
				Short TR ($\leq$ 700 ms) & $T_1$-weighted & not used\\\hline
				Long TR ($\geq$ 2000 ms) & Proton density-weighted & $T_2$-weighted\\
			\end{tabular}
			
			\label{tab::spin_echo_contrast_values}
		\end{table}
			
\begin{figure}[!h]
			\centering
			\includegraphics[width=\linewidth]{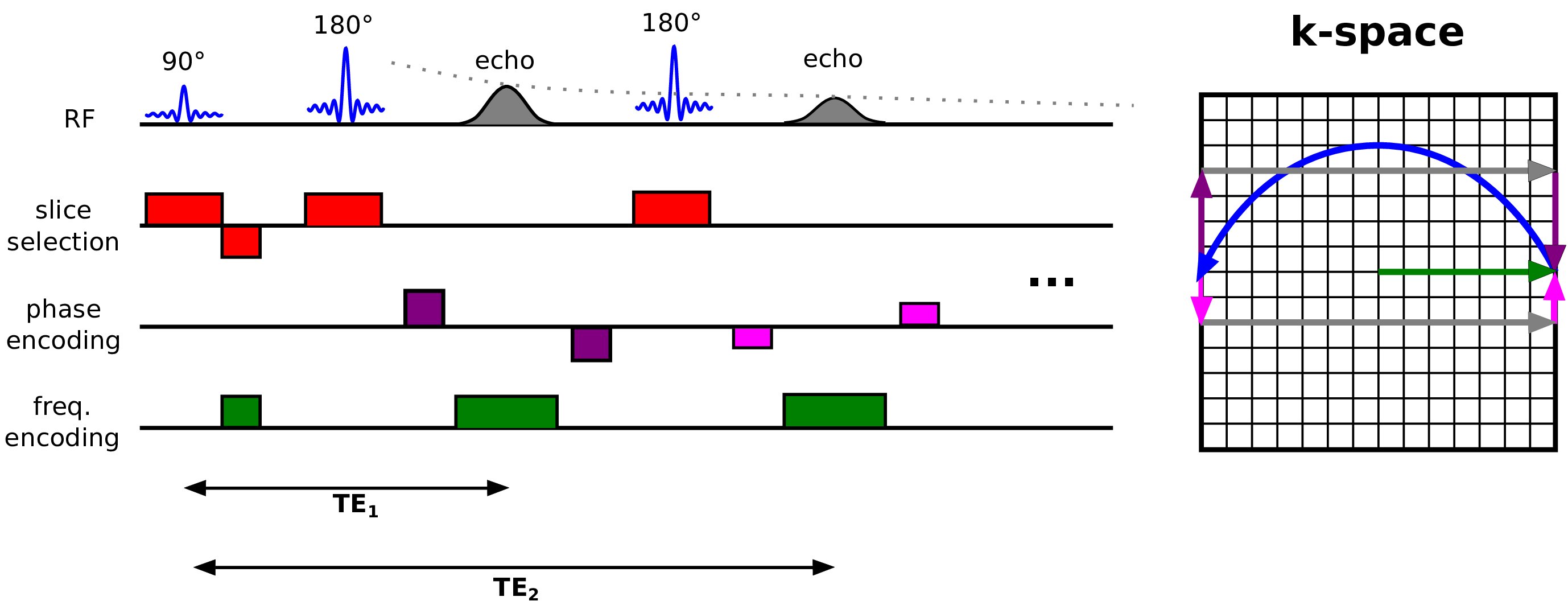}
			\caption{The \textbf{left} side visualizes the gradient switching and RF-pulse timing of a typical Cartesian multi-echo spin-echo sequence. The \textbf{right} side illustrates the evolution of the multi-echo spin-echo sequence analogue to figure \ref{fig::sequence_spin_echo}. }
			\label{fig::sequence_fast_spin_echo}
		\end{figure}

		The echo peak forms at TE because the magnetization is refocused through the 180°-pulse after being defocused during the first frequency encoding gradient.\\		
		In addition to the single-echo spin-echo shown in Figure \ref{fig::sequence_spin_echo}, there are also variants with multiple echoes (compare Figure \ref{fig::sequence_fast_spin_echo}), where each one has its own TE and therefore $T_2$-contrast during the acquisition.
			
		\subsubsection{Gradient-Echo Sequences}
\begin{figure}[!h]
			\centering
			\includegraphics[width=\linewidth]{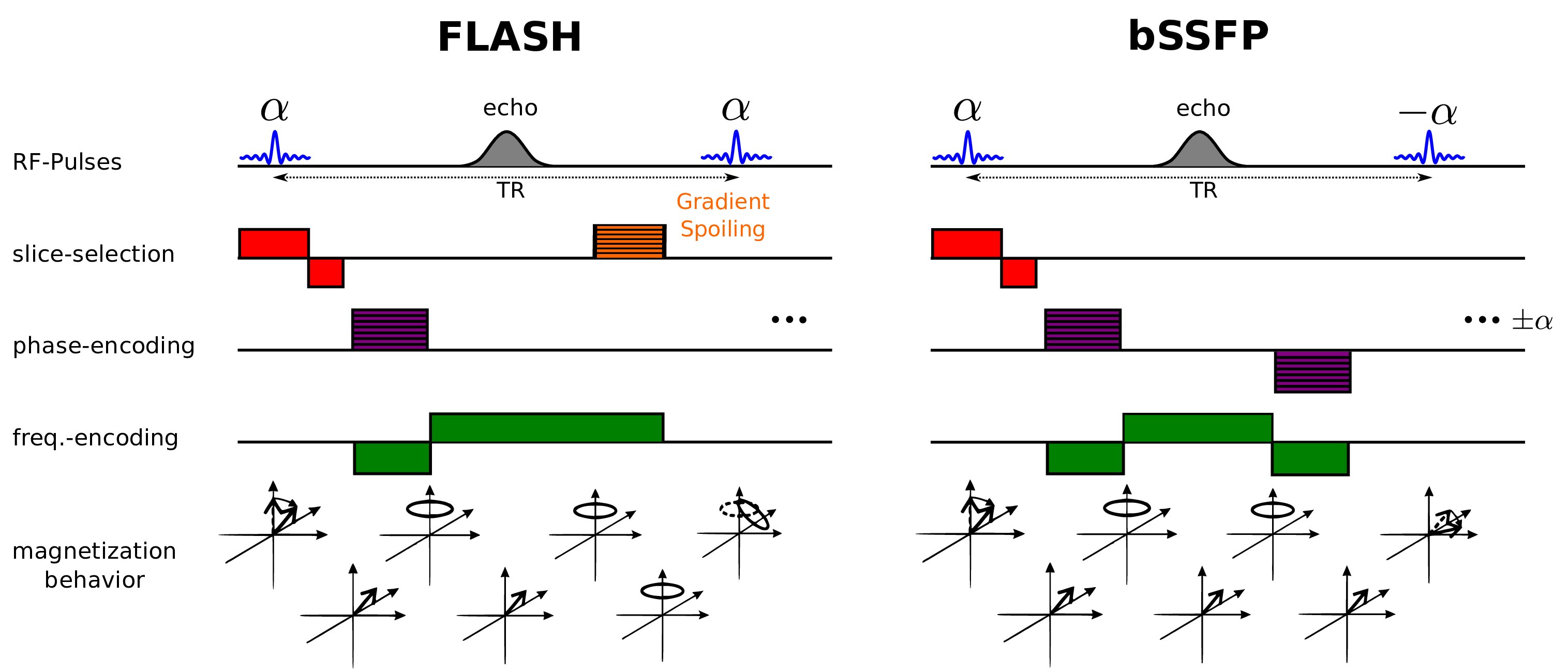}
			\caption{The \textbf{left} side visualizes the gradient switching and RF-pulse timing of a gradient-spoiled FLASH sequence compared to a bSSFP on the \textbf{right}. Additionally, the development of magnetization vector during the sequences adapted from \cite{Scheffler_Eur.Radiol._2003} is added. }
			\label{fig::sequence_flash_bSSFP}
		\end{figure}

		The second class of sequences beside the spin-echo ones is called gradient-echo. Basically, they avoid the radio-frequency based 180°-refocusing pulse by using gradients, like show in Figure \ref{fig::sequence_flash_bSSFP}. This allows to use smaller flip-angles which decreases the injected energy into the patient and reduces the measurement time.\\
		Most gradient-echo techniques are based on a steady-state following their synonym: steady-state-free-precession (SSFP). The equilibrium-state typically needs some repetitions of applied RF-pulses combined with the tissue relaxation to be established but produces a constant signal afterwards. Additionally, the contrasts of these sequences can be modified. With proper selection of TR, TE, gradient balancing, and spoiling type, the resulting images can have $T_1$, $T_2$, $T_2^*$ or even $\frac{T_1}{T_2}$ contrast \cite[p.225]{Liang__1999}.\\
		The two main groups of gradient-echo sequences are the spoiled (by gradient or RF-pulse) as well as the balanced SSFP sequences, which are going to be discussed in the following based on \cite{Hargreaves_J.Magn.Reson.Imaging_2012}. Others are multi-acquisition SSFPs, dual-echo-SSFPs, and combinations of SSFPs with FID signals. For the interested reader an overview including their different names by varying vendors is given in \cite[p.584ff]{Bernstein__2004}.
					
		The \textbf{spoiled SSFP} sequence, like Fast Low-Angle Shot FLASH, has a constant rf-phase, which means that the direction the magnetization is flipped in is constant. Their flip-angles are typically lower than 10° and their repetitions times are under 5 ms \cite{Haase_Magn.Reson.Med._1990}. After each TR, their whole transversal magnetization is spoiled away by either modifying the radio-frequency pulses or applying gradients like shown in Figure \ref{fig::sequence_flash_bSSFP}. Assuming a perfect spoiling, the signal amplitude for a FLASH measurement $S_{\textrm{FLASH}}$ can be derived from the Bloch equations and becomes \cite{Haase_Magn.Reson.Med._1990}
		\begin{align}
			S_{\textrm{FLASH}} = M_0~\frac{\left(1-E_1\right)}{1-\cos\alpha~E_1}\sin\alpha~\textrm{e}^{-\frac{\textrm{TE}}{T_2^*}}
		\end{align}
		with $E_{1}=\textrm{e}^{-\frac{\textrm{TR}}{T_{1}}}$ and the flip-angle $\alpha$. To maximize  $S_{\textrm{FLASH}}$ the flip-angle $\alpha$ is chosen to be the Ernst angle $\alpha_E$:
		\begin{align}
			\alpha_E = \arccos\left(\textrm{e}^{-\frac{\textrm{TR}}{T_1}}\right).
		\end{align}
		\\
		The other important group of gradient-echo sequences is the \textbf{balanced SSFP}, also shown in Figure \ref{fig::sequence_flash_bSSFP} . Because it is used in this work, it is discussed in more detail in the following. 	
			
		\subsubsection{Balanced Steady-State Free Precession}
		\label{ssec::theory_bSSFP}
\begin{figure}[!h]
			\centering
			\includegraphics[width=0.4\linewidth]{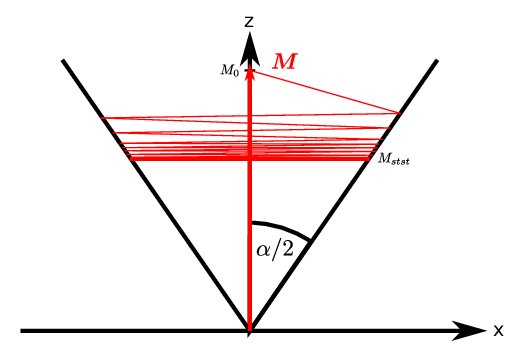}
			\caption{Visualization of the transition of the magnetization into the steady-state during a bSSFP sequence with $\alpha/2$ and TR/2 initialization. Inspired by \cite{Scheffler_Eur.Radiol._2003}.}
			\label{fig::bSSFP_cone}
		\end{figure}

		The balanced steady-state free precession sequence was originally described by Carr in 1958 \cite{Carr_Phys.Rev._1958}.
		Since that, it got several different acronyms like TrueFISP, balanced FFE, and FIESTA, but in the following, we use the short-form bSSFP.\\
		It belongs to the class of gradient-echo sequences, which are characterized by their steady-states. The special feature of bSSFP sequences is their zero gradient moment during a repetition. The balanced gradients, shown in Figure \ref{fig::sequence_flash_bSSFP}, lead to a refocusing of the magnetization $\bm{M}$ at the end of each TR, so that the gradient-induced dephasing of $\bm{M}$ is zero \cite{Scheffler_Eur.Radiol._2003}.\\
		For bSSFP-sequences the RF-pulses have an alternating phase-scheme with $\pi$ between each repetition. Assuming a magnetization starting tilted around $\alpha/2$ from the $B_0$-axis, this rf-pattern flips $\bm{M}$ to the opposing side of the cone, shown in Figure \ref{fig::bSSFP_cone}.
		
		The decrease in signal intensity results from relaxation effects during the acquisition. After a while, a steady-state is reached where relaxation and excitation effects are balanced.\\			

\begin{figure}[!h]
			\centering
			\includegraphics[width=0.9\linewidth]{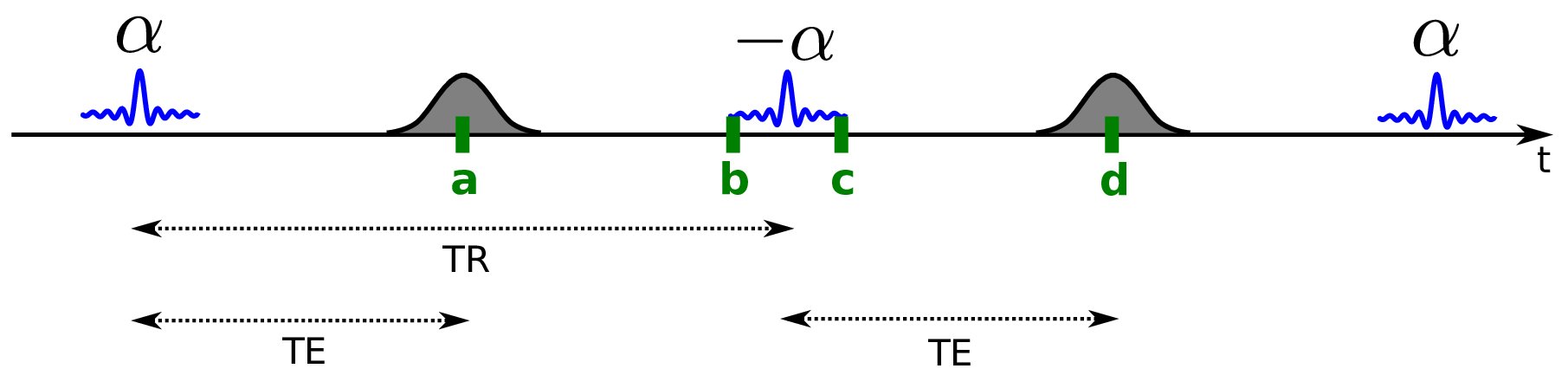}
			\caption{ Visualization inspired by \cite{Hargreaves_Magn.Reson.Med._2001}. It illustrates the time intervals for the derivation of the steady-state magnetization of a bSSFP sequence. The individual time points are marked in \textbf{green}, while the RF-pulses are \textbf{blue} and the echoes are visualized in \textbf{gray}. }
			\label{fig::bSSFP_Mss_derivation_scheme}
		\end{figure}

		To define a model for the time-development of the magnetization during a bSSFP sequence, we follow \cite{Hargreaves_Magn.Reson.Med._2001}
		and use its periodicity. The whole system can be formulated as 
		\begin{align}
		\bm{M}_{k+1} = \bm{AM}_k+\bm{B}
		\label{eq::bSSFP_discrete_time_system_short}
		\end{align}
		with the magnetization $\bm{M}_k$ at the $k$th repetition of TR. $\bm{A}$ is a 3x3 matrix including relaxation-, precession- and rf-excitation effects as well as $\bm{B}$ in form of a 3x1 vector. To derive more precise expressions for $\bm{A}$ and $\bm{B}$, we need to describe their individual components.\\
		We use a hard-pulse approximation\footnote{The hard-pulse approximation assumes, that the RF-pulse just flips the magnetization without any relaxation during this process. More information can be found in section \ref{sec::simulation_fundamental_principles}.} and a $\bm{B}_1$-field along the $x$-axis to reduce the effect of an RF-pulse to the rotational matrix $\bm{R}_{\alpha}$
		\begin{align}
		\bm{R}_{\alpha} = 
		\begin{bmatrix}
		1 & 0 & 0 \\
		0 & \cos\alpha & \sin\alpha\\
		0 & -\sin\alpha & \cos\alpha\\
		\end{bmatrix},
		\end{align}
		and the flip-angle $\alpha$.\\
		Off-resonances can also be simplified to a rotation along the $z$-axis:
		\begin{align}
		\bm{P}(t) = 
		\begin{bmatrix}
		\cos(2\pi\Delta f\cdot t) & \sin(2\pi\Delta f\cdot t) & 0\\
		-\sin(2\pi\Delta f\cdot t) & \cos(2\pi\Delta f\cdot t) & 0\\
		0 & 0 & 1\\
		\end{bmatrix},
		\end{align}
		with the off-resonance $\Delta f$, the time-interval $t$, while the relaxation is included by
		\begin{align}
		\bm{C}(t)=
		\begin{bmatrix}
		e^{-\frac{t}{T_2}} & 0 & 0 \\
		0 & e^{-\frac{t}{T_2}} & 0 \\
		0 & 0 & e^{-\frac{t}{T_1}} \\
		\end{bmatrix},
		\end{align}
		with the relaxation times of the tissue $T_1$ and $T_2$.\\
		Now we start to derive the steady-state magnetization by describing the relation of $\bm{M}$ at different time-points along the acquisition, shown in Figure \ref{fig::bSSFP_Mss_derivation_scheme}.	
		
\begin{figure}[!h]
			\centering
			\includegraphics[width=0.7\linewidth]{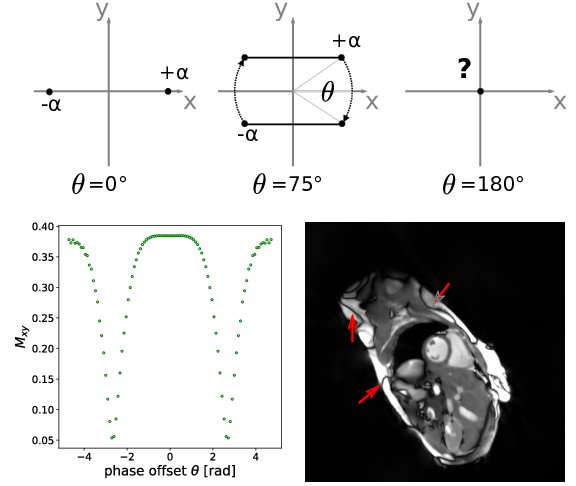}
			\caption{ Overview about the off-resonance effects during a bSSFP sequence. Clockwise from the top: geometrical explanation of the off-resonances during a bSSFP sequence, in-vivo dataset with red-marked banding-artifacts, and signal intensity plotted against the off-resonance for a FA of 70°. The Figure is inspired by \cite{Scheffler_Eur.Radiol._2003} and the in-vivo image is in courtesy from Sebastian Rosenzweig. The relation between $\theta$ and $\Delta\omega$ is: $\theta=\Delta\omega\cdot\textrm{TR}$. }
			\label{fig::bssfp_offresonance_sensitivity}
		\end{figure}

		During the first interval [a,b] only precession and relaxation are present. Therefore we get
		\begin{align}
		\bm{M}_b &= \bm{P}(\textrm{TR} -\textrm{TE})\bm{C}(\textrm{TR} -\textrm{TE})\bm{M}_a + \bm{D}(\textrm{TR} -\textrm{TE})\\
		&= \bm{P}_2\bm{C}_2\bm{M}_a + \bm{D}_2,
		\label{eq::bSSFP_steadystate_derivation_int_AB}
		\end{align}
		with the additional $T_1$-relaxation
		\begin{align}
		\bm{D}(t) = (\bm{1}-\bm{C}(t))
		\begin{bmatrix}
		0\\
		0\\
		M_0\\
		\end{bmatrix},
		\end{align}
		and the original magnetization $M_0$.\\
		The next interval [b,c] captures the effect of the RF-pulse, which leads to
		\begin{align}
		\bm{M}_c = \bm{R}_{\alpha}\bm{M}_b,
		\label{eq::bSSFP_steadystate_derivation_int_BC}
		\end{align}
		while [c,d] is similar to [a,b]:
		\begin{align}
		\bm{M}_d &= \bm{P}(\textrm{TE})\bm{C}(\textrm{TE})\bm{M}_c + \bm{D}(\textrm{TE})\\
		&=\bm{P}_1\bm{C}_1\bm{M}_c + \bm{D}_1.
		\label{eq::bSSFP_steadystate_derivation_int_CD}
		\end{align}
		Now we can combine equation \ref{eq::bSSFP_steadystate_derivation_int_AB}, \ref{eq::bSSFP_steadystate_derivation_int_BC}, and \ref{eq::bSSFP_steadystate_derivation_int_CD} to get a more detailed expression for equation \ref{eq::bSSFP_discrete_time_system_short}:
		\begin{align}
		\bm{M}_{k+1} =
		\bm{P}_1\bm{C}_1\bm{R}_{\alpha}\bm{P}_2\bm{C}_2\bm{M}_k+\bm{P}_1\bm{C}_1\bm{R}_{\alpha}\bm{D}_2 + \bm{D}_1.
		\label{eq::bSSFP_discrete_time_system_long}
		\end{align}		
		For the steady-state, we know that $\bm{M}_{k+1} = \bm{M}_k = \bm{M}^*$ needs to be true, so that we can calculate its magnetization by solving
		\begin{align}
		& \bm{M}^* = \bm{AM}^*+\bm{B}\\
		\Leftrightarrow~~~
		& \bm{M}^*= (\bm{1}-\bm{A})^{-1}\bm{B}.
		\end{align}
		The solution for a bSSFP with a $\pi$-alternating RF-pulse scheme and without off-resonances follows to \cite{Freeman_J.Magn.Reson._1971}
		\begin{align}
		S_{stst} = M_0 \frac{\sin\alpha(1-E_1)}{1-(E_1-E_2)\cos\alpha-E_1E_2} ~\textrm{e}^{-\frac{\textrm{TE}}{T_2^*}},
		\label{eq::signal_bSSFP}
		\end{align}
		with $E_{1,2}=e^{-\frac{\textrm{TR}}{T_{1,2}}}$.\\
		The last factor in equation \ref{eq::signal_bSSFP} $\textrm{e}^{-\frac{\textrm{TE}}{T_2^*}}$ can be simplified to $\textrm{e}^{-\frac{\textrm{TE}}{T_2^*}} = \textrm{e}^{-\frac{\textrm{TE}}{T_2}} = \sqrt{E_2}$ for a central echo time TE=$\frac{\textrm{TR}}{2}$. If TE is moved away from the center of TR, effects belonging to $T_2^*$ occur \cite{Scheffler_Magn.Reson.Med._2003}\cite{Scheffler_Magn.Reson.Med._2003_2}.\\
		The highest signal is reached by using a flip-angle of 
		\begin{align}
		\alpha_{\textrm{max}}=\arccos\left(\frac{T_1-T_2}{T_1+T_2}\right).
		\end{align}
		Beside the $\alpha$/2-magnetization preparation, shown in Figure \ref{fig::bSSFP_cone}, there exist many different transition-styles into the steady-state \cite{Scheffler_Eur.Radiol._2003}\cite{Hargreaves_Magn.Reson.Med._2001}. \\
		Balanced SSFP sequences show a high sensitivity for off-resonance effects. An overview is given in Figure \ref{fig::bssfp_offresonance_sensitivity}. It shows the off-resonance $\Delta\omega$ effects in the rotating frame of reference leading to an additional rotation around the main magnetic field. Therefore, if $\Delta\omega$ reaches a value of $\pi$, the spin is already moved to the position in which it should end after the following RF-pulse. It is therefore moved out of the steady state, so that its signal vanishes, and a banding artifact visualized in figure \ref{fig::bssfp_offresonance_sensitivity} occurs.
		
\section{Image Reconstruction}
	In the previous sections the physical basis of MRI, how the scanner system looks like as well as how the data is collected, encoded with different sequences and finally saved in the k-space is discussed. In the following chapter we focus on the reconstruction of the measured object. 
	
	\subsection[Fourier Reconstruction]{Fourier Reconstruction}
	\textit{This section follows \cite[p.190ff]{Liang__1999}.}\\
	The measured signal is acquired in the frequency domain, represented through the k-space, as discussed in section \ref{sssec: spatial_encoding_kspace_perspective}. This is a representation in a two-dimensional frequency space ($k_x$, $k_y$). Equation \ref{eq::kspace_traj_equation_frequency_encoding} already suggests that there is a Fourier relation between the object $\rho(r)$ and the signal $S(k)$. To get $\rho(r)$, it is therefore be necessary to use the inverse Fourier transformation
	\begin{align}
		\rho(\bm{r}) = \int\limits_{-\infty}^{\infty} S(\bm{k})\textrm{e}^{2\pi i\bm{k} \bm{r}}\textrm{d}\bm{k}.
	\end{align}
	Theoretically this would give the image, but in practice, there are some limitations as will be discussed in the following sections.

	\subsubsection*{Discretization}
\begin{figure}[!h]
			\centering
			\includegraphics[width=0.9\linewidth]{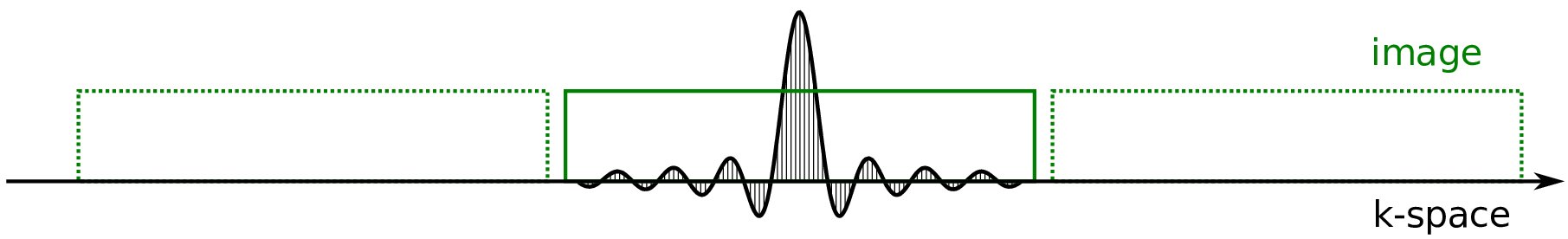}
			\caption{Visualization of an exemplary signal following a sinc-function and illustrates in black. The Fourier transformation leads to the creation of the in green added box as well its sampling-dependent dotted replicas on both sides.}
			\label{fig::discrete_sampling_periodic_extensions}
		\end{figure}

		The first limitation follows with the discretization of the continuous signal. To get an idea of the resulting effects, we simplify the signal equation to be one-dimensional:
		\begin{align}
		S(n\Delta k) = \int\limits_{-\infty}^{\infty} \rho(x)\textrm{e}^{-2\pi in\Delta k x}\textrm{d}x,
		\end{align}
		with an infinite sampling with finit sample-size $\Delta k$ 
		\begin{align}
			D = \{n\Delta k,~-\infty < n <\infty\}.
		\end{align}
		Using the Poisson formula,
		\begin{align}
			\sum\limits_{n = -\infty}^{\infty}S(n\Delta k)\textrm{e}^{2\pi i n \Delta k x} = \frac{1}{\Delta k}\sum\limits_{n = -\infty}^{\infty}\rho\left(x-\frac{n}{\Delta k}\right)
		\end{align}
		can be derived \cite[p.191]{Liang__1999}. This represents the Fourier series with its coefficients $S(n\Delta k)=S[n]$ on the left side as well as the object and its periodically with $\frac{1}{\Delta k}$ occurring replicas to its side, shown in Figure \ref{fig::discrete_sampling_periodic_extensions}.
		
		To separate the image $\rho(x)$ from its replicas, we use the fact that it is always support-limited. Therefore, a limitation of the object by $L$ exists, where signals can come from
		\begin{align}
			\rho(x) = 0,~~|x|>\frac{\textrm{FoV}}{2}.
			\label{eq::support_limited_object}
		\end{align} 
		The framed region given by equation \ref{eq::support_limited_object} is called \textbf{field of view} (FoV).\\
		If the \textbf{Nyquist-sampling-criterion} for MRI
		\begin{align}
			\textrm{FoV} < \frac{1}{\Delta k}
			\label{eq::Nyquist_sampling_theorem}
		\end{align}
		is fulfilled, this ensures that the periodic extensions of $\rho(x)$ do not overlap. Then, no infolding artifacts are observed. Therefore, the reconstruction is simplified to
		\begin{align}
			\rho(x) = \Delta k \sum\limits_{n = -\infty}^{\infty}S(n\Delta k)\textrm{e}^{2\pi i n \Delta k x},~~|x|<\frac{1}{\Delta k}.
		\end{align}

	\subsubsection*{Finite Sampling}
	\label{sssec::finite_sampling}
		A limitation while acquiring data is the acquisition of finite samples because in practice, the signal can not be measured for an infinite time. We assume the signal $S(k)$ to be sampled within $D$:
		\begin{align}
			k \in D = \left\{n\Delta k,~-\frac{N}{2} \leq n <\frac{N}{2}\right\}.
		\end{align}
		This results in a non unique reconstruction of $\rho$ because unmeasured Fourier coefficients $c_n$ occur:
		\begin{align}
			\rho(x) = 
			\Delta k \sum\limits_{n = -\frac{N}{2}}^{\frac{N}{2}-1}S[n]\textrm{e}^{2\pi i n \Delta k x}
			+ \sum\limits_{n<-\frac{N}{2};~n\geq\frac{N}{2}}c_n\textrm{e}^{2\pi i n \Delta k x}.
		\end{align}
		To decide which $c_n$ should be chosen, the minimum-norm constraint is used \cite[p.194]{Liang__1999}. This leads to the vanishing of $c_n$ and the \textbf{Fourier reconstruction formula} of
		\begin{align}
			\rho(x) = 
			\Delta k \sum\limits_{n = -\frac{N}{2}}^{\frac{N}{2}-1}S[n]\textrm{e}^{2\pi i n \Delta k x},~~|x|<\frac{1}{\Delta k},
			\label{eq::fourier_reco_formula}
		\end{align}
		which still does not represent the true image. Through cutting off some Fourier coefficients, the continuity of the data is not longer given. This results in Gibbs ringing artifacts but can be reduced by windowing the data with accepting losses in the spatial resolution. The characterizing feature for useful windowing-functions $w_n$ is their smooth decays to zero at $n=\pm \frac{N}{2}$ \cite[p.195]{Liang__1999}. The most used $w_n$-function is the Hamming window \cite[p.253]{Liang__1999}.

	\subsubsection*{Direct Fourier Reconstruction}
		The finite sampling requires to fulfill the Nyquist criterion, already mentioned in equation \ref{eq::Nyquist_sampling_theorem}, for a unique reconstruction. It also defines the largest voxel size for an uncorrupted image, known as \textbf{Fourier pixel size}:
		\begin{align}
			\Delta x = \frac{1}{N \Delta k},
			\label{eq::theory_fourier_pixel_size}
		\end{align}
		with  $\Delta x = \frac{\textrm{FoV}}{N}$. Following equation \ref{eq::theory_fourier_pixel_size} an increased $\Delta x$ need to go ahead with a small $\Delta k$ for constant $N$ otherwise if the measured objected extends these limit, wrap-around artifacts occur.\\
		The image follows with the \textbf{discrete Fourier transformation}:
		\begin{align}
			\rho(m) = 
			\Delta k \sum\limits_{n = -\frac{N}{2}}^{\frac{N}{2}-1}S[n]\textrm{e}^{\frac{2\pi i n m}{N}},~~-\frac{N}{2} \leq m <\frac{N}{2},
		\end{align}
		with a normalization by $\Delta k = 1$ to
		\begin{align}
			\rho(m) = \sum\limits_{n = -\frac{N}{2}}^{\frac{N}{2}-1}S[n]\textrm{e}^{\frac{2\pi i n m}{N}},
			~~-\frac{N}{2} \leq m <\frac{N}{2}.
			\label{eq::discrete_fourier_transformation}
		\end{align}

	\subsubsection*{Zero-Padded Fourier Reconstruction}
		An extended version of the direct Fourier reconstruction is the zero-padded. It adds a preparation step using a zero-padding to extend the datasets to have $N$ data points, which is an integer power of two. Therefore a radix 2-FFT \cite{Cooley_Math.Comput._1965} can be used and additionally an increased digital resolution follows.

	\subsection[Reconstructions as Linear Inverse Problem]{Reconstructions as Linear Inverse Problem}
	\textit{This section follows \cite{Sumpf__2012}.}\\
		In the previous chapter we have seen that the image reconstruction can be expressed as an inverse Fourier transformation, \ref{eq::kspace_traj_equation_frequency_encoding}, and can be implemented in using a DFT, \ref{eq::discrete_fourier_transformation}. For more complex reconstructions, the interpretation of the problem as a inverse problem is superior. Therefore, the image is understood as a pixel-vector $\bm{x}$ and the signal follows with the sampling vector $\bm{y}$. The latter is represented as a diagonal-matrix of a binary sampling mask $\mathfrak{P}$, and the Fourier transformation is represented as matrix of Fourier coefficients $\mathfrak{F}$. The \textbf{forward problem} is
		\begin{align}
			\bm{y}= \mathfrak{P}\mathfrak{F}\bm{x} = \bm{A}\bm{x},
			\label{eq::linear_inverse_problem}
		\end{align}
		with system-matrix $\bm{A}$. Equation \ref{eq::linear_inverse_problem} is a linear system if $\bm{A}$ is linear. This representation nicely shows why the reconstruction of $\bm{x}$ is called an \textbf{inverse problem}.\\
		Inverting $\bm{A}$ would allow to directly determine $\bm{x}$ from $\bm{y}$. In practice, this has some disadvantageous: determining $\bm{A}^{-1}$ is not possible if not all k-space points are sampled and the solution for $\bm{x}$ is not unique and a regularization becomes necessary to solve the inverse problem. The Moore-Penrose pseudo inverse can be calculated, which is related to a Tikhonov-regularization. To allow even more advanced regularization terms $\bm{R}$, like the $l_1$ or Sobolev-norm, it is more common to describe the problem using a forward operator and solving it iteratively \cite{Wang_Magn.Reson.Med._2018}\cite{Uecker_Magn.Reson.Med._2008}:
		\begin{align}
			\hat{x} := \underset{\bm{x}}{\textrm{argmin}}\|\bm{A}\bm{x}-\bm{y\|_2^2}+\alpha\bm{R}
			\label{eq::Linear_inverse_problem_optimization}
		\end{align}
		with the regularization parameter $\alpha$.
		
	\subsection[Undersampling]{Undersampling}
\begin{figure}[!h]
			\centering
			\includegraphics[width=0.8\linewidth]{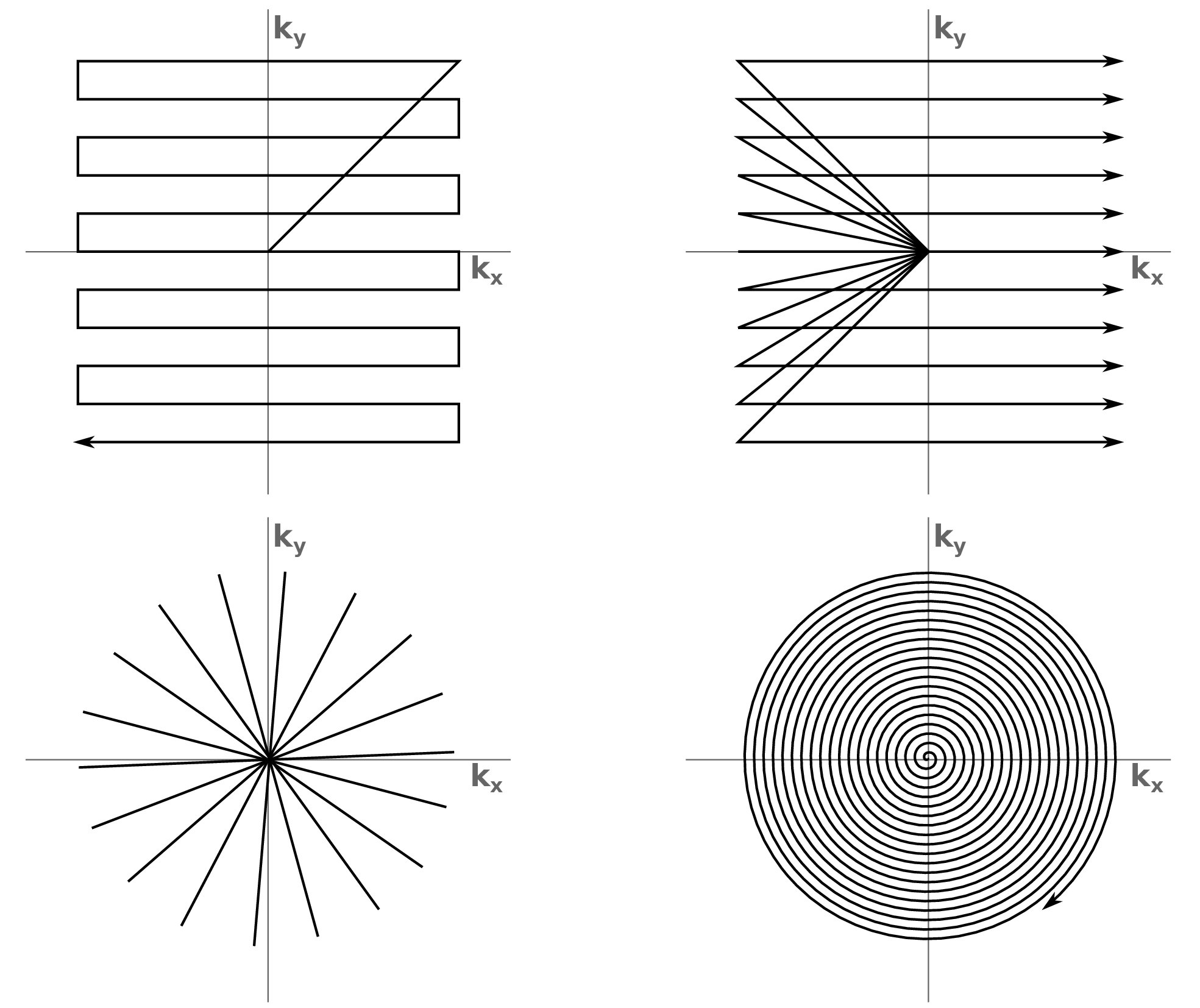}
			\caption{ Visualization of various sampling schemes. In clockwise order from upper left: echo-planar, Cartesian, spiral and radial encoding. Inspired by \cite{Uecker__2009}.}
			\label{fig::theory_sampling_schemes}
		\end{figure}

		If the k-space is fully-sampled, the system-matrix $\bm{A}$ is squared and has full rank, so it can be solved uniquely. This behavior changes with skipping some k-space lines. This effect, also called \textbf{undersampling}, leads to a speed up of the measurement because its time linearly depends on the sampled k-space lines. The way of how the not acquired data is skipped changes with different sampling schemes like radial-, echo-planar- and spiral-ones (shown in figure \ref{fig::theory_sampling_schemes}). The speed-up factor is called the acceleration- or reduction factor $R$. 
		
		Staying in the forward model described in equation \ref{eq::linear_inverse_problem} undersampling leads to an under-determined system matrix $\bm{A}$ and the solutions for $\bm{x}$ are no longer unique. They vary based on the point spread function of the sampling pattern. For a reduction factor of $R = 2$ and a Cartesian sampling the signal intensity can be distributed over two virtual objects. For reaching uniqueness of the solution additional information is necessary. Therefore, parallel imaging is used.


	\subsection[Parallel Imaging]{Parallel Imaging}
		
		\textit{This section follows \cite[p.522]{Bernstein__2004}.}\\
		Parallel Imaging (PI) allows to get aliasing-free images by just acquiring a smaller fraction of the original fully-sampled data. It works with \textbf{phased array coils}, also known as multi-coil arrays. They replace one single large coil by overlapping each other and offering additional spatial information. This can only work when different coils detect varying signals from one point. For example for Cartesian sampling, the coil profiles need to be different in phase-direction.\\
		There are some methods that exploit these additional information. The techniques are typically categorized in k-space and image-space methods because the first ones fill up k-space before using a Fast-Fourier transformation FFT, and the latter need to have the reconstructed images with artifacts and remove them afterwards. Typical k-space methods are GRAPPA and SMASH, while SENSE is working in the image domain. An overview about different methods is given in \cite{Blaimer_Top.Magn.Reson.Imaging_2004}. Because some of these techniques already need the coil profiles, they are typically measured before (calibration scan) or are extracted from the images after sampling fully sampling of their k-space center and using its reconstruction as low-resolution estimate. Others are directly using the k-space data for coil profile estimation \cite{Uecker_Magn.Reson.Med._2014}, or combining it with the image reconstruction \cite{Uecker_Magn.Reson.Med._2008}\cite{Ying_Magn.Reson.Med._2007}.\\
		The usage of phased-array coils also offers noise benefits. In MRI, most of the detected noise comes from the patient itself. The contribution of the coils is much smaller and can be neglected. By having smaller overlapping coils, their sensitive volume is reduced compared to one large antenna, which leads to a reduced noise amplitude. This results from the weighting of the detected noise by the coils profile.



\section{Quantitative MRI}
	In classical MRI, the image quality of reconstructed images is influenced by the used sequences as well as by their individual parameters. They allow to adjust the contrast for different tissues but assumes special knowledge of sequences and their parameters and their effect on the image contrast. In the end, the physical parameters can only be described qualitatively and tissue differences are referred to be \textit{hypo-}, \textit{iso-} or \textit{hyper-intense}. \\
	The idea behind quantitative MRI is to measure the physical properties of the tissue directly by determining their specific parameters like relaxation times, proton density, and off-resonances. This does not only allow to differentiate between varying types of tissues, and probably locate abnormal ones like scars and edema \cite{Okur_Diagn.Interv.Radiol._2014}\cite{Taylor_JACC:CardiovascularImaging_2016}, but also may speed up the measurement procedure for the patient by enabling to simulate any contrast with knowledge of the physical parameters of the tissue, which is called synthetic magnetic resonance imaging \cite{Warntjes_Magn.Reson.Med._2008}\cite{Tanenbaum_AJNR_2017}.\\
	This section will give a short overview about the different concepts of quantitative MRI, which are partly used in this work.\\
	Classical quantitative MRI uses approximations of the Bloch equations to derive sequence specific signal models. The sequences are chosen to be sensitive to special parameters but are insensitive to non-wished ones. This makes them robust against other influences and enables the precise calculation of the wished tissue characteristics from the determined images or included into the reconstruction from the k-space data itself.
	
	\subsubsection*{Gold-Standard $T_1$ Quantification}
	\label{ssec::gold_standard_t1_quantification}
\begin{figure}[!h]
		\centering
		\includegraphics[width=0.9\linewidth]{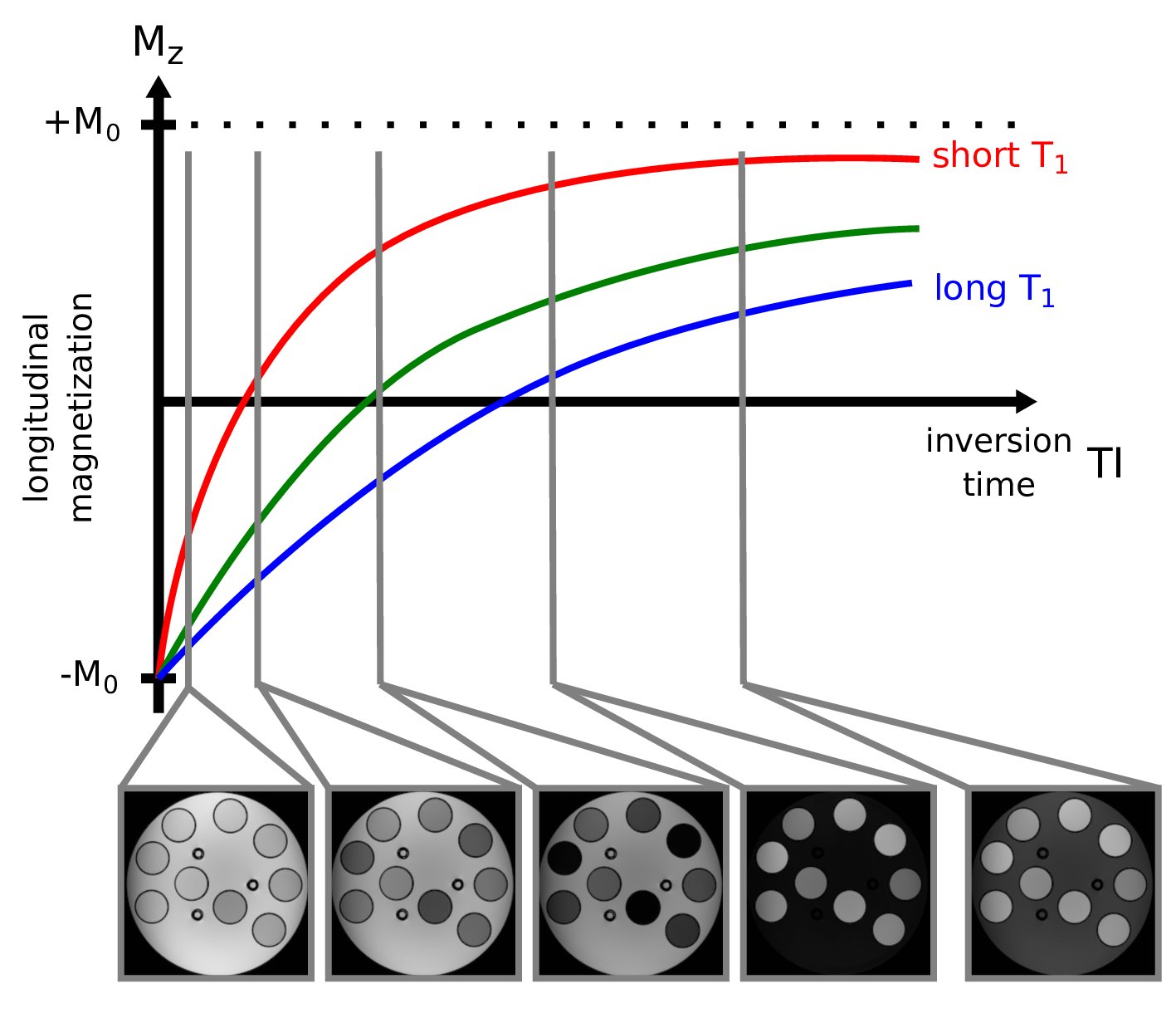}
		\caption{Illustration of the regrowing, longitudinal magnetization after an inversion during a gold-standard $T_1$-quantification. The evolution of the signal follows equation \ref{eq:signal_T1_relaxation} and varies for the tubes visualized in the acquired dataset below. The measured object is a custom-built phantom discussed in chapter \ref{chap::phantom_creation}. }
		\label{fig::theory_GoldStandartT1}
	\end{figure}

	The gold-standard method for quantification of $T_1$ is based on a fully-sampled spin-echo sequence. First, a 180°-pulse is used to invert the magnetization from $M_0$ to $-M_0$. Data is then acquired using varying inversion times TI and an inverse Fourier transformation to reconstruct them. Afterwards, snap-shots are provided, which represent different magnetization states along the inversion curve of the signal, like visualized in Figure \ref{fig::theory_GoldStandartT1}.
	
	To estimate $T_1$, the previously mentioned equation \ref{eq:signal_T1_relaxation} for the $T_1$-relaxation $M_z(t) = M_0\cdot \left(1-\exp\left(-\frac{t}{\textrm{T}_1}\right)\right)$
	is fitted pixel-wise to the dataset. This model assumes a constant $M_0$ during the k-space acquisitions. For a standard fully-sampled spin-echo measurement, this results in breaks after each k-space line to ensure that the magnetization of the slice has reached its original state. Even for low resolution two-dimensional images, it leads to a highly increased acquisition time. While three-dimensional datasets can be speeded up by interleaved acquisition schemes, they are still slow ($>$1 h) and not well realizable in clinical practice. Nevertheless, they are the most precise existing techniques to determine $T_1$.\\
	
	\subsubsection*{Gold-Standard $T_2$ Quantification}
	\label{ssec::gold_standard_t2_quantification}
\begin{figure}[!h]
		\centering
		\includegraphics[width=0.9\linewidth]{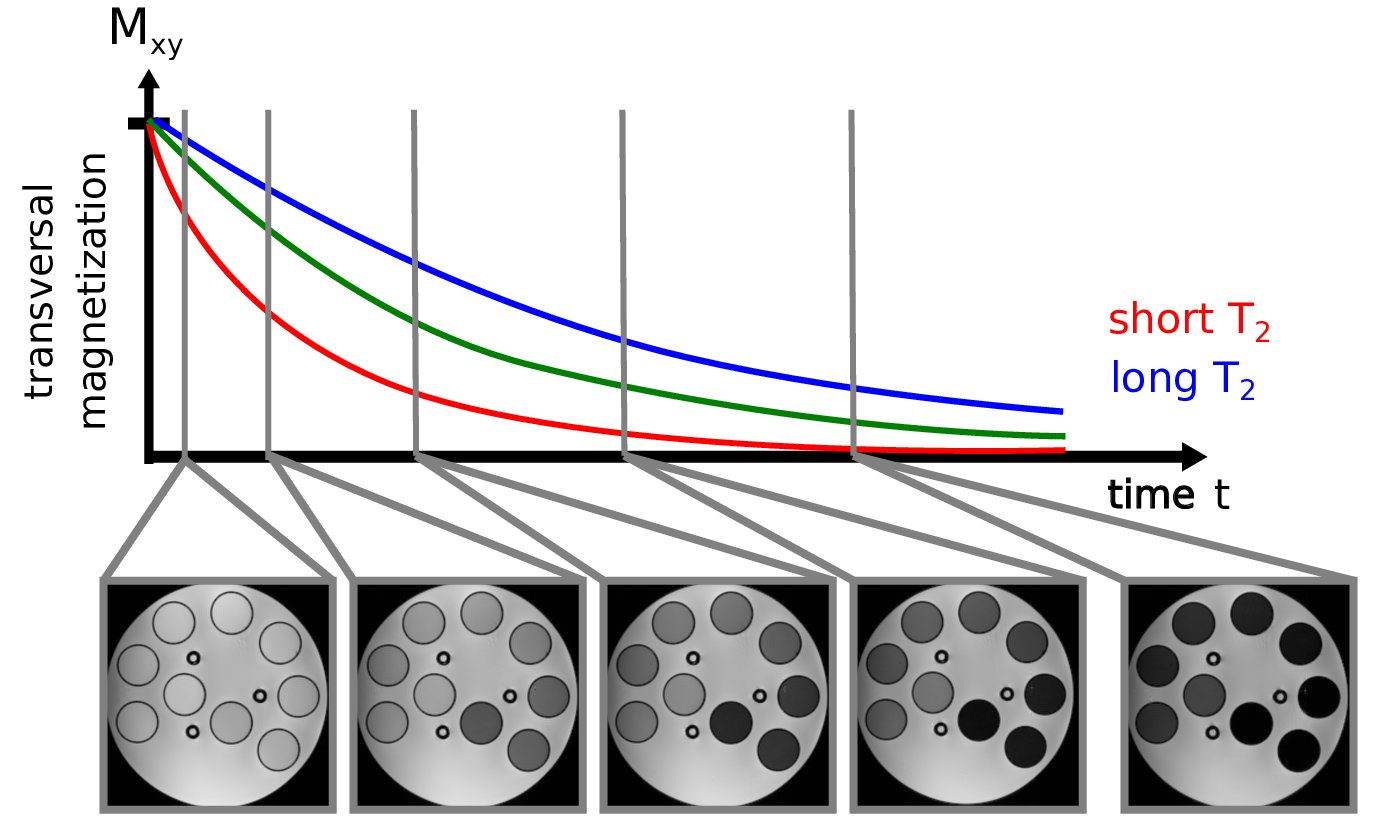}
		\caption{Illustration of the decay of the transversal magnetization during a gold-standard $T_2$-quantification. The evolution of the signal follows equation \ref{eq:signal_T2_relaxation} and varies for the tubes visualized in the acquired dataset below. The measured object is a custom-built phantom discussed in chapter \ref{chap::phantom_creation}. }
		\label{fig::theory_GoldStandartT2}
	\end{figure}

	The gold-standard $T_2$-quantification is very similar to the $T_1$ version. It is based on a spin-echo sequence but is not using an inversion pulse. The echo-time is varied between different acquisitions, which provides snaps-shots along the decay of the transverse magnetization, like shown in Figure \ref{fig::theory_GoldStandartT2}. 
	
	The parameters are estimated using a pixel-wise fit with the model mentioned in equation \ref{eq:signal_T2_relaxation} $
	M_{xy}(t) = M_0\exp\left(-\frac{t}{\textrm{T}_2}\right).
	$
	The original sequence includes waiting periods until the $M_0$ is reached again, which increases the acquisition time. To speed it up Turbo-Spin-Echo measurements are established. They acquire multiple echoes, the \textit{echo train}, after each 90°-pulse \cite[p.291ff]{Liang__1999}. Therefore the fitting model needs to be adjusted to reduce the effect of stimulated echoes \cite{Milford_PLOSONE_2015}.

	\subsubsection*{Model-based $T_2$-Mapping}
	Model-based methods rely on special signal models, which are included into the reconstruction process. Various methods exist for mapping of single relaxation parameter. One of the earliest uses a non-linear inverse reconstruction for estimation of $T_2$-maps from undersampled k-space data acquired using a fast-spin-echo sequence \cite{Block_IEEETrans.Med.Imaging_2009}.
	Therefore, the simplified signal model for the exponential decay, while using a fast-spin-echo sequence, is given by
	\begin{align}
		\bm{M}_{\textrm{TE}}(\bm{\rho},\bm{R}_2) = \bm{\rho}(\bm{r})\cdot e^{-\bm{R}_2(\bm{r})\cdot{\textrm{TE}}},
	\end{align}
	with the magnetization $\bm{M}_{\textrm{TE}}$ at the echo time TE, the proton-density map $\bm{\rho}$, and the $\bm{R}_2$-map as entry-wise inverse of the $T_2$-one, is combined with the discrete Fourier transformation $\mathfrak{F}$ and a pattern operator $\mathfrak{P}$ to get synthetic data. Afterwards, this is compared with the acquired data $\bm{s}_{\textrm{TE}}$ by a cost function $\Phi$:
	\begin{align}
		\hat{x} = \underset{\bm{x}}{\textrm{argmin}}\left[\Phi(\bm{x})\right],	
	\end{align}
	with
	\begin{align}
		\Phi(\bm{x}) = \frac{1}{2}\sum\limits_{\textrm{TE}} \|\mathfrak{P}\mathfrak{F}\bm{M}_{\textrm{TE}}(\bm{x})-\bm{s}_{\textrm{TE}}\|_2^2,
	\end{align}
	and
	\begin{align}
		\bm{x} = 
		\begin{pmatrix}
		\bm{\rho}\\
		\bm{R}_2\\
		\end{pmatrix}.
	\end{align}
	Finally, the minimization problem is solved using a Conjugate-Gradient-Descent algorithm \cite{Hager_SIAMJ.Optim._2005}.

	\subsubsection*{Model-based $T_1$-Mapping}
	An other fast and accurate model-based method is focused on $T_1$-mapping \cite{Wang_Magn.Reson.Med._2018}\cite{Volkert_Int.J.Imag.Syst.Tech._2016}.
	An inversion prepared FLASH sequence is used to acquire highly undersampled data. A non-linear inverse problem is set up using the signal equation
	\begin{align}
		\bm{M}_{t}(\bm{r})=\bm{M}_{ss}(\bm{r}) - (\bm{M}_{ss}(\bm{r}) + \bm{M}_{0}(\bm{r}))\cdot e^{-t\cdot \bm{R}_1^*(\bm{r})},
	\end{align}
	with the steady-state magnetization $\bm{M}_{ss}$, the original offset $\bm{M}_0$, the location $\bm{r}$, as well as the effective relaxation map $\bm{R}_1^*$ influenced by the RF-pulses of the continuous FLASH readout. The real $T_1$ map follows with the entry-wise calculation of
	\begin{align}
		T_1 = \frac{M_0}{M_{ss}\cdot R_1^*}.
	\end{align} 
	The non-linear problem is solved using an Iterative-Regularized-Gauss-Newton-Method based on FISTA to solve the linearized equation.
\chapter{Creation of $T_1$-$T_2$ Phantom}
\label{chap::phantom_creation}
	
	\section{General Idea of Relaxation Phantom}
\begin{figure}[!h]
		\centering
		\includegraphics[width=0.6\linewidth]{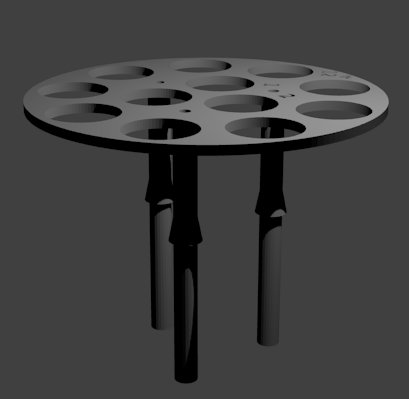}
		\caption{Visualization of the 3D-printed holder for the tubes with varying $T_1$ and $T_2$. }
		\label{fig::holder}
	\end{figure}

	The $T_1$-$T_2$ phantom is build to allow a precise quantification of techniques for quantitative MRI. Here the relaxation parameter are designed to be in the range of typical heart tissue at 3 T. The created nine tubes should fit into a later 3D-printed rack (see Figure \ref{fig::holder}). The tubes have different $T_1$ and $T_2$ values so that they form a grid in $T_1$-$T_2$-space. To mainly modify the $T_1$ time, gadolinium-chloride GdCl$_3$ is used. Agarose is used to change $T_2$. Additionally, carragenaan leads to a gelling-effect, natrium-chlorid NaCl to the preferred conductivity, and sodium-azide NaN$_3$ to a resistance against mould.

	\section{Methods of Creating the Phantom}
\begin{figure}[!h]
		\centering
		\includegraphics[width=0.6\linewidth]{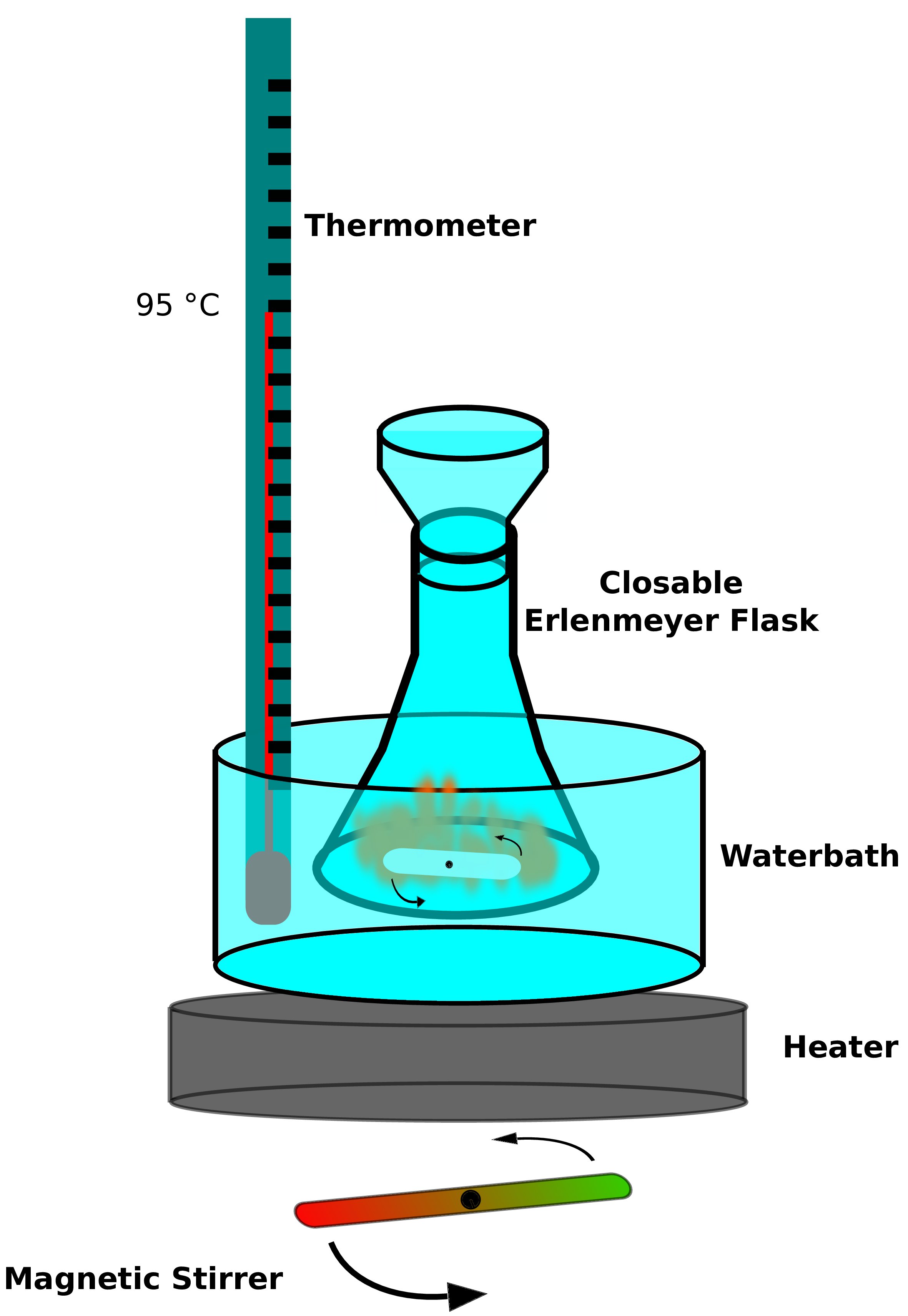}
		\caption{Drawing of the experimental setup necessary for the creation of the desired tube stuffing. }
		\label{fig::setup}
	\end{figure}

	The setup for creating the tube stuffing is shown in Figure \ref{fig::setup} and the used materials are listed in Table \ref{tab::ingredients}. Additional dead volumes of the pipettes (see GdCl$_3$) are measured to be 200 $\mu$l per usage and are added to Table \ref{tab::ingredients}. The tube number one has not been evaluated because it is equal to tube 2 with just  a larger amount of salt, which does not influence the relaxation parameter. 
	
	In the first step, a  GdCl$_3$ Stock-Solution is created. The necessary weight of GdCl$_3$ in the phantom tubes to reach $T_1$ values of heart tissue are in the unity micro grams \cite{Hattori_Med.Phys._2013}.
	
	A stock solution with a concentration of 180 $\frac{\mu\textrm{mol}}{100~\textrm{ml}}$ is created to reduce the error of weight of these little masses. Following the manufacture information, the molar mass of GdCl$_3\cdot$6 H$_2$O is 371.7 $\frac{\textrm{g}}{\textrm{mol}}$. This leads to a mass of 0.066906 $\frac{\textrm{g}}{100~\textrm{ml}}$. To further reduce the weighting error, the volume of the stock solution is increased to 1 l in which approximately 0.669 g of GdCl$_3\cdot$6 H$_2$O is solved. Now, 5 ml of the stock-solution in \mbox{100 ml} final phantom tube volume result in a concentration of 90 $\frac{\mu\textrm{mol}}{\textrm{l}}$.\\
	
	In the next step 85 ml of distilled water are filled into an Erlenmeyer flask. This is advantageous to a beaker because its neck is tighter, which leads to a smaller boundary-surface between the solution and the air. Additionally, a smaller air-volume inside the flask can store less steam, thus a smaller amount of water evaporates during the short time-intervals when the flask has to be opened to add substances.\\
	When the solution is almost boiling, the salt as well as the carragenaan mixed with agarose can be added slowly. The mixing process has to happen at about 90 to 95 degrees because the materials start to form clusters in a colder environment which are difficult to dissolve. For the same reason the whole procedure needs to be done under continuous mixing of a magnetic stirrer. If nevertheless some clusters are inside of the solution, it is sometimes necessary to speed up and slow down the stirring frequency so that the stirring fish destroys the bulbs.\\
	When everything is dissolved, the calculated amount of gadolinium-stock solution can be added as well as the NaN$_3$. The remaining difference of distilled water to reach a volume of 100 ml \footnote{During the experiments some self made errors occurred. They result in varying volumes regarding to the 100 ml volume of liquids. They are discussed in section \ref{ssec::t1t2phantom_discussion}.} is added and used to clean the bottleneck from immobile substances.\\
	If everything is dissolved, the mixture can be filled in Falcon tubes. Remaining air bubbles soar up, which led to a homogeneous filling. Otherwise a vacuum pump can remove last air arrears.
	\begin{table}[!h]
		\centering
		\caption{This table lists the ingredients of the phantom tubes. Additional dead volumes of the pipettes (see GdCl$_3$) are measured to be 200 $\mu$l per usage and are added.}
		\begin{tabular}{|c|c|c|c|c|c|c|}\addlinespace[2ex]\hline
			Number & Carragenaan [g] & H20 [ml]& GdCl$_3$ [ml]\footnotemark& Agarose [g] & NaCl [g]& NaN3 [g]\\\hline\hline
			1 & 3 & 95.00 & 5.00	(+1.00) & 1.0 & 0.7 & 0.00\\\hline
			2 & 3 & 95.00 & 5.00	(+1.00) & 1.0 & 0.3 & 0.03\\\hline
			3 & 3 & 97.50 & 2.50	(+0.60) & 1.0 & 0.3 & 0.03\\\hline
			4 & 3 & 103.75 & 1.25	(+0.40) & 1.0 & 0.3 & 0.03\\\hline
			5 & 3 & 95.00 & 5.00	(+1.00) & 0.2 & 0.3 & 0.03\\\hline
			6 & 3 & 97.50 & 2.50	(+0.60) & 0.2 & 0.3 & 0.03\\\hline
			7 & 3 & 98.75 & 1.25	(+0.40) & 0.2 & 0.3 & 0.03\\\hline
			8 & 3 & 95.00 & 5.00	(+1.00) & 0.0 & 0.3 & 0.03\\\hline
			9 & 3 & 97.50 & 2.50	(+0.60) & 0.0 & 0.3 & 0.03\\\hline
			10 & 3 & 98.75 & 1.25	(+0.40) & 0.0 & 0.3 & 0.03\\\hline
		\end{tabular}
		
		\label{tab::ingredients}
	\end{table}
	\footnotetext{from stock solution.}

	\section{Gold-Standard Phantom Characterization}
	
	The parameter analysis of the $T_1$-$T_2$ phantom is performed with Python 3.5 as well as with the Berkeley Advanced Reconstruction Toolbox (BART) \cite{Uecker__2013} and includes several steps to extract the relaxation values from the maps, which are explained in the following. 
	
	\subsection*{Data Acquisition and Mapping}
	
\begin{figure}[!h]
		\centering
		\includegraphics[width=0.8\linewidth]{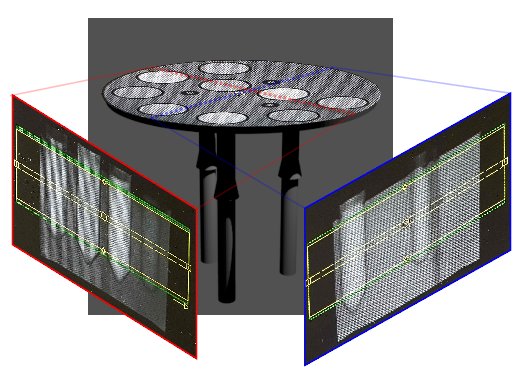}
		\caption{Visualization of the layer positioning for the gold-standard relaxation parameter quantification measurements. The covered area is visualized in yellow, while the calibration region is green. The yellow arrow marks the phase-encoding direction. The underlying images are localizer acquisitions for planning the measurement.  }
		\label{fig::t1t2phantom_GS_meas_slices}
	\end{figure}

	To characterize the $T_1$- and $T_2$-values of the produced phantom, the measurements are performed with two sequences at constant room- and phantom-temperatures of 21 °C. This is checked before and after each measurement. As the sensitivity of $T_1$ for changes in the temperature lies around 1 \%/degree \cite{Hynynen_Magn.Reson.Med._2000}, and as $T_2$ is assumed to be even lower \cite{Nelson_Magn.Reson.Imaging_1987}, the intermediate time interval do not have to be checked. \\
	The first one is a gold-standard $T_1$ quantification, like explained in section \ref{ssec::gold_standard_t1_quantification}, using a spin-echo sequence with a previous non-selective inversion pulse. The inversion-time-intervals are varied in 250 ms increments from 30 ms to 2530 ms, and the echo-times are constant over all measurements to record snapshots at different relaxation states of the magnetization. The full protocol is added to Table \ref{tab::sequence_protocol_T1_gold_standard}. The images are then reconstructed using an inverse Fast-Fourier transformation because only fully sampled datasets are acquired. The fitting is performed pixel-wise with the signal model described in equation \ref{eq:signal_T1_relaxation}.\\
	The second sequence is a spin-echo without inversion pulse for gold-standard $T_2$ quantification, like described in section \ref{ssec::gold_standard_t2_quantification}. The echo-time is increased in 40 ms increments from 15 ms up to 455 ms (see Table \ref{tab::sequence_protocol_T2_gold_standard}).\\
	A full k-space with a base-resolution of 256 is recorded per TE. The images are reconstructed with an inverse Fast-Fourier transformation. The signal is then fitted pixel-wise with the model already presented in equation \ref{eq:signal_T2_relaxation}.\\
	The three acquired layers are chosen to cover the entire wider part of the tube (see Figure \ref{fig::t1t2phantom_GS_meas_slices}). 
	
	\subsection*{Thresholding of Parameter Maps}
	The thresholding of the parameter maps works with upper and lower limits, which have to be determined manually. Because of the visual homogeneity of both maps it is easy to find such limits. They are shown in Table \ref{tab::threhLim}.
	
	\begin{table}[!h]
		\centering
		\caption{The table lists the thresholding limits for the automatic segmentation of the fitted parameter maps during the gold-standard characterization of the $T_1$-$T_2$ phantom.}
		\begin{tabular}{|c|c|c|}\addlinespace[2ex]\hline
			Map & Lower limit & Upper limit \\ \hline\hline
			$T_1$ [s] & 0.5 & 1.6 \\\hline
			$T_2$ [s]& 0 & 0.9 \\\hline
		\end{tabular}
		
		\label{tab::threhLim}
	\end{table}
	
	The resulting maps are segmented into background and tubes. For the following analysis only the tubes are used.

	\subsection*{Labeling and Filtering of Masks}
	After thresholding of the parameter maps some residual noise remains in the resulting segmented masks. Before the individual objects can be labeled, a size-selection operator is used. It deletes all particles inside the mask which are not between 100 px and 700 px large. Afterwards, the remaining particles are eroded to remove artifacts of the plastic border and are labeled to allow a differentiation between the individual tubes. Finally, their $T_{1}$ and $T_{2}$ is determined from the original data.

	\subsection*{Quantification of the Tubes}
	In the following, the segmented images with labeled tubes are analyzed. The determination of the relaxation parameters of each tube starts with a calculation of an arithmetic mean $\overline{{T}}_i$:
	\begin{align}
	\overline{{T}}_i = \frac{1}{N}\sum\limits_{x=0}^{N-1} \textrm{T}_{i,x}
	\end{align}
	for $i=1,2$ and all $N$ pixel of one segmented tube within one slice. The error is assumed to be the standard deviation $\textrm{std}_i$ of all included pixel-values $\textrm{T}_{i,x}$:
	\begin{align}
	\textrm{std}_i= \sqrt{\left(\frac{\sum\limits_{x=0}^{N-1}({T}_{i,x}-\overline{{T}}_i)^2}{N-1}  \right)},
	\end{align}
	because the fitting error is much smaller and does not need to be taken into account. This is performed for all three slices over the whole range of the tubes.\\
	The mean parameter for each tube $\overline{{T}}_{i,tot}$ can then be determined over all single slices $s$:
	\begin{align}
	\overline{{T}}_{i,tot} = \frac{1}{N_s}\sum\limits_{s=0}^{N_s-1} {T}_{i,s},
	\end{align}
	with the number of slices $N_s$ and the magnetization in each layer $\overline{{T}}_{i,s}$. The weighted mean is not used because it is not ensured that the errors between the slices are uncorrelated. Therefore, an arithmetic mean is chosen. Its error follows the Gaussian error propagation and becomes
	\begin{align}
	\sigma_{\overline{{T}}_{i,tot}} =\frac{1}{N_s} \sqrt{ \sum\limits_{s=0}^{N_s-1}\textrm{std}_{i,s}^2},
	\end{align}
	with the standard derivation $\textrm{std}_{i,s}$ of tube $i$ within the slice $s$.

	\subsection*{Results of Phantom Characterization}
\begin{figure}[!h]
		\centering
		
		\begin{subfigure}{.45\textwidth}
			\centering
			\includegraphics[width=\linewidth]{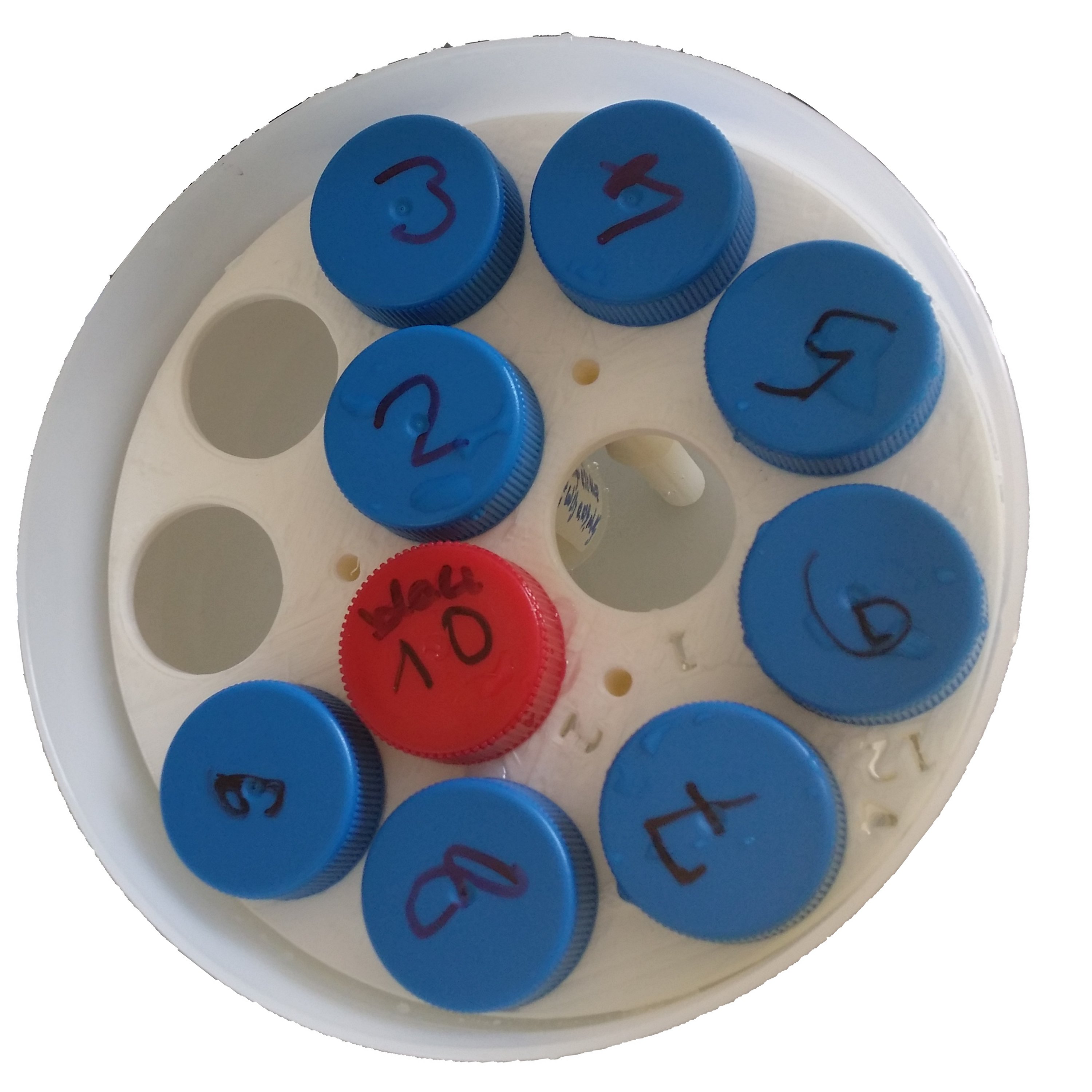}
		\end{subfigure}
		\hfill
		\begin{subfigure}{.45\textwidth}
			\centering
			\includegraphics[width=\linewidth]{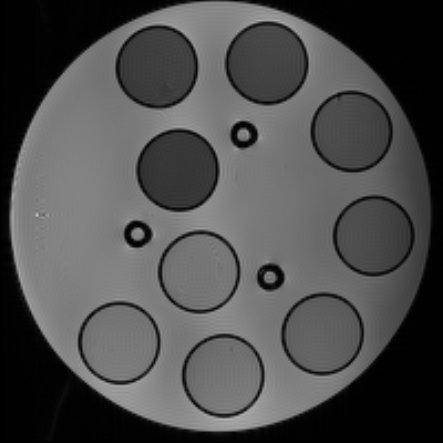}
		\end{subfigure}
		\caption{The \textbf{left} side visualizes the numbering of the individual tubes. The \textbf{right} presents a spin-echo acquisition using the same protocol, like listed in Table \ref{tab::sequence_protocol_T2_gold_standard}, with a TE of 15 ms. It is acquired on the 26.11.2018.}
		\label{fig::t1t2phantom_tubeOrdering}
	\end{figure}

	An image of the used ordering of the tubes inside of the $T_1$-$T_2$ phantom is shown in Figure \ref{fig::t1t2phantom_tubeOrdering} combined with a fully-sampled spin-echo image.
	
	\subsubsection*{Spatial Homogeneity of the Phantom}
\begin{figure}[!h]
		\centering
		\begin{subfigure}{.45\textwidth}
			\centering
			\includegraphics[width=1\linewidth]{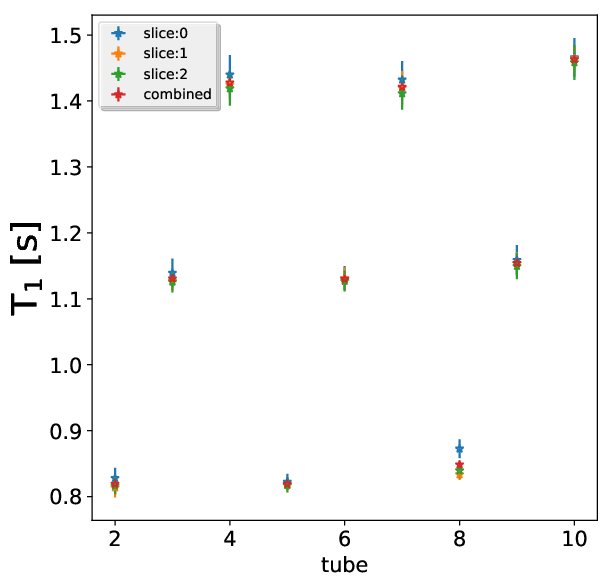}
			\label{fig::t1t2phantom_homogeneity_t1}
		\end{subfigure}
		\hfill
		\begin{subfigure}{.45\textwidth}
			\centering
			\includegraphics[width=1\linewidth]{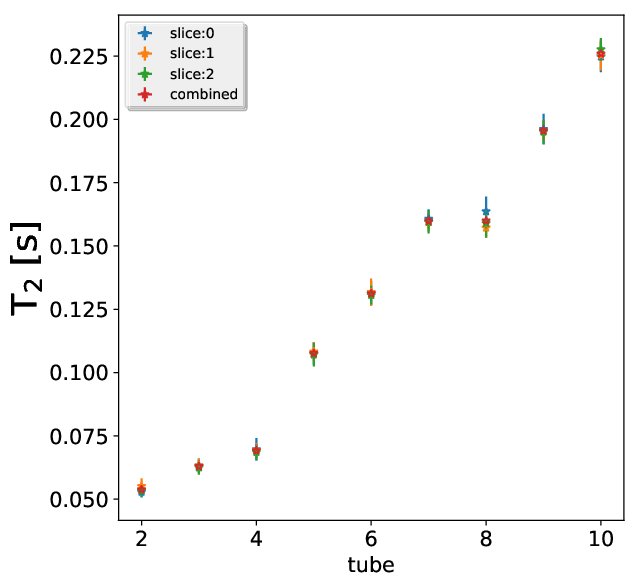}
			\label{fig::t1t2phantom_homogeneity_t2}
		\end{subfigure}
		\caption{Visualization of the homogeneity of the $T_1$-$T_2$ phantom over the three acquired slices for $T_1$ ion the \textbf{left} side and for $T_2$ on the \textbf{right} side. The dataset is acquired at the 17.12.2018. The different data-points are plotted with error bars. The error of the combined point is calculated using the Gaussian error propagation of the individual standard-derivations of the regions of interest ROIs. The numbering of the tubes refers to Figure \ref{fig::t1t2phantom_tubeOrdering}. }
		\label{fig::t1t2phantom_homogeneity}
	\end{figure}

	To determine the spatial homogeneity of the phantom within each tube, three slices are measured, like illustrated in Figure \ref{fig::t1t2phantom_GS_meas_slices}. After calculating their relaxation parameters, they are combined and plotted for every experiment. A representative visualization from the 17.12.2018 is plotted in Figure \ref{fig::t1t2phantom_homogeneity} and the relaxation times are listed in Table \ref{tab::t1t2phantom_relaxation_times_20181217}.

	All acquired datasets are showing a good homogeneity for all three slices, like visualized in Figure \ref{fig::t1t2phantom_homogeneity}. Only in tube\footnote{Numbering referring to Figure \ref{fig::t1t2phantom_tubeOrdering}.} 4 and 8 some slight variations appear. Therefore the acquired relaxation parameter from the individual day of measurement are used as a reference.
	
	\begin{table}[!h]
		\footnotesize
		\centering
		\caption{Table with averaged parameters of each tube over all slices. The tube numbering is consistent with Figure \ref{fig::t1t2phantom_tubeOrdering}. The data is acquired at the 17.12.2018.}
		\begin{tabular}{|c|c|c|c|c|c|c|c|c|c|}\addlinespace[2ex]\hline
			Tube &	2&	3&	4&	5&	6&	7&	8&	9&	10\\ \hline\hline
			$T_1$ [ms] &819(3)	&1131(4)	&1428(5)	&820(3)	&1130(4)	&1421(6)	&849(3) &1154(4)	&1463(5)	\\ \hline
			$T_2$ [ms] &54(1)	&63(1)	&69(1)	&108(1)	&131(1)	&160(1)	&160(2)	&196(1)	&226(2)	\\ \hline
		\end{tabular}
		\label{tab::t1t2phantom_relaxation_times_20181217}
	\end{table}

	\subsubsection*{Temporal Stability of the Phantom}
\begin{figure}[!h]
		\centering
		\includegraphics[width=0.9\linewidth]{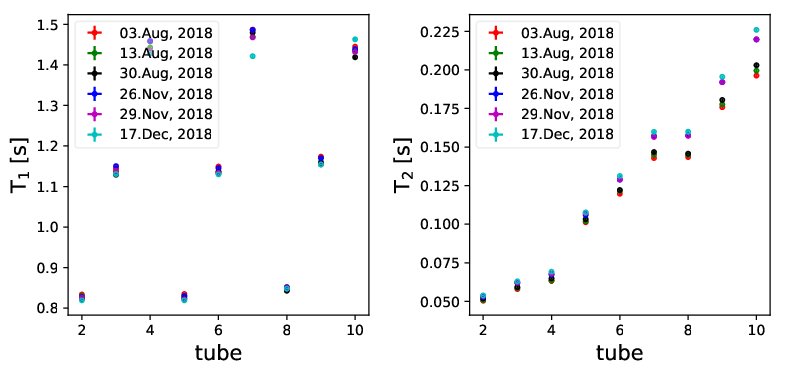}
		\caption{Visualization of the temporal stability of the relaxation parameter $T_1$ on the \textbf{left} and of $T_2$ on the \textbf{right} averaged over all three acquired slices including errors as standard-derivations propagated with the Gaussian error rule through all further calculations. }
		\label{fig::t1t2phantom_temporal stability}
	\end{figure}

	To learn more about the temporal stability of the phantom, the gold-standard characterization with exactly the same protocols over different months is repeated. At first three measurements are acquired within one month to get an idea of the parameter variations on smaller time-scales. Afterwards, three others are acquired approximately three month later to validate the stability on even larger time scales.\\
	The temporal stability within the tubes is good, while the combined $T_2$ values are changing depending on the time difference between those acquisitions. Therefore, it is necessary to acquire a new phantom $T_2$-characterization, if the last one is older than a week. The results for the stability analysis are visualized in Figure \ref{fig::t1t2phantom_temporal stability}.

	Here a decreasing of $T_2$, especially for higher values, can be observed. The measurements within 3 days (26. and 29. November 2018) are overlapping strongly for $T_2$. $T_1$ in comparison shows fluctuations during all measurements without any observable trend.

	\section{Discussion of Phantom Characterization}
	\label{ssec::t1t2phantom_discussion}
	Figure \ref{fig::t1t2phantom_temporal stability} shows the final range of $T_1$- and $T_2$-values covert by the custom-built phantom. The grid-spacing is very regular and the range between 819(3) ms and 1463(5) ms for $T_1$ as well as 54(1) ms and 226(2) ms for $T_2$ is good for representing torso tissue like heart and kidney \cite{Stanisz_Magn.Reson.Med._2005}. \\
	The homogeneity between different slices is high. This is represented as small errors for the over all slices averaged parameters of each tube. The corresponding visualization is representatively visualized by Figure \ref{fig::t1t2phantom_homogeneity}.\\
	The temporal stability is good for $T_1$ but not given for $T_2$ in time intervals longer than one week. Those require a repeated $T_2$-characterization.\\
	
	Nevertheless, there are sources of errors during the preparation of the phantom.
	First of all, a miscalculation of the amount of water occurred in the fourth tube. Additionally, the pipette was used incorrectly leading to an erroneously increased GdCl$_3$ concentration in the stock solution affecting all tubes systematically. The additional volumes are marked in Table \ref{tab::ingredients}. Referring to \cite{Hattori_Med.Phys._2013},
	the higher concentration of GdCl$_3$ results in lower $T_1$-values. The higher the concentrations are, the smaller becomes the error because the changes of the relaxation parameter get smaller.\\
	The miscalculation of the water component in tube 4 results in an error which is small compared to the effects of water evaporation during the filling of the Erlenmeyer flask in the preparation step. The necessary temperature for solving agarose and carragenaan was high enough to lead to strong evaporation of water. One possible way to reduce the error is to cook the solution with to much water and shrink it to 100 ml afterwards\footnote{It was not done, because the influence of the evaporation was expected to be lower.}.  This does not work in the cases in which the substances are not dissolved at that time-point, but for future experiments it is still highly recommendable.\\
	All in all, the values of the phantom have a good range for quantitative MRI experiments and are homogeneous within the wider part of the Falcon tubes. Even if the quantification has to be repeated regularly, the phantom allows to test quantitative MRI methods even for longer acquisition times and is a good representation for different torso tissues.\\
	
	The reference values, used in the following chapters, are gold-standard characterizations directly measured after the other sequences are applied. The phantom is not moved in between and the temperature can be assumed to be constant. Therefore, the accuracy of the developed reconstruction techniques can be determined on the basis of very precise reference maps.

\chapter{Sequence}
\label{ch::sequence}

\section{Inversion-Prepared bSSFP Sequence}
	\label{sec::sequence_ibSSFP}
\begin{figure}[!h]
		\centering
		\includegraphics[width=0.8\linewidth]{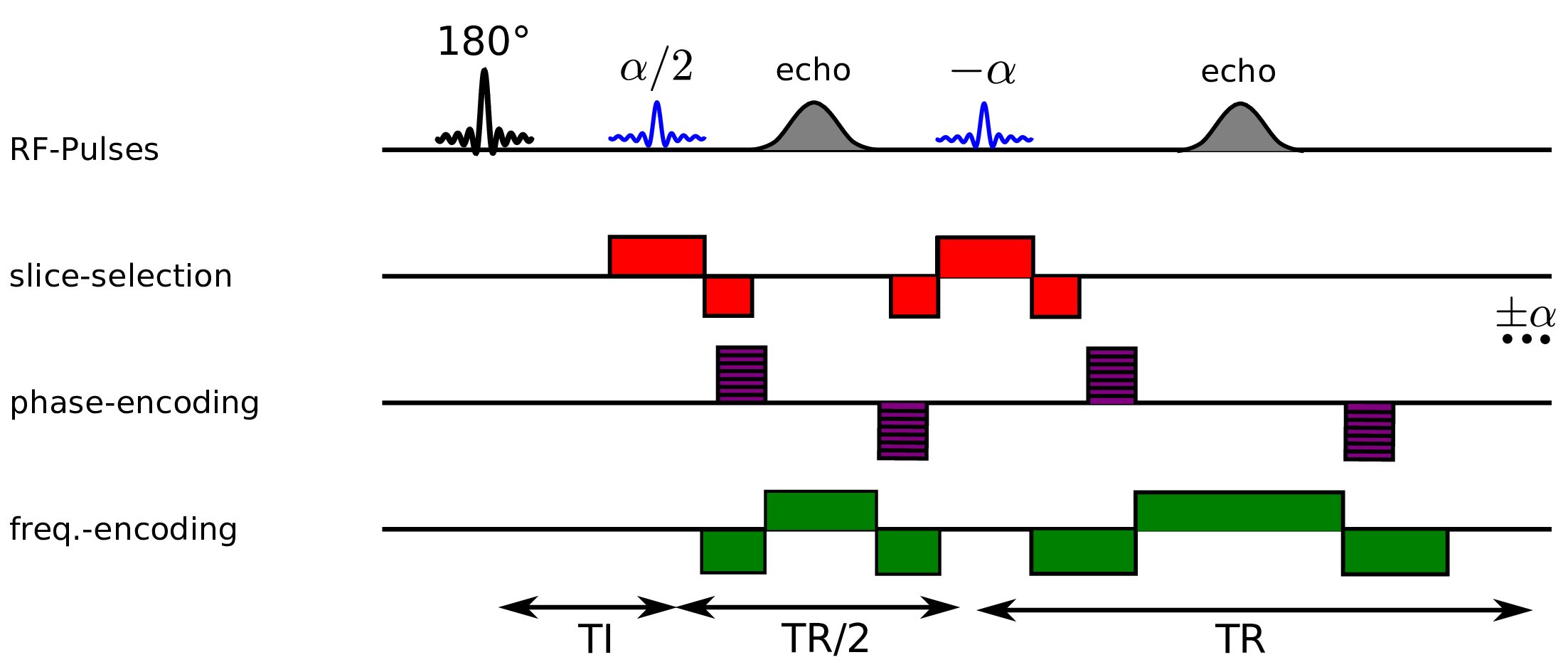}
		\caption{Visualization of an inversion-prepared bSSFP sequence with its gradient switching and RF-pulses. The magnetization preparation is added with an $\alpha$/2 pulse and a shortened TR. The $\pi$-switching phase is represented as sign of the flip-angle.}
		\label{fig::sequence_design_ibSSFP}
	\end{figure}

	To test the developed reconstruction technique, an inversion-prepared bSSFP sequence is selected. Recent work \cite{Schmitt_Magn.Reson.Med._2004}\cite{Ma_Nature_2013}\cite{Scheffler_Magn.Reson.Med._2001}
	has shown its sensitivity to $T_1$ and to $T_2$, it has a strong tissue contrast as well as a high signal intensity can be expected. Additionally, theoretical model is known and the simple behavior of the magnetization during one TR is an advantage. A typical sequence diagram is visualized in Figure \ref{fig::sequence_design_ibSSFP}. 

	It extends the bSSFP already explained in section \ref{ssec::theory_bSSFP} with an inversion-pulse which flips the net magnetization $\bm{M}$ around 180°. The images are then acquired during the relaxation of $\bm{M}$ until a steady-state is reached. While $\bm{M}$ is in a transition state during the acquisition, preparation-pulses like they are used in normal bSSFP imaging to reach a steady-sate are not applied. To avoid strong oscillations of the signal during the transition, the first flip-angle is reduced to $\alpha/2$ and its repetition time to TR/2 \cite{Scheffler_Eur.Radiol._2003}.
	This sets the starting point of the transition to the edge of a cone, illustrated in Figure \ref{fig::bSSFP_cone}, and allows the evenly regrowing magnetization.\\	
	
\begin{figure}[!h]
		\centering
		\includegraphics[width=0.8\linewidth]{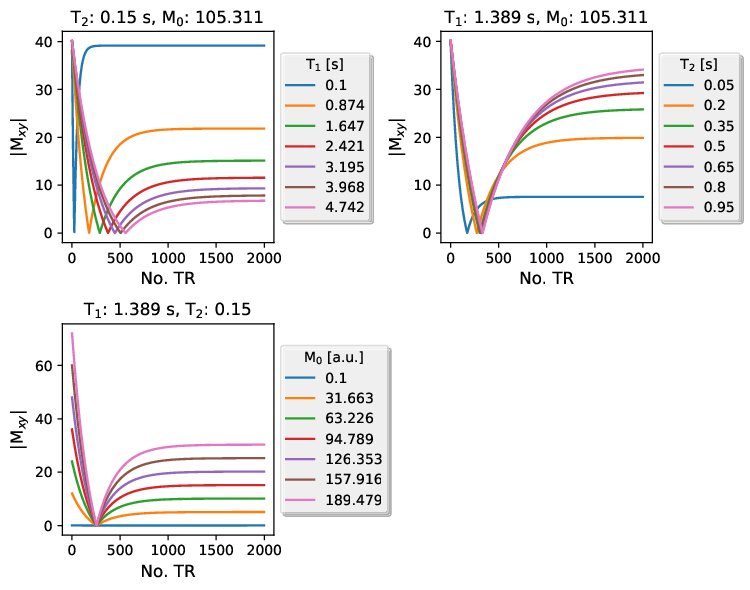}
		\caption{Visualization of the signal-development during an inversion-prepared bSSFP sequence simulated with varying relaxation parameter and offsets. The simulation defining parameters are TR/TE= 3.0/1.5 s and a flip-angle of 45°. }
		\label{fig::simulation_uniqueness_ibSSFP}
	\end{figure}

	To get an idea of the influence of the $T_1$, $T_2$, and $M_0$ on the signal behavior, Figure \ref{fig::simulation_uniqueness_ibSSFP} visualizes the signal development based on the theoretical expectations following the later introduced equation \ref{eq::ibSSFP_signal_model} for various combinations of relaxation- and density-parameters. While $M_0$ scales, $T_1$ and $T_2$ influence both the zero-crossing and the final steady-state signal. Their effect differ especially in the time interval after the zero-crossing on the way to the final steady state. The transient phases before are very similar.

\section{Measurement}
\label{ssec::measurements}

	All measurements are performed on a SIEMENS Skyra 3T MRI scanner. The used inversion-prepared bSSFP sequence is implemented in the integrated development environment for (MR) applications IDEA framework. To reduce slice-profile effects during the inversion, a non-slice-selective adiabatic pulse with a length of 10 ms is selected. 	
	The sequence is based on a radial sampling scheme because it is more robust to object-motion during the measurement and allows high undersampling of the k-space \cite{Lauterbur_Nature_1973} \cite{Trouard_J.Magn.Reson.Imaging_1996}.
	Two different types of experiments are performed. Both are explained in the following.

	\subsection{Fully-Sampled Acquisitions}
\begin{figure}[!h]
		\centering
		\includegraphics[width=0.9\linewidth]{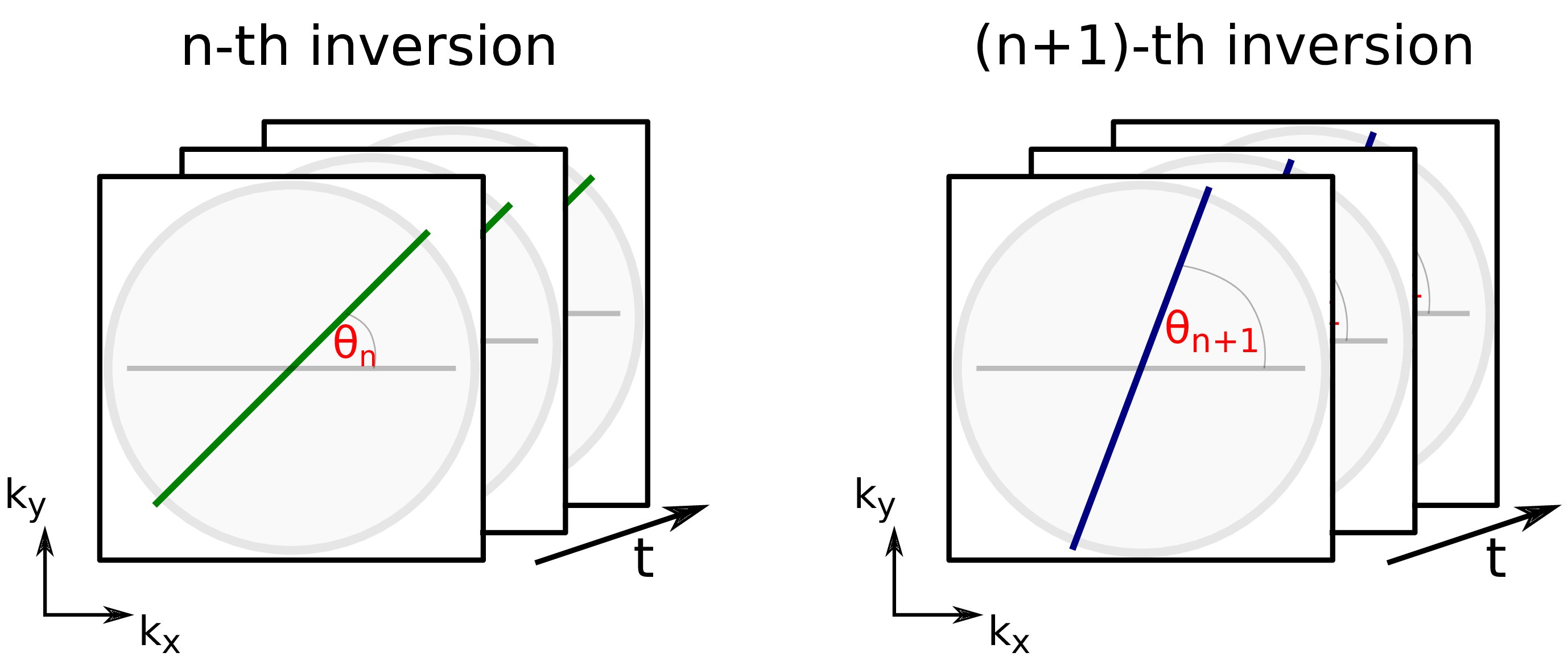}
		\caption{Visualization of an aligned sampling scheme during two inversion repetitions using an inversion-prepared bSSFP sequence implementation. The rotation angle $\theta$ follows equation \ref{eq::sequence_fully_sampled_rotation_angle}.}
		\label{fig::sequence_sampling_schemes_aligned}
	\end{figure}

	The aim of the fully-sampled acquisitions is to sample a dataset with full k-spaces at each repetition time. This is reached by acquiring 191 spokes for each TR during the recovering of the magnetization after inversion. Therefore, every k-space includes enough information to be reconstructed without artifacts using a Fast-Fourier transformation and is referring to the gold standard for future undersampled methods.\\
	A fully-sampled k-space per repetition is only possible if just one spoke is acquired per TR. This requires multiple inversion experiments with always one spoke measured during the relaxation of $\bm{M}$. To ensure that the same signal curve is sampled over the different inversion experiments, $\bm{M}$ needs to be in the same initial magnetization state before each inversion. This is realized by including a break of 10 s after each measurement to allow a full relaxation of $\bm{M}$ and assume a static phantom. For the clinical practice with volunteers and patients these acquisition times of about 39:07 min are too long.\\
	In the implementation used in this work the rotation angle $\theta$ of the spoke is constant per inversion experiment:
	\begin{align}
		\theta = \frac{n_{\textrm{inv}}}{N_{\textrm{inv}}}\cdot 2\pi
		\label{eq::sequence_fully_sampled_rotation_angle}
	\end{align}
	with the total number of inversions $N_{\textrm{inv}}$ and the actual one $n_{\textrm{inv}} = 0,\dots,N_{\textrm{inv}}-1$. Finally, this leads to an aligned sampling pattern within each frame, visualized in Figure \ref{fig::sequence_sampling_schemes_aligned}.
	
	An exemplary protocol is added to Table \ref{tab::sequence_protocol_FS_ibSSFP}.

	\subsection{Single-Shot Acquisitions}
	The goal of a single-shot acquisition is to speed up the measurement and to acquire a dataset of the whole relaxation curve during one single inversion experiment. This avoids unused time during breaks of the measuring procedure that are required to ensure a full relaxation of the magnetization. This makes multiple inversions obsolete. It also leads to smaller amounts of energy inserted into the measured objects and allows a clinically useful acquisition time of 7 s. The trade-off comes with less acquired information for each repetition time, which is solved using a model-based reconstruction.\\
	In practice the implementation of the single-shot acquisition works with a golden-angle based sampling pattern. This allows to combine data from consecutive repetitions while still regular sampling of k-space. The rotation angle is adjusted for each TR following \cite{Winkelmann_IEEETrans.Med.Imag._2007}\cite{Wundrak_IEEETransMedImag_2015}\cite{Wundrak_Magn.Reson.Med._2016_2}:
	\begin{align}
		\theta = \frac{\pi}{\tau + N_{\textrm{tGA}} - 1}\cdot n_{\textrm{TR}},
		\label{eq::sequence_single_shot_rotation_angle}
	\end{align}
	with the golden ratio $\tau = \frac{1 + \sqrt{5}}{2}$, the number of the used tiny golden-angle $N_{\textrm{tGA}}$, and of the current repetition $ n_{\textrm{TR}}$.
	
	\begin{figure}[!h]
		\centering
		\includegraphics[width=0.9\linewidth]{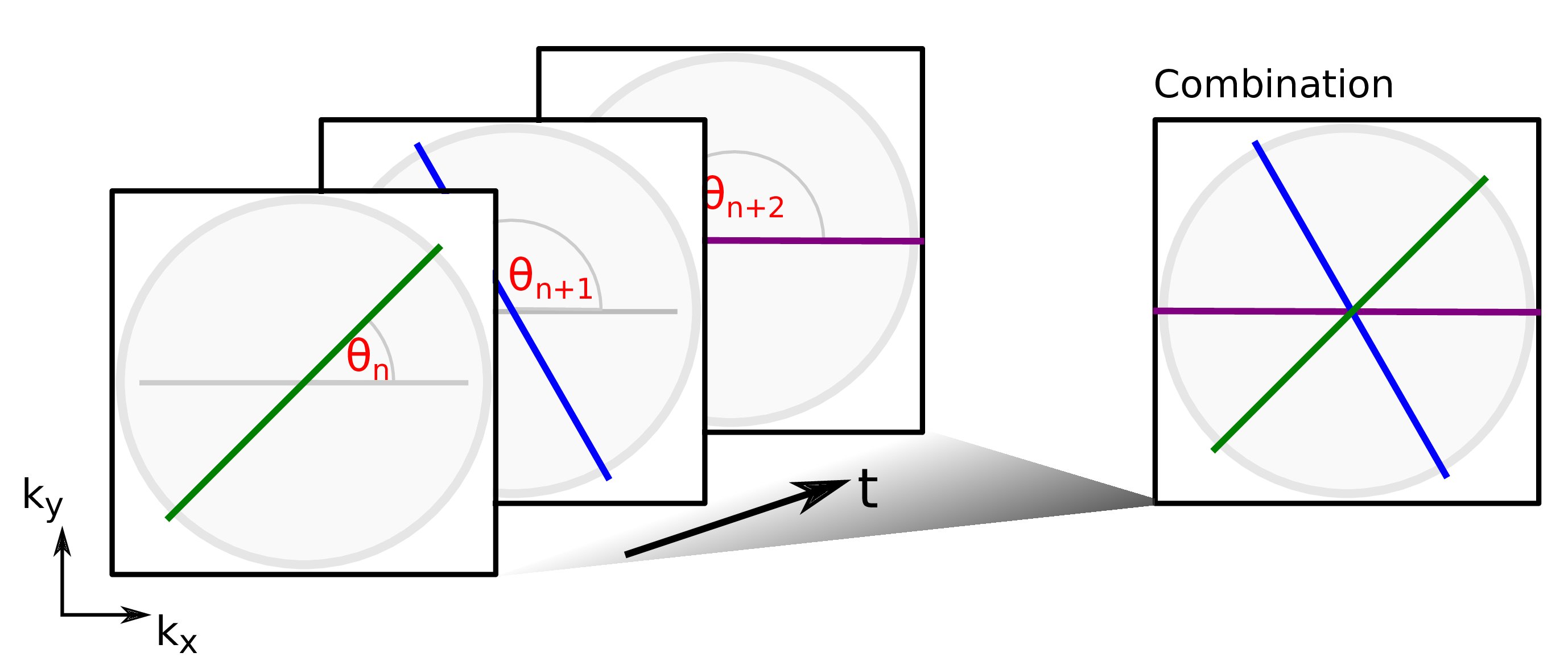}
		\caption{Visualization of a golden-angle based sampling scheme using just one inversion of an inversion-prepared bSSFP sequence implementation. The rotation angle $\theta$  follows equation \ref{eq::sequence_single_shot_rotation_angle}.}
		\label{fig::sequence_sampling_schemes_ga}
	\end{figure}

	An exemplary protocol is added to Table \ref{tab::sequence_protocol_SS_ibSSFP}.

\chapter{Setting up a Simulation Environment}

\section{Fundamental Principles for Simulation}
\label{sec::simulation_fundamental_principles}
	Different methods exist to set up a simulation environment for MRI experiments. They are all based on the Bloch equations in equation \ref{eq::Bloch-Equation} but differ in the degree of assumptions, which are used to solve it for the time-development of the magnetization $\bm{M}$.

	\subsection{Rotation-based Method}
	\label{ssec::rot_based_simu}
\begin{figure}[!h]
		\centering
		\includegraphics[width=\linewidth]{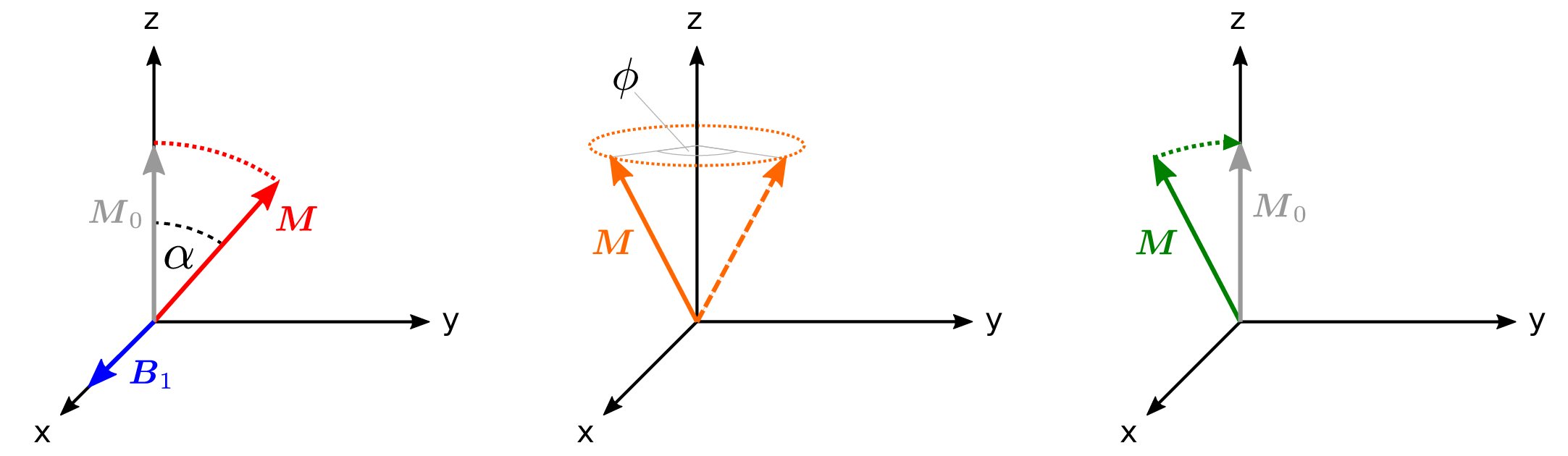}
		\caption{Visualization of the rotating effects of the excitation (\textbf{left}), off-resonance/gradient (\textbf{center}) and relaxation (\textbf{right}). }
		\label{fig::simulation_matrix_rotations}
	\end{figure}

	Many simulation in MRI are based on rotation-matrices $\bm{R}_i$ for $i=x,~y,~z$ to describe the effect of RF-pulses using hard-pulses, relaxation, and off-resonances, visualized in \mbox{Figure \ref{fig::simulation_matrix_rotations}}.
	
\begin{figure}[!h]
		\centering
		\includegraphics[width=0.7\linewidth]{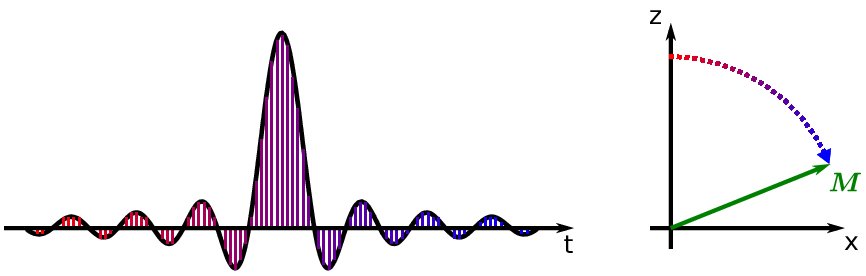}
		\caption{Sketch of discretized RF-pulse envelope (left). The colors are related to the different flip-angles of the magnetization visualized on the right.}
		\label{fig::simulation_discretized_rf_pulse}
	\end{figure}

	A simplified form is exploited in section \ref{ssec::theory_bSSFP} to derive the equation for a bSSFP steady-state signal. The time-development of the magnetization $\bm{M}$ is discretized into equally spaced intervals, which are chosen to be short enough to avoid relaxation effects in between. An RF-pulse then becomes a series of delta pulses, illustrated in Figure \ref{fig::simulation_discretized_rf_pulse}, which rotate $\bm{M}$ following
	
	\begin{align}
		\bm{M}_+ = \bm{R}_y(\theta)\bm{R}_z(\phi)\bm{R}_x(\Delta\alpha')\bm{R}_z(-\phi)\bm{R}_y(-\theta)\bm{M}_-,
	\end{align}
	with the magnetization before ($\bm{M}_-$) and after ($\bm{M}_+$) the RF-pulse, its phase $\phi$, the tipping angle of the effective field \mbox{$\theta = \arctan\left(\frac{\Delta\omega\cdot\Delta t}{\Delta\alpha}\right)$}, its flip-angle \mbox{$\Delta\alpha'=\sqrt{\Delta\omega^2+\left(\frac{\Delta\alpha}{\Delta t}\right)^2}$}, the small time-step $\Delta t$, and the off-resonance frequency $\Delta\omega=\gamma B_0-\omega_{\textrm{rf}}$.
	
	Relaxation effects are following with
	\begin{align}
		\bm{M}_+=\bm{R}_{1,\textrm{rel}}(\Delta t)\bm{M}_-+\bm{R}_{2,rel}(\Delta t),
	\end{align}
	and 
	\begin{align}
		\bm{R}_{1,rel}(\Delta t)=
		\begin{pmatrix}
		E_2 & 0 & 0\\
		0 & E_2 & 0\\
		0 & 0 & E_1\\
		\end{pmatrix},
		~~~~~~~~
		\bm{R}_{2,rel}(\Delta t)=
		(1-E_1)
		\begin{pmatrix}
		0 \\
		0 \\
		1 \\
		\end{pmatrix},
	\end{align}
	assuming $M_0=1$ and $E_{1,2}=e^{-\frac{\Delta t}{T_{1,2}}}$.\\
	Free-precession effects are rotating the magnetization in this model around the $z$-axis. The angle depends on the two main effects, which can result in free-precession: gradients
	\begin{align}
		\theta_g=\gamma\bm{r}\int\limits_t^{t+\Delta t}\bm{G}(\tau)\textrm{d}\tau,
	\end{align}
	which depend on the location $\bm{r}$, the time $t$, and field inhomogeneities
	\begin{align}
		\theta_i=\gamma\Delta \bm{B}(\bm{r})\Delta t,
	\end{align}
	which depend on the position only. At the end, the free-precession rotational matrix becomes
	\begin{align}
		\bm{M}_+=\bm{R}_z(\theta_g)\bm{R}_z(\theta_i)\bm{M}_-.
	\end{align}
	Using this matrix-based method, allows for calculation of the RF-pulse energy using
	\begin{align}
		E = \sum\limits_{n=1}^{N}|A_n|^2\Delta t
		\label{eq::Energy_rf_pulse_discrete}
	\end{align}
	with the amplitude $A_n$ of the $n$th discrete delta peak, the number of all delta peaks $N$ and the sample size $\Delta t$.

	\subsection[Runge-Kutta Method]{Runge-Kutta Method\footnote{The main source of this chapter is \cite[p.710ff]{Press__1992}.}}
	
	The simulation based on rotational matrices, presented in section \ref{ssec::rot_based_simu}, has the disadvantage that the sample-size has to be constant over the whole simulation. While it needs to be small for using RF-pulses, it leads to unnecessary calculations in regions with no large changes, like in relaxation intervals. Therefore, a simulation, that automatically adjusts its step-size would be desired. This leads to an other class of possible Bloch simulations: the ordinary differential equation (ODE) solver based methods. Those are returning the time-development of $\bm{M}$ directly by approximating the Bloch equations stepwise and using information of the gradients at different time-points. Combined with an adaptive step-size control, the time-development of $\bm{M}$ can be calculated very efficiently.\\
	The selected Runge-Kutta method (RK) is an iterative procedure to solve systems of ODEs like
	\begin{align}
		\frac{\textrm{d}y_i(x)}{\textrm{d}x}=f_i(x,y_1,...,y_N)
	\end{align}
	with $i$=1,...,$N$. To translate this to the Bloch equations in \ref{eq::Bloch-Equation} $y_{1,2,3} = M_{x,y,z}$, $x=t$ and $f_{1,2,3}$ corresponds to the known equations on the right side of \ref{eq::Bloch-Equation}.\\
	The Runge-Kutta method extends the Euler method. While the latter calculates the $n+1$ step from the previous $n$ by using information just from the beginning of the interval $h$,
	\begin{align}
		y_{n+1} = y_n + hf(x,y) + O(h^2),~~\textrm{with}~~x_{n+1} = x_n + h,
	\end{align}
	the RK-method exploits gradient information of intermediate steps within $h$. As it is only the midpoint of $h$, it is called the midpoint-method \cite{Press__1992}:
	\begin{align}
		k_1 &= hf(x,y)\\
		k_2 &= hf(x_n+\frac{1}{2}h, y_n+\frac{1}{2}k_1)\\
		y_{n+1}& = y_n + k_2 + O(h^3).
	\end{align}
	It increases the order of the error term to $ O(h^3)$. The basic idea of the RK method is that by adding up more gradient information of intermediate steps with the right coefficients $k_i$ the error can be even further reduced. This allows a higher generalization to arbitrary stage numbers $s$:
	\begin{align}
		y_{n+1} &= y_n + \sum\limits_{i=1}^{s}c_ik_i\\
		k_i &= h_nf\left(t_n+a_ih_n, y_n+\sum\limits_{j=1}^s b_{ij}k_j\right),~~\textrm{with}~~i=2,\dots,s
	\end{align}
	with the coefficients $a$, $b$ and $c$ typically saved in a Butcher tableau shown in Table \ref{tab::simulation_butcher_tableau_overview}.\\
	As an example, a short view on the workhorse of solving ODEs, the RK4 algorithm, is following. To increase its efficiency, extensions to an adaptive step-size control are possible. The two common implementations are step-doubling and the embedded formula solution. Here the focus is on the latter, because it is two times more efficient and can be realized by a classical RK5 algorithm \cite{Press__1992}:
	\begin{align}
		k_1 &= hf(x_n,y_n)\nonumber\\
		k_2 &= hf(x_n+a_2 h, y_n+b_{21}k_1)\nonumber\\
		&~\dots\label{eq::equations_rk4}\\
		k_6 &= hf(x_n+a_6 h, y_n+b_{61}k_1+\dots+b_{65}k_5)\nonumber\\
		y_{n+1} &= y_n + c_1k_1+c_2k_2+c_3k_3+c_4k_4+c_5k_5+c_6k_6+O(h^6),\nonumber
	\end{align}
	with embedded RK4:
	\begin{align}
		y_{n+1}^* = y_n + \sum\limits_{i=1}^{s=6}c_i^*k_i + O(h^5).
	\end{align}
	Both are evaluated at the same time-points and the coefficients can be visualized in a Butcher tableau in Table \ref{tab::simulation_butcher_tableau_overview} following their positions in equations \ref{eq::equations_rk4}.
	
	\begin{table}[!h]
		\centering
		\caption{Butcher tableau of the RK4 algorithm with the coefficients $a$, $b$ and $c$ following their positions in equations \ref{eq::equations_rk4}.}
		\begin{tabular}{c|c c c c c}\addlinespace[2ex]
			0 & & & & & \\
			$a_2$ & $b_{21}$ & & & & \\
			$a_3$ & $b_{31}$ & $b_{32}$ & & & \\
			$\vdots$ & $\vdots$ & & $\ddots$ & & \\
			$a_s$ & $a_{s1}$ & $a_{s2}$ & $\dots$ & $a_{s,s-1}$ & \\ \hline
			& $c_1$ & $c_2$ & $\dots$ & $c_{s-1}$ & $c_s$ \\			
		\end{tabular}
		\label{tab::simulation_butcher_tableau_overview}
	\end{table}
	
	The error between the fifth and the fourth order methods $\textrm{err} = y_{n+1}-y_{n+1}^*$ can then be used to update the step-size $h$:
	\begin{align}
		h_{n+1}=h_n\theta\left(\frac{\delta}{\textrm{err}}\right)^{\frac{1}{5}},
	\end{align}
	with the tolerated error $\delta$ and a safety factor $\theta$.\\
	In this work a RK5(4) algorithm, designed by Dormand and Prince, also called Dopri54, is selected because of its accuracy and stability \cite{Dormand_J.Comput.Appl.Math._1980}. Its coefficients can be found in the butcher tableau in Table \ref{tab::simulation_butcher_tableau_Dopri54}.
	
	\begin{table}[!h]
		\setlength{\tabcolsep}{12pt}
		\renewcommand{\arraystretch}{1.3} 
		\centering
		\caption{Butcher tableau with RK5(4) coefficients determined by Dormand and Prince \cite{Dormand_J.Comput.Appl.Math._1980}.}
		\begin{tabular}{c|c c c c c c c}\addlinespace[2ex]
			0 & & & & & & & \\
			$\frac{1}{5}$ & $\frac{1}{5}$ & & & & & & \\
			$\frac{3}{10}$ & $\frac{3}{40}$ & $\frac{9}{40}$ & & & & & \\
			$\frac{4}{5}$ & $\frac{44}{45}$ & $-\frac{56}{15}$ & $\frac{32}{9}$ & & & & \\
			$\frac{8}{9}$ & $\frac{19372}{6561}$ & $-\frac{25360}{2187}$ & $\frac{64448}{6561}$ & $-\frac{212}{729}$ & & & \\
			1 & $\frac{9017}{3168}$ & $-\frac{355}{33}$ & $\frac{46732}{5247}$ & $\frac{49}{176}$ & $-\frac{5103}{18656}$ & & \\
			1 & $\frac{35}{384}$ & 0 & $\frac{500}{1113}$ & $\frac{125}{192}$ & $-\frac{2187}{6784}$ & $\frac{11}{84}$ & \\ \hline
			& $\frac{35}{384}$ & 0 & $\frac{500}{1113}$ & $\frac{125}{192}$ & $-\frac{2187}{6784}$ & $\frac{11}{84}$ & \\
			& $\frac{5179}{57600}$ & 0 & $\frac{7571}{16695}$ & $\frac{393}{640}$ & $-\frac{92097}{339200}$ & $\frac{187}{2100}$ & $\frac{1}{40}$\\			
		\end{tabular}
		\label{tab::simulation_butcher_tableau_Dopri54}
	\end{table}

\section{Structure of Simulation}
	
	\begin{figure}[!h]
		\centering
		\includegraphics[width=0.8\linewidth]{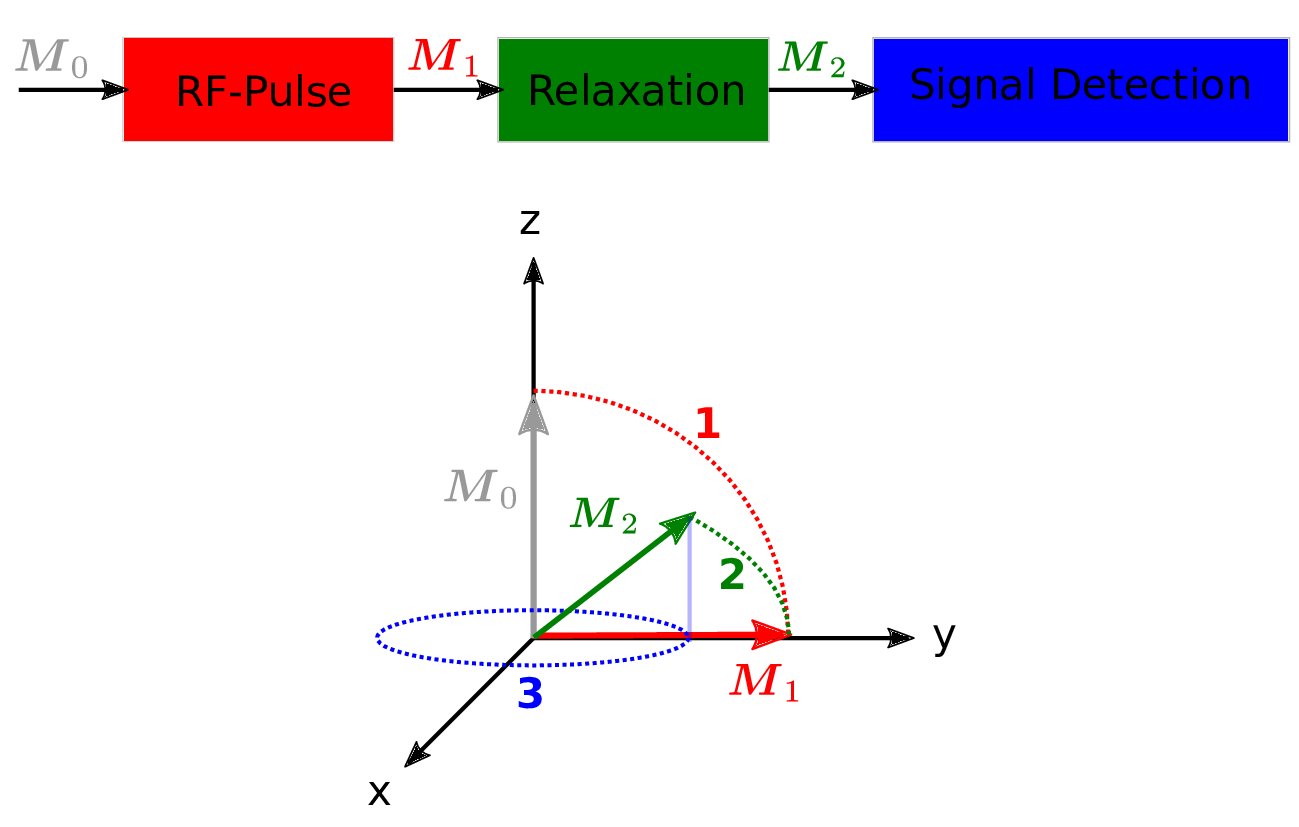}
		\caption{Overview of the block structure of the developed simulation. The three blocks RF-pulse (red), relaxation (green), and signal detection (blue) are sketched with their effects on the magnetization. The colored magnetization states $\bm{M}_{0,1,2}$ are representing the states that are passed to the following blocks. }
		\label{fig::simultation_structure}
	\end{figure}
	
	The main structure of the simulation is illustrated in Figure \ref{fig::simultation_structure}. It consists of different blocks for RF-pulses, relaxation events, and readouts. The basic idea is to solve the ODE-representation of the Bloch equations \ref{eq::Bloch-Equation} in time using a Dopri54-algorithm with partially embedded step-size control. The actual magnetization is passed from one sequence event block to the other. This allows to set up almost every sequence like on the MRI scanner. The simulation can work with multiple spins, so slice-profiles as well as off-resonance effects are captured by simulating multiple spins with varying flip-angles or different phase offsets and by averaging them.

	\subsubsection*{RF-Pulse Block}
\begin{figure}[!h]
		\centering
		\includegraphics[width=0.5\linewidth]{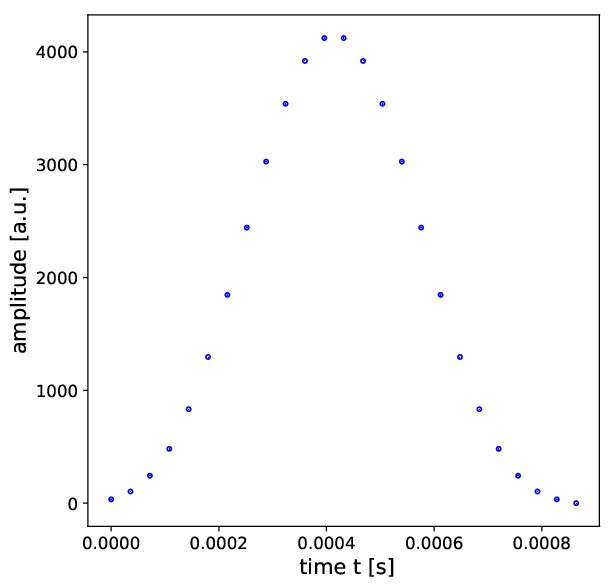}
		\caption{Visualization of an RF-pulse simulated with the Dopri54 ODE-solver. The pulse length is 0.9 ms and the FA is 45°. }
		\label{fig::simultation_rfPulsePlot}
	\end{figure}

	The RF-pulse block is based on a $B_1$-field, which is added to the Bloch equations \ref{eq::Bloch-Equation}. Its amplitude $A$ is calculated using equation \ref{eq::rf_pulse_shape_sinc}. The phase spreads the rotating effect on the $x$- and the $y$-component of the magnetization using triangular functions analogue to equation \ref{eq:labFrame_B1}. While the pulse is sampled homogeneously with an interval size of $\Delta t$, the pulse-energy is calculated with equation \ref{eq::Energy_rf_pulse_discrete}. The sampled RF-pulse is sketched in Figure \ref{fig::simulation_discretized_rf_pulse} and a real simulation is visualized in Figure \ref{fig::simultation_rfPulsePlot}. 
	
	\subsubsection*{Relaxation Block}
	The relaxation block solves the Bloch equations without additional magnetic fields or gradients. As the changes of $\bm{M}$ in time are much smaller than during excitation, the adaptive step-size control decreases computational costs efficiently.

	\subsubsection*{Readout Block}
	Using an embedded RK-method, it has to be ensured that the readout acts at the correct echo time TE. This is realized by splitting the relaxation block at TE and saving the magnetization state before passing $\bm{M}$ to the second relaxation block.

\section{Simulated Sequences}
\begin{figure}[!h]
		\centering
		\includegraphics[width=\linewidth]{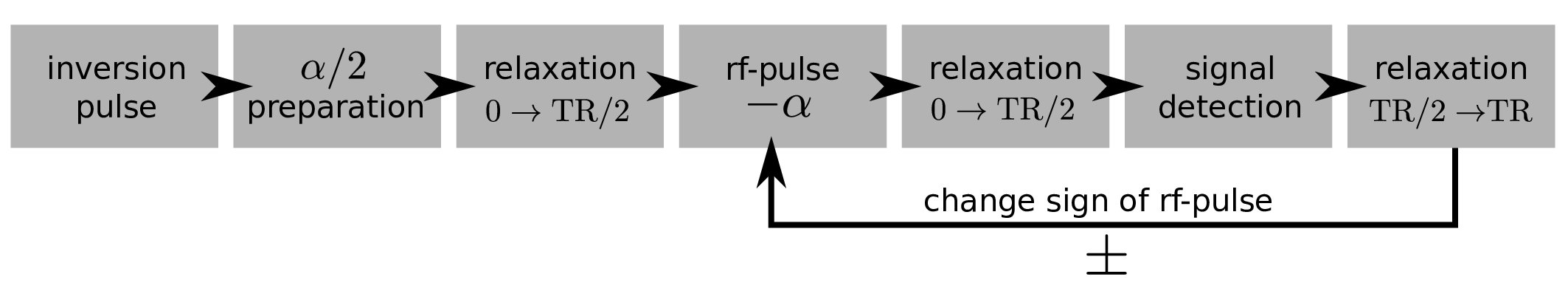}
		\caption{Overview of the event-blocks for simulating an inversion-prepared bSSFP sequence.}
		\label{fig::simulation_ibSSFP_event_blocks}
	\end{figure}

	Through its block-wise design the simulation tool can work with any kind of sequence. The actual implementation mainly provides gradient-echo sequences like bSSFP, phase-cycled bSSFP, FLASH, and their inversion-prepared versions, but there is even an option for simulating a multi-echo spin-echo one. In the following, the inversion-prepared bSSFP is used, which is already mentioned in section \ref{sec::sequence_ibSSFP}. A visualization of its event-blocks is illustrated in Figure \ref{fig::simulation_ibSSFP_event_blocks}.
	
\begin{figure}[!h]
		\centering
		\includegraphics[width=0.7\linewidth]{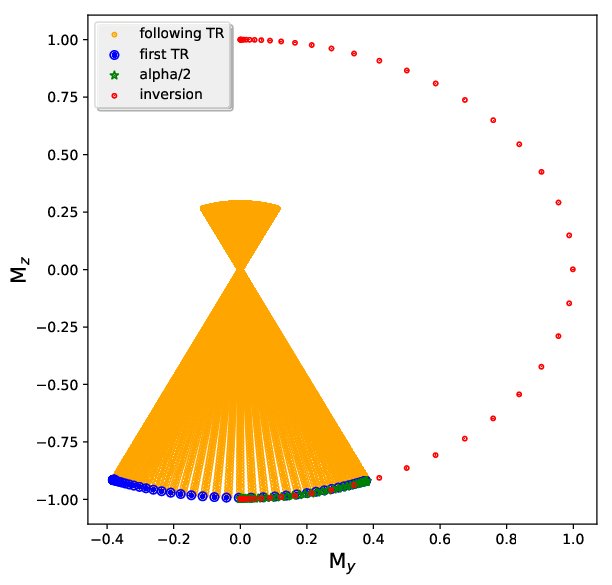}
		\caption{Simulation of an inversion-prepared bSSFP sequence using the Dopri54 ODE-solver and the structure illustrated in Figure \ref{fig::simulation_ibSSFP_event_blocks}. The defining simulation parameters are $N_{\textrm{TR}}$ = 2000, TR/TE= 3.0/1.5 ms, $T_1/T_2$=1.60/0.09 ms, $t_{RF}$=0.9 ms, $t_{inv}$=0.9 ms, $\alpha=$45°, and $M_0$=1. The points are plotted in samples so the time differences between them are not homogeneous (except during RF-pulses). Therefore, less points are plotted during relaxation. The step width is increased in those regions. }
		\label{fig::simulation_ibSSFP_magnetization_behavior}
	\end{figure}

	A block for inversion- and magnetization-preparation through an $\alpha/2$-pulse and a shortened TR of TR/2 is presented. The latter consists of an rf-event and a relaxation. Afterwards, the magnetization is flipped by an $\alpha$-pulse with a $\pi$-phase and then passed to an other relaxation event, which ends at TE. The first signal is collected and $\bm{M}$ is passed to the last relaxation block until the repetition ends at TR. Then, the next $\alpha$-pulse with a $\pi$ larger phase starts and the procedure is repeated until the last TR is finished.\\
	A visualization of $\bm{M}$ during that type of inversion-prepared bSSFP sequence is visualized in Figure \ref{fig::simulation_ibSSFP_magnetization_behavior}. 

	\FloatBarrier
	
\begin{figure}[!h]
		\centering
		\includegraphics[width=0.8\linewidth]{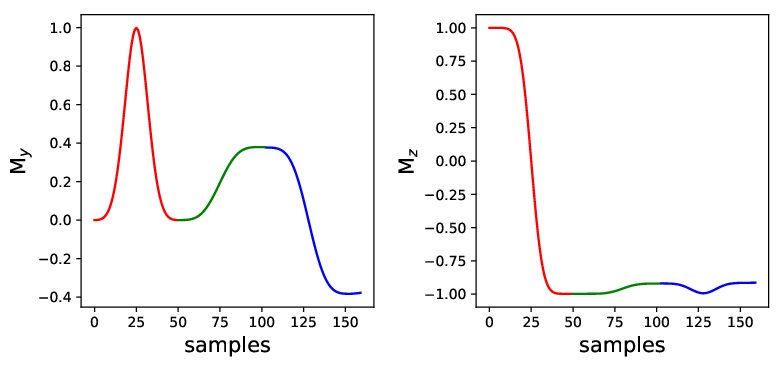}
		\caption{Visualization of the $M_y$ and $M_z$ component of the magnetization during the first colored phases of an inversion-prepared bSSFP sequence illustrated in Figure \ref{fig::simulation_ibSSFP_magnetization_behavior}. The simulation parameters are the same and the colors correspond to the phases in Figure \ref{fig::simulation_ibSSFP_magnetization_behavior}. Samples are favored over homogeneous time points for reasons named above. }
		\label{fig::simulation_ibSSFP_magentization_components}
	\end{figure}

	The colors symbolically represent different phases of the sequence. The red phase is the transition of $\bm{M}$ during a 0.9 ms long inversion-pulse, until $M_z$ is inverted. Afterwards a 0.9 ms long $\alpha/2$-pulse is visualized in green. It prepares the magnetization to be at the cones edge. The included 1.5 ms long relaxation is not visible, because it is not changing the magnetization much. After the preparation of $\bm{M}$ the train of RF-pulses with $\pi$-phase difference starts. It is illustrated in orange, while the first TR is visualized in blue. It is an RF-pulse of 0.9 ms length with a flip-angle FA $\alpha$ and a $\pi$-phase including inversion until a TR of 3 ms is reached. The following 1999 TRs are shown in orange until the steady-state at approximately $M_z$=0.26 is reached. The RF-pulses are aligned with the x-axis and no off-resonances of the spins are assumed. Therefore, the x-component of $\bm{M}$ is always 0 and the image is reduced to a two-dimensional representation.\\
	A development of $M_y$ and $M_z$ during the first three color-labeled sequence events is plotted in Figure \ref{fig::simulation_ibSSFP_magentization_components}. Here the samples rather than time-points are favored for reasons of clarity. They correspond to the time, but through the adaptive step-size control of the ODE-solver the sampling size is not homogeneous over the whole TR. It is much larger for relaxation, than for RF-pulses. 

	\FloatBarrier

\section{Numerical Phantom}
\label{sec::nummerical_phantom}
\begin{figure}[!h]
		\centering
		\includegraphics[width=0.8\linewidth]{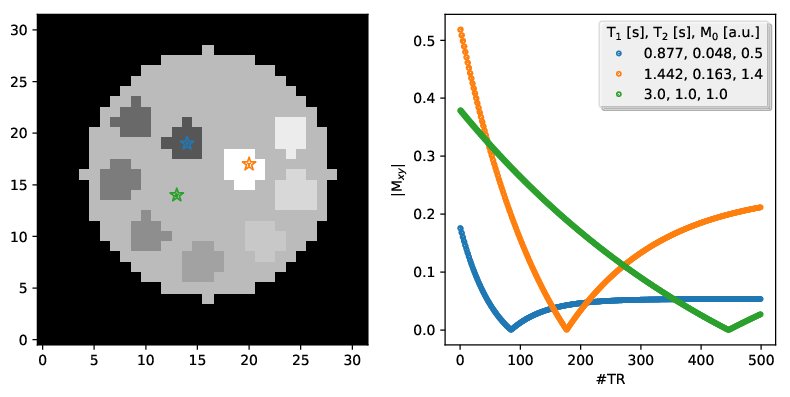}
		\caption{Visualization of a simulated phantom with a resolution of 32x32 px on the \textbf{left}. The colored stars are representing the positions of the signal traces on the \textbf{right}. The legend includes the belonging relaxation- and proton-density values. An inversion-prepared bSSFP with TR/TE= 4.5/2.25 ms, $t_{RF}$=0.9 ms, $t_{inv}$=10 ms, FA=45°, and 500 repetitions is simulated.}
		\label{fig::simulation_visual_phantom_signal.pdf}
	\end{figure}

	The numerical phantom in Figure \ref{fig::simulation_visual_phantom_signal.pdf} is based on a pixel-wise simulation of previously defined and time-constant relaxation parameters and proton-density maps. They are located in elliptical regions. For the interested reader even more complex shaped structured phantoms based on B\'ezier regions are possible \cite{Guerquin-Kern_IEEETrans.Med.Imag._2012}. 
	
\section{Verification of Simulation}
	In this section, it is described how the individual blocks for the relaxation, the RF-pulse events as well as the whole sequence using an inversion-prepared bSFFP are verified. The general strategy is to simulate scenarios where an exact analytical solution is known. In the end a more realistic sequence is compared to its simplified analytical model.

	\subsection{Verification of Relaxation Event-Block}
\begin{figure}[!h]
		\centering
		\includegraphics[width=0.6\linewidth]{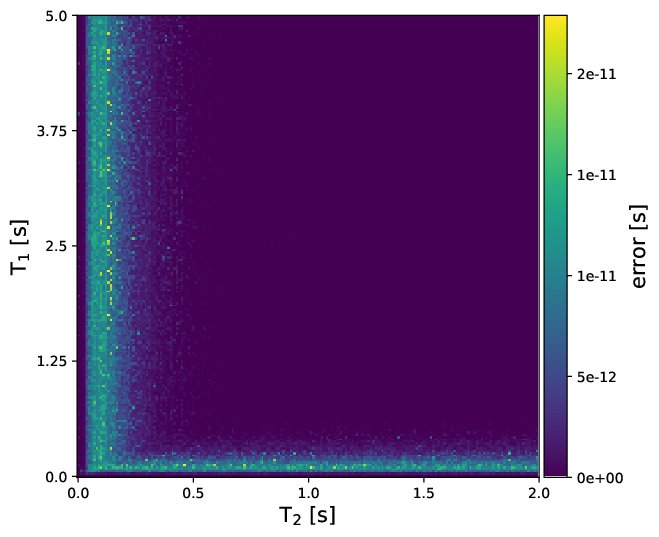}
		\caption{Visualization of the accuracy errors $\delta_{\textrm{rel}}$ of the relaxation event-block using an analytical model and the ODE-solver. $\delta_{\textrm{rel}}$ is color-coded and determined for 200 different $T_1$ and $T_2$, respectively. Fixed parameters are $\Delta t$ = 0.2 s and $\bm{M_0} = (0,1,0)^T$. }
		\label{fig::simultation_verif_relax}
	\end{figure}

	The relaxation event-block is verified using the Bloch equations \ref{eq::Bloch-Equation} as an analytical model. First, a static field in the rotating frame of reference $\bm{B}(t) = G_zz \bm{\hat{e}}_z = \frac{\omega_z}{\gamma}\bm{\hat{e}}_z$ is assumed and equation \ref{eq::Bloch-Equation} to
	\begin{align}
		\frac{\textrm{d}M_x}{\textrm{d}t} &= \frac{M_0-M_z}{T_1}\nonumber\\
		\frac{\textrm{d}M_y}{\textrm{d}t} &= \omega_z M_y - \frac{M_x}{T_2}\\
		\frac{\textrm{d}M_z}{\textrm{d}t} &= -\omega_z M_x -\frac{M_y}{T_2}\nonumber
	\end{align}
	simplified.	Afterwards, the differential equations are solved resulting in \cite[p.60]{Brown__2014}:
	\begin{align}
		M_x(t) &= \left[M_{0,x}\cos(\omega_z t)+M_{0,y}\sin(\omega_z t)\right]\cdot\textrm{e}^{-\frac{t}{T_2}}\nonumber\\
		M_y(t) &= \left[M_{0,y}\cos(\omega_z t)-M_{0,x}\sin(\omega_z t)\right]\cdot\textrm{e}^{-\frac{t}{T_2}}
		\label{eq::simulation_verification_relaxation_bloch_solution}\\
		M_z(t) &= M_{0,z} + (M_0 - M_{0,z})\cdot\textrm{e}^{-\frac{t}{T_1}}.\nonumber
	\end{align}
	Then, choosing a starting magnetization $\bm{M}_0$ and a time-interval $\Delta t$, the final $\bm{M}_{\Delta t}$ and the error $\delta_{\textrm{rel}}$ compared to the solution of the ODE-solver $\bm{M}_{\textrm{ODE}}$ can be calculated:
	\begin{align}
		\delta_{\textrm{rel}} = \sqrt{\sum\limits_{i}(M_{\Delta t,i}-M_{\textrm{ODE},i})^2}~~\textrm{for}~~i=(x,y,z).
	\end{align}
	The result is visualized for varying $T_1$ and $T_2$ and constant $\Delta t$ = 0.2 s and $\bm{M}_0 = (0,1,0)^T$ in Figure \ref{fig::simultation_verif_relax}.
	
	The ODE-solver reproduces the analytically expected results very accurately. The error shown in figure \ref{fig::simultation_verif_relax} is a numerical one and can not be avoided. Although it is too small to have a relevant impact on the simulations. The increased error for small relaxation parameters results from their stronger effect on the signal curve, but is still small enough to be ignored.

	\subsection{Verification of RF-Pulse Event-Block}
\begin{figure}[!h]
		\centering
		
		\begin{subfigure}{.45\textwidth}
			\centering
			\includegraphics[width=1.1\linewidth]{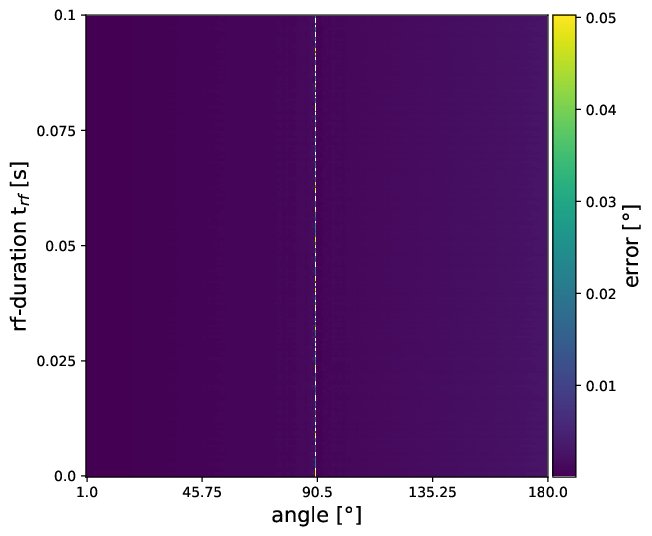}
			\caption{Hard-pulse approximation allows no relaxation effects during  $t_{\textrm{rf}}$.}
			\label{fig::simultation_verif_rf_noRelax}
		\end{subfigure}
		\hfill
		\begin{subfigure}{.45\textwidth}
			\centering
			\includegraphics[width=1.1\linewidth]{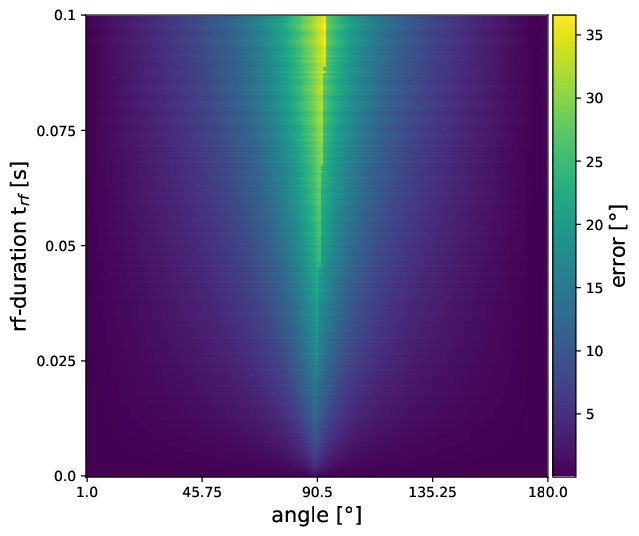}
			\caption{Relaxation effects during  $t_{\textrm{rf}}$ are included by using the ODE-solver.}
			\label{fig::simultation_verif_rf_Relax}
		\end{subfigure}
		\caption{Visualization of $\delta_{\alpha}$ for 200 different flip-angles $\alpha$ and RF-pulse duration times $t_{\textrm{rf}}$ with starting magnetization $\bm{M_0} = (0,0,1)^T$. }
	\end{figure}

	The RF-pulse event-block adds an additional $B_1$-field with phase $\phi$ to the external static field: $\bm{B}(t) = G_zz \bm{\hat{e}}_z + \cos(\phi) B_1\bm{\hat{e}}_x + \sin(\phi) B_1\bm{\hat{e}}_y$. As a result, $\bm{M}$ is turned by the flip-angle $\alpha$. It is tested if the expected $\alpha$ is reached for varying flip-angles and RF-pulse duration times $t_{\textrm{rf}}$ to check whether the simulated RF-pulses are working or not. Therefore, the ODE-solver is applied to solve the Bloch equations with an additional $B_1$ field, followed by a calculation of the actual flip-angle $\alpha_{\textrm{eff}}$, using
	
	\begin{align}
		\alpha_{\textrm{eff}} = 
		\begin{cases}
			\arcsin\left(\frac{\sqrt{(M_{t_{\textrm{rf}},x})^2+(M_{t_{\textrm{rf}},y})^2}}{|\bm{M}|}\right),~~\textrm{for } M_{t_{\textrm{rf}},z} \geq 0,\\
			\arccos\left(\frac{\sqrt{(M_{t_{\textrm{rf}},x})^2+(M_{t_{\textrm{rf}},y})^2}}{|\bm{M}|}\right),~~\textrm{for } M_{t_{\textrm{rf}},z} < 0.
		\end{cases}
	\end{align}
	The two cases are necessary because the range on which the triangular functions are defined is limited. The difference $\delta_{\alpha}$ between the wished $\alpha$ and the resulting $\alpha_{\textrm{eff}}$ are determined and visualized in Figure \ref{fig::simultation_verif_rf_noRelax} for varying $\alpha$ and $t_{\textrm{rf}}$ and without the influence of relaxation during $t_{\textrm{rf}}$.\\
	The largest error $\delta_{\alpha}$ is still in the range of 10$^{-2}$ degree for $\alpha_{\textrm{eff}}$ close to 90°, which leads to a good precision for the simulation. The central line of randomly alternating error values results from numerical errors during the calculation of $\alpha_{\textrm{eff}}$. The argument of the trigonometric functions exceeds its defined limits of [-1;1], while $\alpha_{\textrm{eff}}$ is chosen to be close to 90°. It has no influence on the accuracy of the simulation tool.\\
	A similar plot to Figure \ref{fig::simultation_verif_rf_noRelax} is added in \ref{fig::simultation_verif_rf_Relax}, where relaxation during $t_{\textrm{rf}}$ is allowed. It appears that the expected result, an increased error with larger $t_{\textrm{rf}}$ is present. The relaxation effect can influence the result for a longer time. Additionally, the larger effect of the $T_1$-relaxation on the flip-angle for a position around $\alpha$ = 90° leads to larger $\delta_{\alpha}$. 

	\subsection{Verification of the Sequence Simulation}
	\label{ssec::simulation_verification_sequence}
\begin{figure}[!h]
		\centering
		\includegraphics[width=0.8\linewidth]{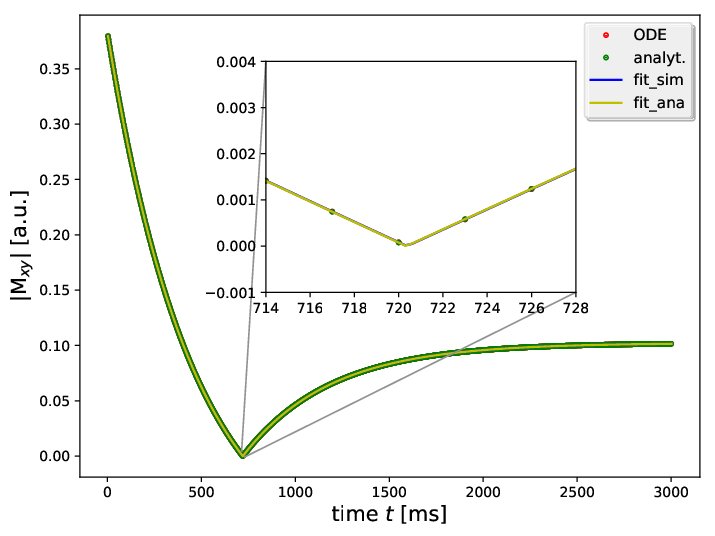}
		\caption{Visualization of the signal-development during an inversion-prepared bSSFP sequence simulated with the previously described Dopri54 ODE-solver, but assuming hard pulses. Additionally, the theoretically expected signal is plotted. Therefore, two non-linear fits with equation \ref{eq::ibSSFP_signal_model} are carried out for the analytical and the ODE-solver simulation. The determined parameters are added to Table \ref{tab::fitting_results_sequence_verification}. The simulation parameters are TR/TE= 3.0/1.5 s, $T_1$/$T_2$=1.60/0.09 s, FA=45°, $N_{\textrm{TR}}$ = 1000, and $M_0$=1. }
		\label{fig::simultation_sequence_test_hardpulse}
	\end{figure}

	The verification of the simulation for whole inversion-prepared bSSFP sequences is visualized in Figure \ref{fig::simulation_ibSSFP_event_blocks}. After determining the steady-state signal of a bSSFP sequence in equation \ref{eq::signal_bSSFP}, the signal model can be extended to an inversion-prepared model.\\
	It is assumed that the initial signal $S_0$ lies on the $\alpha/2$-cone shown in Figure \ref{fig::bSSFP_cone} and can therefore be approximated by
	\begin{align}
	S_0 = M_0\sin\frac{\alpha}{2}.
	\end{align}
	Additionally, $S_{stst}$ is always lower than $S_0$ and thus the signal follows a restricted growth model \cite{Schmitt_Magn.Reson.Med._2004}:
	\begin{align}
	S(n\textrm{TR})=S_{stst} - (S_{stst} + S_0)e^{-\frac{n\textrm{TR}}{T_1^*}},
	\label{eq::ibSSFP_signal_model}
	\end{align}
	with the apparent relaxation time \cite{Schmitt_Magn.Reson.Med._2004} \cite{Scheffler_Magn.Reson.Med._2003_2}:
	\begin{align}
	T_1^*=\left(\frac{1}{T_1}\cos^2\left(\frac{\alpha}{2}\right)+\frac{1}{T_2}\sin^2\left(\frac{\alpha}{2}\right)\right).
	\end{align}
	After including the model assumptions to the simulation, its output can be approximated using a non-linear least square fitting function implemented in python with model in equation \ref{eq::ibSSFP_signal_model}.
	Afterwards, the time development of $S$ can be compared to the theoretical and the differences between the fitted and the input relaxation parameters can be quantified.\\
	To determine $T_1$, $T_2$ and $M_0$ from the fitted parameters \cite{Schmitt_Magn.Reson.Med._2004},
	\begin{align}
		T_1 &= T_1^*\frac{S_0}{S_{stst}}\cos\left(\frac{\alpha}{2}\right),\\
		T_2 &= T_1^*\sin^2\left(\frac{\alpha}{2}\right)\left[1-\frac{S_{stst}}{S_0}\cos\left(\frac{\alpha}{2}\right)\right],~~~\textrm{and}\\
		M_0 &= \frac{S_0}{\sin\left(\frac{\alpha}{2}\right)}
	\end{align}
	has to be used. The errors follow the Gaussian error propagation and become
	\begin{equation}
	\resizebox{.9\hsize}{!}{$
		\sigma_{T_1} = \sqrt{
			\sigma_{T_1^*}^2\left(\frac{S_0}{S_{stst}}\cos\left(\frac{\alpha}{2}\right)\right)^2 +
			\sigma_{S_0}^2\left(\frac{T_1^*}{S_{stst}}\cos\left(\frac{\alpha}{2}\right)\right)^2 +
			\sigma_{S_{stst}}^2\left(-\frac{S_0 T_1^*}{S_{stst}^2}\cos\left(\frac{\alpha}{2}\right)\right)^2
		}\\
	$},
	\end{equation}
	\begin{equation}
	\resizebox{.9\hsize}{!}{$
		\sigma_{T_2} = \sqrt{
			\sigma_{T_1^*}^2\left(\frac{\sin^2\left(\frac{\alpha}{2}\right)}{1-\frac{S_{stst}}{S_0}\cos\left(\frac{\alpha}{2}\right)}\right)^2 +
			\sigma_{S_0}^2\left(-\frac{T_1^*\sin^2 \left(\frac{\alpha}{2}\right)\left[\frac{S_{stst}}{S_0^2}\cos\left(\frac{\alpha}{2}\right)\right]}{\left[1-\frac{S_{stst}}{S_0}\cos\left(\frac{\alpha}{2}\right)\right]^2}\right)^2 +
			\sigma_{S_{stst}}^2\left(-\frac{T_1^*\sin^2 \left(\frac{\alpha}{2}\right)\left[-\frac{\cos\left(\frac{\alpha}{2}\right)}{S_0}\right]}{\left[1-\frac{S_{stst}}{S_0}\cos\left(\frac{\alpha}{2}\right)\right]^2}\right)^2			
		}
	$}
	\end{equation}
	and
	\begin{equation}
		\sigma_{M_0} = \frac{\sigma_{S_0}}{\sin\left(\frac{\alpha}{2}\right)}.
	\end{equation}
	One example for TR/TE= 3.0/1.5 s, $T_1$/$T_2$=1.60/0.09 s and $M_0$=1 is presented in Figure \ref{fig::simultation_sequence_test_hardpulse}.
	
	Visually, it is not possible to see any difference between the theoretically expected and the simulated signal development. For higher zoom factors some differences can be observed, but they are to small too require a further visualization. Even the equal fitting parameters (see Table \ref{tab::fitting_results_sequence_verification}) for both methods point out that only some numerical errors occur.
	
	\begin{table}[!h]
		\centering
		\caption{Table presents the fitting results of the signal verifications in Figure \ref{fig::simultation_sequence_test_hardpulse} and \ref{fig::simultation_sequence_test_ODE}. The ODE-Simulation relies to the signal curve in which the relaxation during $t_{rf}$ is turned on, while it is off for the hard-pulse approach.}
		
		\begin{tabular}{c|c|c|c}\addlinespace[2ex]
			Parameters	&	Hardpulse Simulation	&	ODE-Simulation	&	Theoretical Parameters\\\hline
			$T_1$ [s]	& 1.600(4)	&	1.541(293)	&	1.600(1)	\\
			$T_2$ [s]	& 0.090(1)	&	0.097(17)	&	0.090(1)	\\
			$M_0$ [a.u.]& 1.000(1)	&	0.971(1)	&	1.000(1)	\\
		\end{tabular}
		\label{tab::fitting_results_sequence_verification}
	\end{table}

\begin{figure}[!h]
		\centering
		\includegraphics[width=0.8\linewidth]{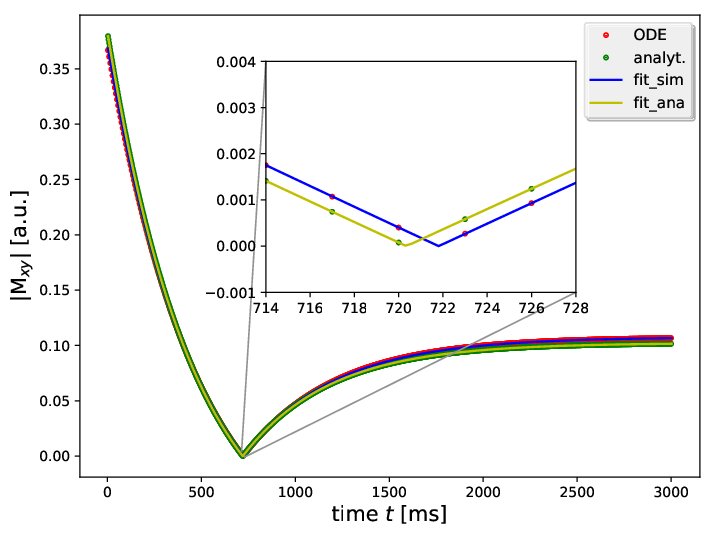}
		\caption{Visualization of the signal-development during an inversion-prepared bSSFP sequence simulated with the previously described Dopri54 ODE-solver enabling relaxation during RF-pulses. Additionally, the theoretically expected signal is plotted. Therefore, two non-linear fits with equation \ref{eq::ibSSFP_signal_model} are carried out for the analytical and the ODE-solver simulation. The determined parameters are added to Table \ref{tab::fitting_results_sequence_verification}. The simulation parameters are TR/TE= 3.0/1.5 s, $T_1$/$T_2$=1.60/0.09 s, FA=45°, $N_{\textrm{TR}}$ = 1000, and $M_0$=1.}
		\label{fig::simultation_sequence_test_ODE}
	\end{figure}

\begin{figure}[!h]
		\centering
		
		\begin{subfigure}{.45\textwidth}
			\centering
			\includegraphics[width=1.1\linewidth]{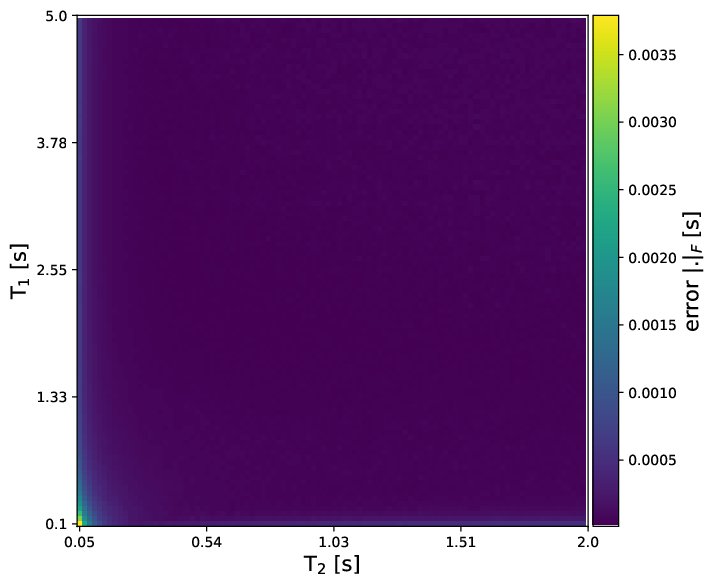}
			\caption{Hard-pulse approximation is used. }
			\label{fig::simultation_sequence_test_hardpulse_range}
		\end{subfigure}
		\hfill
		\begin{subfigure}{.45\textwidth}
			\centering
			\includegraphics[width=1.1\linewidth]{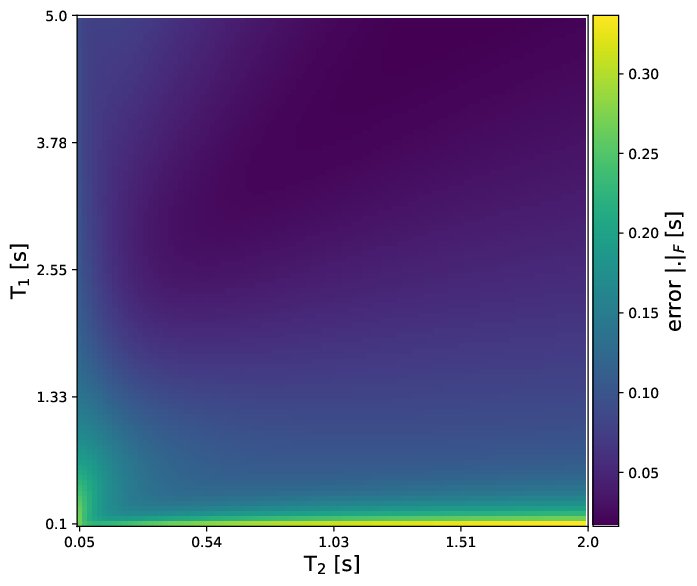}
			\caption{ Relaxation during $t_{\textrm{rf}}$ is allowed. }
			\label{fig::simultation_sequence_test_ode_range}
		\end{subfigure}
		\caption{Visualization of the error between simulated and analytical time development of an inversion-prepared bSSFP sequence. The error is calculated following the Frobenius norm of the difference of both curves. The simulation parameters are: TR/TE= 3.0/1.5 s, FA=45°, $N_{\textrm{TR}}$ = 1000, and $M_0$=1.}
	\end{figure}

	To prove this observations, the analogue comparison is performed for various $T_1$ and $T_2$ values. The Frobenius-norm $|\cdot|_F$ of the difference between the analytical and simulated signal development is calculated, visualized in Figure \ref{fig::simultation_sequence_test_hardpulse_range}.\\
	There are still some higher differences for very small relaxation parameters, but they are practically not relevant because no human tissue has corresponding $T_1$ and $T_2$ combinations. $M_0$ does not need to be tested because it works as a scaling constant and does not influence the time development of the signal. Finally, the simulation reproduces the theoretical signal model for inversion-prepared bSSFP sequences well.\\
	To get an idea about the influence of the hard-pulse assumptions to derive equation \ref{eq::ibSSFP_signal_model}, some relaxation during the RF-pulses is turned on. The inverting-, the preparation-, and the imaging RF-pulse durations are therefore set to 0.5 ms for the same simulation parameter as used in Figure \ref{fig::simultation_sequence_test_hardpulse}. The result is plotted in Figure \ref{fig::simultation_sequence_test_ODE}.
	
	The simulated curve is shifted compared to the analytical curve and additionally, an offset is present for the first echoes. Because the $y$-axis describes the absolute value of $M_{xy}$ and the relaxation allowing ODE-simulation has a smaller offset compared to the analytical model, the relaxation, (mostly $T_2$) is assumed to be the reason for this difference. The higher steady-state signal might also be an effect of the relaxation but, with Figure \ref{fig::simulation_uniqueness_ibSSFP} in mind, it seems to be a shared effect of $T_1$ and $T_2$. The signal variations manifest themselves also in the fitting parameters in Table \ref{tab::fitting_results_sequence_verification} and in the error plot in Figure \ref{fig::simultation_sequence_test_ode_range}. Especially the regions of small relaxation parameters show higher variations because the influence of the magnetization during an RF-pulse is stronger.\\
	The differences in the fitting results in Table \ref{tab::fitting_results_sequence_verification} show how important it is to accurately model the scanners behavior for a correct estimation of the parameter maps. Using only the theoretical model and fit it pixel-wisely, should lead to systematic errors, because the signal model does not capture the real behavior of the magnetization well enough. A more detailed comparison with \cite{Schmitt_Magn.Reson.Med._2004} is necessary.\\
	
	The accuracy check of the simulation tool makes the test of the numerical phantom obsolete because every pixel gets its characteristic parameters and is simulated with the simulation-tool like shown in the upper section \ref{ssec::simulation_verification_sequence}. It therefore implies the accuracy of the numerical phantom.

	\section{Discussion of Simulation Tool}
	The simulation tool developed here reproduces the theoretical expected behavior of an inversion-prepared bSSFP as long as the same approximations are assumed. Therefore, Figure \ref{fig::simultation_sequence_test_hardpulse} presents a perfect similarity between both signal curves, and the fitting parameters in Table \ref{tab::fitting_results_sequence_verification} are the same for the simulation and the theoretical model.\\
	On the other hand, there is a mismatch between the two signals in Figure \ref{fig::simultation_sequence_test_ODE} if the hard-pulse approximation is removed from the simulation. Therefore, the fitting parameters in Table \ref{tab::fitting_results_sequence_verification} are not the same anymore. It is hard to verify the whole sequence in a realistic physical way because the theoretical model does not include the relaxation completely. As the method reproduces the results using the approximations and as the individual blocks of the simulation are verified, it can be assumed that the simulation shows a realistic physical behavior.\\
	For future work, it would be desirable to integrate the RF-pulses directly in the Bloch equations so that the adaptive stepsize can be used for them. This would make the simulation more efficient than the matrix-based simulations. In the actual state, it would need to be bench-marked to be compared to it. Further the simulated phantom should be moved from the image to the frequency domain to show a more realistic behavior. Additional options for added $B_1$-maps and slice-profiles would increase its realistic behavior.

\chapter{Model-based Quantification of Relaxation Parameters}

This section presents the main part of this project. The Bloch (equation) model-based reconstruction is introduced including its theoretical ideas. After verification of the implemented tools, the technique is tested on simulated data followed by measured ones from phantoms using the sequence already mentioned in section \ref{sec::sequence_ibSSFP}.

\section{Development of the Bloch Model Operators}

	\subsection{Estimation of Relaxivity as Non-Linear Inverse Problem}
	\label{ssec::bloch_operator}
	To start the development of the Bloch model-based reconstruction, the signal model is set up. This is realized by treating the fit as non-linear inverse problem, where
	\begin{align}
		\hat{x} = \textrm{argmin}\|\mathfrak{B}(\bm{x_p})-\bm{\widetilde{y}}\|_2^2 + \alpha\|\bm{x_p}\|^2_2
		\label{eq::BlochModel_inverse_problem}
	\end{align}
	is minimized including the desired parameter maps $\bm{x}_p$, the acquired data $\bm{\widetilde{y}}$ in pixel-domain, and the Bloch forward operator of the model $\mathfrak{B}$. To include the Bloch equations as a model, the magnetization $\bm{M}_{\bm{x_p}, t_i}$ for the parameter $\bm{x_p} = \left(R_1,~R_2,~M_0\right)^T$ for all individual timepoints $t_i$ with $i = 1, ... , n$ needs to be determined. It is obtained by solving the Bloch ODEs using the Dopri54 algorithm with partially adaptive step-size control, which is based on adapted implementations in BART. The Bloch forward operator becomes
	\begin{align}
		\mathfrak{B}:\bm{x_p} = 		
		\begin{pmatrix}
		R_1\\
		R_2\\
		M_0
		\end{pmatrix}
		\longmapsto
		\begin{pmatrix}
		M_{t_1}\\
		\vdots\\
		M_{t_n}\\
		\end{pmatrix}.
	\end{align}

	\subsection{Solving the Inverse Problem: IRGNM-FISTA}
	\label{ssec:Devel_Bloch_Operator_SolvingInverseProblem}
	
	To solve the inverse minimization problem proposed in equation \ref{eq::BlochModel_inverse_problem} the Iteratively Regularized Gauss Newton Method (IRGNM) is selected \cite{Bakushinsky__2005}. In a first step, it linearizes equation \ref{eq::BlochModel_inverse_problem} using an initial guess $\bm{x_p}^k$:
	\begin{align}
		\mathfrak{B}(\bm{x_p}^k + \textrm{d}\bm{x_p})\approx
		\bm{D}\mathfrak{B}(\bm{x_p}^k)\textrm{d}\bm{x_p}+\mathfrak{B}(\bm{x_p}^k),
	\end{align}
	with the Jacobian $\bm{D}\mathfrak{B}(\bm{x_p}^k)$, the iteration counter $k$ and the update of the parameter maps $\textrm{d}\bm{x_p}$. The linearized equation
	\begin{align}
		\bm{D}\mathfrak{B}(\bm{x_p}^k)\textrm{d}\bm{x_p}+\mathfrak{B}(\bm{x_p}^k) = \bm{y}
		\label{eq::Devel_Bloch_Operator_LinearizedEquation}
	\end{align}
	is solved for $\textrm{d}\bm{x_p}$ using a fast	iterative shrinkage/thresholding algorithm (FISTA), which is applied to the symmetric matrix by transforming equation \ref{eq::Devel_Bloch_Operator_LinearizedEquation} into
	\begin{align}
		\bm{D}\mathfrak{B}^H(\bm{x_p}^k)\bm{D}\mathfrak{B}(\bm{x_p}^k)\textrm{d}\bm{x_p} = \bm{D}\mathfrak{B}^H\left(\bm{y}- \mathfrak{B}(\bm{x_p}^k) \right)
		\label{eq::Devel_Bloch_Operator_LinearizedEquationSymetric}
	\end{align}
	including the adjoint of the derivative $\bm{D}\mathfrak{B}^H$.\\ 
	FISTA has the advantage to work on non-smooth convex functions, too and it allows a simple implementation of a non-negativity constraint, which is useful for the positive parameter maps and the necessary constraints \cite{Beck_SIAMJ.Img.Sci._2009}.
	To solve the still ill-conditioned problem a regularization needs to be included into equation \ref{eq::Devel_Bloch_Operator_LinearizedEquationSymetric}, which penalizes high norm values and stabilizes the solution. For a quadratic regularization this lead to
	\begin{align}
		\left(\bm{D}\mathfrak{B}^H(\bm{x_p}^k)\bm{D}\mathfrak{B}(\bm{x_p}^k) + \alpha^k\bm{1}\right)\textrm{d}\bm{x_p} = \bm{D}\mathfrak{B}^H\left(\bm{y}- \mathfrak{B}(\bm{x_p}^k) \right)
		\label{eq::Devel_Bloch_Operator_LinearizedEquationSymetricRegularized}
	\end{align}
	with the regularization parameter $\alpha^k$ and the identity matrix $\bm{1}$. Using the relation of the Gauss normal equation\footnote{ $\bm{A}^H\bm{A}\bm{u}=\bm{A}^H\bm{v}$ holds, if $\bm{u}$ is a least-square solution of $\bm{A}\bm{u}=\bm{v}$}
	\begin{align}
		\bm{A} = 
		\begin{pmatrix}
		\bm{D}\mathfrak{B}(\bm{x_p}^k)\\
		\sqrt{\alpha^k}\bm{1}\\
		\end{pmatrix}
		~~~~
		\bm{u} = \textrm{d}\bm{x_p}
		~~~~
		v = 
		\begin{pmatrix}
		\bm{y}-\mathfrak{B}(\bm{x_p}^k)\\
		0\\
		\end{pmatrix}
	\end{align}
	one can show that the minimization in equation \ref{eq::Devel_Bloch_Operator_LinearizedEquationSymetricRegularized} becomes:
	\begin{align}
		\bm{x_p}^{k+1}&=\underset{\bm{x_p}}{\textrm{argmin}}\|\bm{D}\mathfrak{B}(\bm{x_p}^k)\textrm{d}\bm{x_p}+\mathfrak{B}(\bm{x_p}^k)-\bm{y}\|_2^2+\alpha^k\|\bm{x_p}^k\|^2_2 \label{eq::Bloch_optimization_equation}\\
		&= \underset{\bm{x_p}}{\textrm{min}}\left\{f(\bm{x_p}))\right\}.
	\end{align} 
	To solve equation \ref{eq::Bloch_optimization_equation} with FISTA $f(\bm{x_p})$ has to be continuous, convex and smooth. Additionally, its gradient $\nabla f$ has to be Lipschitz continuous. Following its definition
	\begin{align}
		\| \nabla f(\bm{x_p}) - \nabla f(\bm{x'_p})\|\leq L(f)\|\bm{x_p}-\bm{x'_p}\|,
	\end{align}
	the Lipschitz constant $L(f)$ can be calculated as largest eigenvalue of the Hessian matrix 
	\begin{align}
		\bm{H}(\bm{x_p})=\bm{D}\mathfrak{B}^H \bm{D}\mathfrak{B}(\bm{x_p})
		\label{eq:Hessian-matrix}
	\end{align}
	using the power iteration algorithm, if it exists \cite{Mises_ZAMM_1929} and can be used for calculating the stepsize of FISTA.\\
	Afterwards the parameter maps are constrained to be real- and non-negative-valued and are as \cite{Beck_SIAMJ.Img.Sci._2009}
	\begin{align}
		\bm{a_p}^{k} = p_L(\bm{x_p}^k),
	\end{align}
	with the included ravine step \cite{Nesterov_SovietMathematicsDoklady_1983}
	\begin{align}
	t^{k+1} &= \frac{1+\sqrt{1+4(t^{k})^2}}{2}\\
	\bm{x_p}^{k+1} &= \bm{a_p}^{k} + \left(\frac{t^{k}-1}{t^{k+1}}\right)\left(\bm{a_p}^{k}-\bm{a_p}^{k-1}\right),
	\end{align}
	the iterative shrinkage operator
	\begin{align}
		p_L(\bm{x_p}^k)= \bm{x_p}^k - \frac{1}{L}\nabla f(\bm{x_p}^k)
	\end{align}
	and projected parameter maps $\bm{a_p}^{k}$.\\
	Equation \ref{eq::Bloch_optimization_equation} describes a least-square optimization because the $l_2$-norm is selected for regularization. It can only be solved as long as the derivative and adjoint-derivative operators are known. The current implementation is set up on a modified BART version \cite{Wang_Magn.Reson.Med._2018}.

%
%
%

	\subsection{Derivative Operator and Its Adjoint}
	Following the linearized equation \ref{eq:Hessian-matrix}, the derivative- and the adjoint-operators are necessary to solve the linear system, using an IRGNM. Knowing that the forward operator is $\mathfrak{B}$ and the definition of the Jacobian is
	\begin{align}
	\bm{D}\mathfrak{B}(\bm{x_p}) = 
	\begin{pmatrix}
	\frac{\partial \mathfrak{B}_1(\bm{x_p})}{\partial R_1} & \frac{\partial \mathfrak{B}_1(\bm{x_p})}{\partial R_2} & \frac{\partial \mathfrak{B}_1(\bm{x_p})}{\partial M_0} \\
	\frac{\partial \mathfrak{B}_2(\bm{x_p})}{\partial R_1} & \frac{\partial \mathfrak{B}_2(\bm{x_p})}{\partial R_2} & \frac{\partial \mathfrak{B}_2(\bm{x_p})}{\partial M_0} \\
	& \vdots &\\
	\end{pmatrix},
	\end{align}
	the operator for the derivative has to follow
	\begin{align}
		\bm{D}\mathfrak{B}(\bm{x_p})
		\begin{pmatrix}
		\textrm{d}R_1\\
		\textrm{d}R_2\\
		\textrm{d}M_0
		\end{pmatrix}
		= 
		\begin{pmatrix}
		\frac{\partial M_{t_1}}{\partial R_1}\textrm{d}R_1+
		\frac{\partial M_{t_1}}{\partial R_2}\textrm{d}R_2+
		\frac{\partial M_{t_1}}{\partial M_0}\textrm{d}M_0\\
		\vdots\\
		\frac{\partial M_{t_n}}{\partial R_1}\textrm{d}R_1+
		\frac{\partial M_{t_n}}{\partial R_2}\textrm{d}R_2+
		\frac{\partial M_{t_n}}{\partial M_0}\textrm{d}M_0
		\end{pmatrix}
	\end{align}
	as well as the adjoint-derivative
	\begin{align}
		\bm{D}\mathfrak{B}^H(\bm{x_p})
		\begin{pmatrix}
		y_{1}\\
		y_{2}\\
		\vdots\\
		y_{n}
		\end{pmatrix}
		=
		\begin{pmatrix}
			\textrm{d}R_1\\
			\textrm{d}R_2\\
			\textrm{d}M_0
		\end{pmatrix}
		=
		\begin{pmatrix}
			\sum\limits_{k=1}^n
			\overline{\left(\frac{\partial M_{t_k}}{\partial R_1}\right)}
			\cdot y_{k}\\
			\sum\limits_{k=1}^n
			\overline{\left(\frac{\partial M_{t_k}}{\partial R_2}\right)}
			\cdot y_{k}\\
			\sum\limits_{k=1}^n
			\overline{\left(\frac{\partial M_{t_k}}{\partial M_0}\right)}
			\cdot y_{k}\\
		\end{pmatrix}
	\end{align}
	with $\overline{(\cdot)}$ as notation for complex conjugated. These operators include the derivatives of the magnetization depending on the mapped parameter $\frac{\partial \bm{M}}{\partial \bm{x_p}}$. They are determined using a direct sensitivity analysis of the Bloch equations, which will be introduced in the following section.

	\subsection{Direct Sensitivity Analysis}
	The direct sensitivity analysis, or forward sensitivity method (FSM) \cite{Lakshmivarahan__2017}, allows to calculate the gradients and therefore to solve the non-linear inverse problem presented in equation \ref{eq::Bloch_optimization_equation}. 
	FSM describes the evolution of the sensitivities, which is the time development of $\bm{M}$ under small parameter deviations.\\
	In the following, the FSM for the Bloch equations is introduced and it is explained how it is integrated in this project. Furthermore, the calculated gradients are verified using an approximation of a difference quotient.

	\subsubsection*{Direct Sensitivity Analysis of Bloch Equations}
	
	Assuming the Bloch equations are
	\begin{align}
	\frac{\textrm{d}\bm{M}(\bm{x_p}, t)}{\textrm{d}t} = \bm{f}(t, \bm{M}(\bm{x_p}, t), \bm{x_p}),~~\textrm{with}~~\bm{x_p}(t) = \textrm{const},
	\label{eq::DSA_Bloch_equation}
	\end{align}
	and the parameter maps $\bm{x_p} = \left(R_1,~R_2,~M_0\right)^T$, its sensitivities are described by
	\begin{align}
		\bm{S}(t)|_{\bm{x}_{p_i}} = \frac{\partial \bm{M}(\bm{x_p}, t)}{\partial \bm{x}_{p_i}} = \nabla_{\bm{x}_{p_i}}\bm{M}(\bm{x_p}, t).
		\label{eq::DSA_sensitivity_expression}
	\end{align}
	Calculating the time derivative of equation \ref{eq::DSA_sensitivity_expression} and inserting \ref{eq::DSA_Bloch_equation} after switching of time- and parameter-derivative leads to	
	\begin{align}
	\frac{\textrm{d}\bm{S}(t)}{\textrm{d}t}  = \nabla_{\bm{x_p}}\bm{f}(t, \bm{M}(\bm{x_p}, t),\bm{x_p}).
	\label{eq::DSA_sens_time_derivative_1}
	\end{align}
	The vector of the partial derivative $\nabla_{\bm{x}_{p_i}}$ in equation \ref{eq::DSA_sensitivity_expression} is reformulated to a matrix $\nabla_{\bm{x_{p}}}$ including all parameters in $\bm{x_p}$.\\
	Using the chain rule, the temporal change of the sensitivity in equation \ref{eq::DSA_sens_time_derivative_1} becomes
	\begin{align}
		\frac{\textrm{d}\bm{S}(t)}{\textrm{d}t} =
		 \bm{D}_{{\bm{M}}}\bm{f}(t, \bm{M}(\bm{x_p}, t),\bm{x_p})\cdot \nabla_{\bm{x_p}}\bm{M}(\bm{x_p}, t) + \bm{D}_{{\bm{x_p}}}\bm{f}(t, \bm{M}(\bm{x_p}, t),\bm{x_p}),
	\end{align}
	which further leads to the differential equation for the sensitivities
	\begin{align}
	\frac{\textrm{d}\bm{S}(t)}{\textrm{d}t} = \bm{J}_{\bm{M}, \bm{f}} \cdot \bm{S}(t) + \bm{J}_{\bm{x_p}, \bm{f}},
	\label{eq::DSA_ODE}
	\end{align}
	with the Jacobi matrices
	\begin{align}
		\bm{J}_{\bm{M}, \bm{f}} = 
		\begin{pmatrix}
		-R_2 & \gamma B_z & -\gamma B_y\\
		-\gamma B_z & -R_2 & \gamma B_x\\
		\gamma B_y & -\gamma B_x & -R_1
		\end{pmatrix}
	\end{align}
	and
	\begin{align}
	\bm{J}_{\bm{x_p}, \bm{f}} = 
	\begin{pmatrix}
	0 & -M_x \\
	0 & -M_y \\
	M_0-M_z & 0
	\end{pmatrix},
	\label{eq::DSA_jacobian_xpf}
	\end{align}	
	as well as the current magnetization $\bm{M}=\left(M_x,~M_y,~M_z\right)^T$, and the external field $\bm{B}=\left(B_x,~B_y,~B_z\right)^T$.\\
	Because the Bloch equations \ref{eq::DSA_Bloch_equation} are linear in $M_0$, the time derivative of the sensitivity $\left.\bm{S}(t)\right|_{M_0}$ is
	\begin{align}
		\left.\frac{\textrm{d}\bm{S}(t)}{\textrm{d}t}\right|_{M_0} 
		&= \nabla_{M_0}\bm{f}(t, \bm{M}(\bm{x_p}, t),\bm{x_p})\\
		&= \bm{f}(t, \left.\bm{S}(t)\right|_{M_0=1}, ,\bm{x^*_p}),
		\label{eq::differential_equation_m0_FSM}
	\end{align}
	with $\bm{x^*_{p}} = (R_1,~R_2,~M_0=1)^T$. With equation \ref{eq::differential_equation_m0_FSM} the sensitivity to $M_0$ becomes the time development of the signal in the normalized form and $M_0$ is set to 1 in matrix \ref{eq::DSA_jacobian_xpf}.
	
	\subsubsection*{Verification of Sensitivities in the Simulation}
	To verify the sensitivities $\bm{S}$ of each tissue characteristic parameter $\bm{x}_{p_i}$, the approximation of a gradient through the difference quotient is used
	\begin{align}
		\bm{S}_{\bm{x}_{p_i}} = \frac{\partial \bm{M}}{\partial \bm{x}_{p_i}} = \frac{\bm{M}(\bm{x_{p^*}},t) - \bm{M}(\bm{x_{p}},t)}{\delta},
		\label{eq::sensitivities_verification_difference_quotient}
	\end{align}
	with a small distortion $\delta$ in one of the parameters in $\bm{x_{p}}$ leading to $\bm{x_{p*}}$.\\	
	Therefore, a correct estimation of the sensitivities minimizes the error 
	\begin{align}
		\epsilon = |\delta\cdot\bm{S}_{\bm{x}_{p_i}} - [\bm{M}(\bm{x_{p^*}},t) - \bm{M}(\bm{x_{p}},t)]|.
		\label{eq::sensitivities_verification_error}
	\end{align}

\begin{figure}[!h]
		\centering
		\includegraphics[width=0.9\linewidth]{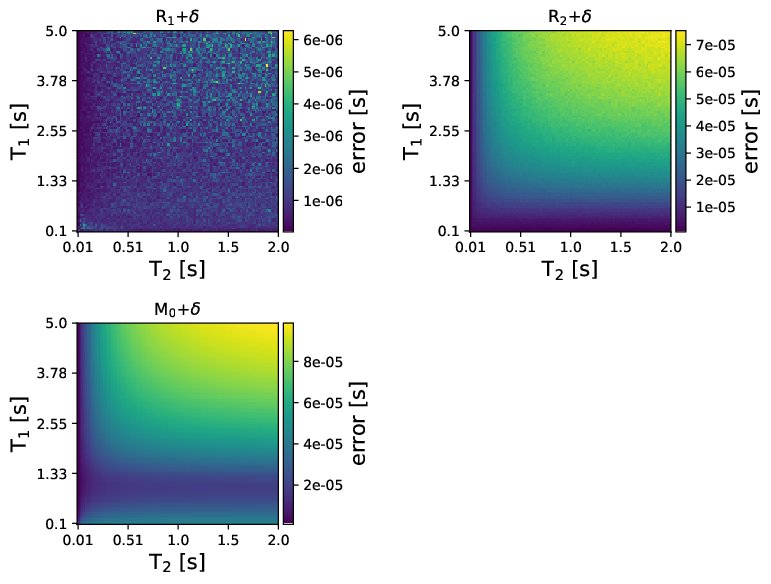}
		\caption{Visualization of the error between the product of the distortion with the calculated gradient and the difference between both magnetization positions, presented in equation \ref{eq::sensitivities_verification_error}. The error is calculated as arithmetic mean of all echo-times and dimensions. The simulation parameters are: $N_{\textrm{TR}}$ = 2000, TR/TE= 3.0/1.5 ms, $t_{rf}$=0.9 ms, $t_{inv}$=0.9 ms, and $M_0$=1. }
		\label{fig::simultation_sequence_verification_sens}
	\end{figure}

	In practice, a temporal evolution of the magnetization $\bm{M}(\bm{x_{p}},t)$ for one parameter combination $\bm{x_p}$ is simulated. By solving equation \ref{eq::DSA_ODE} simultaneously to the Bloch equations \ref{eq::DSA_Bloch_equation}, the sensitivities $\bm{S}_{\bm{x}_{p_i}}$ are directly estimated. Following the difference quotient, a small distortion $\delta$ is added to one parameter in $\bm{x_p}$, leading to $\bm{x_{p^*}}$.\\
	After simulating the sequence for $\bm{M}(\bm{x_{p^*}},t)$, one can determine the error $\epsilon$ following equation \ref{eq::sensitivities_verification_error} for each dimension $(x,~y,~z)$ and for each echo-time during the sequence. Afterwards, the arithmetic mean over the dimensions and time-steps of the error is calculated.\\
	This verification is performed for all parameters in $\bm{x_p}$.
	Figure \ref{fig::simultation_sequence_verification_sens} shows examples of $\epsilon$ for various combinations of $T_1$ and $T_2$ and a distortion of $\delta_{R_1}$= 0.001 s, $\delta_{R_2}$= 0.0001 s, and $\delta_{M_0}$= 0.0001 s.	
	
	Figure \ref{fig::simultation_sequence_verification_sens} demonstrates that the determination of the gradients is correct. The appearing errors are very low and may result from numerical noise.

	\subsection{Verification of the Derivative Operator and its Adjoint}
	\label{ssec::scalar_prod_verification_bloch_operator}
	The numerical verification of the implemented derivative operator and its adjoint exploits the scalar product. According to the definition of the adjoint operator $\bm{D}\mathfrak{B}$ and $\bm{D}\mathfrak{B}^H$ are implemented correctly, if
	\begin{align}
		\langle\bm{D}\mathfrak{B}\bm{x},\bm{y}\rangle = \langle\bm{x},\bm{D}\mathfrak{B}^H\bm{y}\rangle.
		\label{eq::Verification_Bloch_Operator_Scalar_product}
	\end{align}
	holds.\\
	Two arrays $\bm{x_1}$ and $\bm{x_2}$ are filled with Gaussian-distributed random numbers and the operators are applied, leading to
	\begin{align}
		\bm{D}\mathfrak{B}\bm{x_1} &= \bm{y_1}\\
		\bm{D}\mathfrak{B}^H\bm{x_2}&= \bm{y_2}.
	\end{align}
	Determining the scalar product for each side of equation \ref{eq::Verification_Bloch_Operator_Scalar_product} leads to the error
	\begin{align}
		\delta = \left|\langle\bm{x_1},\bm{y_2}\rangle - \langle\bm{x_2},\bm{y_1}\rangle\right|.
	\end{align}
	The error for an inversion-prepared bSSFP simulation\footnote{10 ms inversion pulse, $t_{RF}$= 0.90 ms, TR=4.50 ms, TE=2.25 ms, FA=45°, 500 repetitions} with random $T_1$ and $T_2$ values, and an image size of 16x16 px becomes $\delta<10^{-3}$ and the test is fulfilled.

	\subsection{Verification of the Optimization Algorithm}
	\label{ssec::verification_optimization_simulated_phantom}
	
\begin{figure}[!h]
		\centering
		\includegraphics[width=\linewidth]{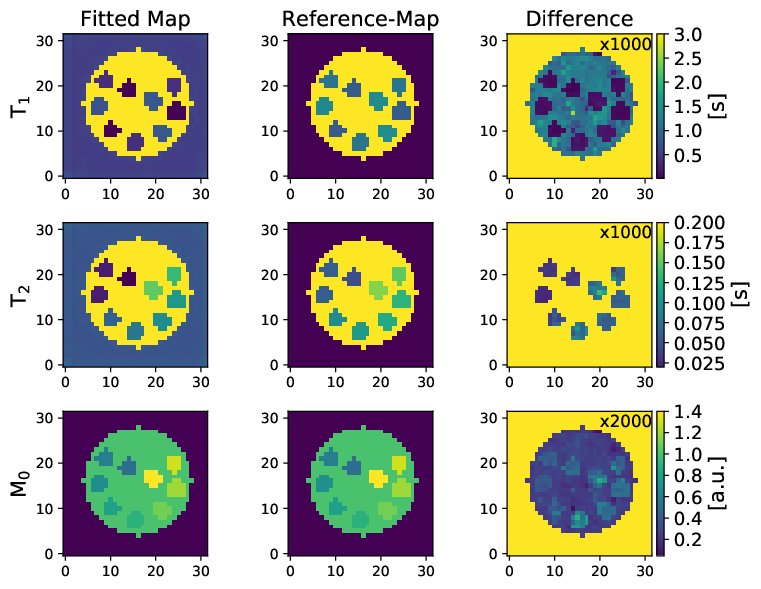}
		\caption{Visualization of the fitting results (\textbf{left}) using a simulated phantom in the image domain (\textbf{center}) with the simulation parameters: inversion-prepared bSSFP sequence, 10 ms inversion pulse, $t_{RF}$= 0.90 ms, TR=4.5 ms, TE=2.25 ms, FA=45°, and $N_{\textrm{TR}}$ = 500.  The estimated differences between simulation and approximation are plotted on the \textbf{right} and multiplied by the added factor to improve the visualization. The reconstruction model just includes the Bloch operator.}
		\label{fig::optimization_verification_just_Bloch_Operator}
	\end{figure}

\begin{figure}[!h]
		\centering
		\includegraphics[width=0.5\linewidth]{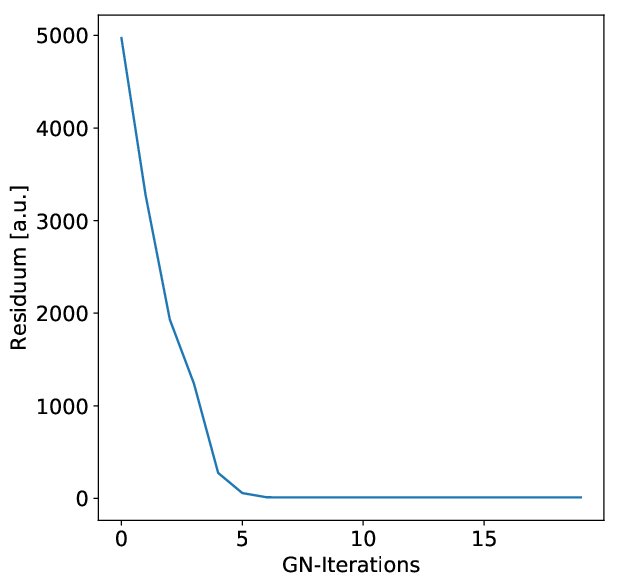}
		\caption{Visualization of the residuum during the Gauss-Newton steps of the optimization to create Figure \ref{fig::optimization_verification_just_Bloch_Operator}. }
		\label{fig::optimization_verification_Bloch_Operator_residuum}
	\end{figure}

	The optimization algorithm is verified with a simulated phantom. In principle both the simulation and the forward operator are based on the same code. Therefore, it can only be checked if the implementation of the optimization works, but leads to no information about its accuracy\footnote{Otherwise, an inverse crime would arise \cite[p.154]{Colton__2013}. }.\\
	To start testing the optimization algorithm, the Bloch operator $\mathfrak{B}$, introduced in section \ref{ssec::bloch_operator}, is selected. The input data $\bm{y}$ is given in the image domain and equation \ref{eq::BlochModel_inverse_problem} is solved using the IRGNM-FISTA algorithm, mentioned in section \ref{ssec:Devel_Bloch_Operator_SolvingInverseProblem}. Therefore, this test corresponds to a pixel-wise fitting of the Bloch signal-model to the data $\bm{y}$.\\
	The fully-sampled simulated phantom is created with an inversion-prepared bSSFP sequence with a 10 ms inversion pulse, $t_{RF}$ of 0.9 ms, TR=4.5 ms, TE=2.25 ms, FA=45°, and 500 repetitions. An image is determined for each TR.\\
	Because the solved inverse-problem is non-linear, the initial guesses for the parameter maps are essential. To simplify the signal-model, the inverse relaxation coefficients are estimated. The starting guesses are  $\bm{R}_1=\frac{0.8}{s_{R_1}}$, $\bm{R}_2=\frac{11}{s_{R_2}}$, and $\bm{M}_0=\frac{1}{s_{M_0}}$ including the scaling factors $s_{R_1}$, $s_{R_2}$, and $s_{M_0}$, which are used to balance the numerical gradients of the individual parameters. They are chosen by try and error depending on the input data. This is one limitation of the proposed method, but literature like \cite{Maier_Magn.Reson.Med._2018} provides solutions which could be adapted here. For the simulation, $s_{R_1}=1$, $s_{R_2}=30$, and $s_{M_0}=15$ are estimated by try and error. Additionally, a scaling factor is introduced to keep the range of the input data constant and therefore increase the stability of the reconstruction algorithm. It scales the initial $l_2$-norm of the k-space dataset to 5000, which is found to work well.\\
	The regularization parameter $\alpha$ is reduced in each step by a reduction factor $R$=2 and the power iteration algorithm uses 20 iterations to find the largest eigenvalue of the Hessian matrix for FISTA. The latter increases its iterations to solve the linearized problem following $N_{\textrm{iter},k}=\textrm{min}\left(300, 10\cdot 2^{k}\right)$ with the current $k$th Gauss-Newton step.\\		
	A visualization of the resulting reconstructions and the original phantom data is visualized in Figure \ref{fig::optimization_verification_just_Bloch_Operator}. The evolution of the residuum is plotted against the Gauss-Newton steps of the optimization in Figure \ref{fig::optimization_verification_Bloch_Operator_residuum}. The starting offset of 5000 comes from the initial scaling of the input data.
	
	The results point out a high similarity of the calculated to the original map. Their differences are small for the tubes. As they cover the most relevant area in $T_1$-$T_2$ space, the plotted colorbars are optimized for them and not for the water around. The matching of the water produces accurate results but their difference is larger than for the tubes. This occurs mostly through the tube optimized initial guesses of the parameter maps. Therefore, it can be assumed that the optimization algorithm is accurately implemented for pixel-wise fitting.\\

\begin{figure}[!h]
		\centering
		\includegraphics[width=\linewidth]{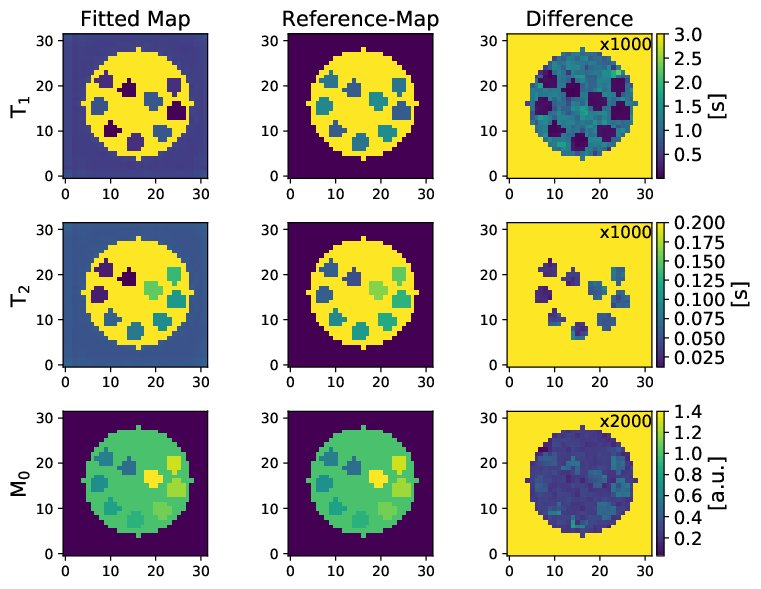}
		\caption{Visualization of the fitting results (\textbf{left}) using a simulated phantom (\textbf{center}) with the simulation parameters: inversion-prepared bSSFP sequence, 10 ms inversion pulse, $t_{RF}$= 0.90 ms, TR=4.5 ms, TE=2.25 ms, FA=45°, and $N_{\textrm{TR}}$ = 500. It is created pixel-wise and transformed to the frequency domain using a Fourier transformation to test the reconstruction. The estimated differences between simulation and approximation are plotted on the \textbf{right} and multiplied by the added factor to improve the visualization. The reconstruction model includes the chained Bloch and Fourier operator.}
		\label{fig::optimization_verification_Fourier_Bloch_Operator}
	\end{figure}

\begin{figure}[!h]
		\centering
		\includegraphics[width=0.5\linewidth]{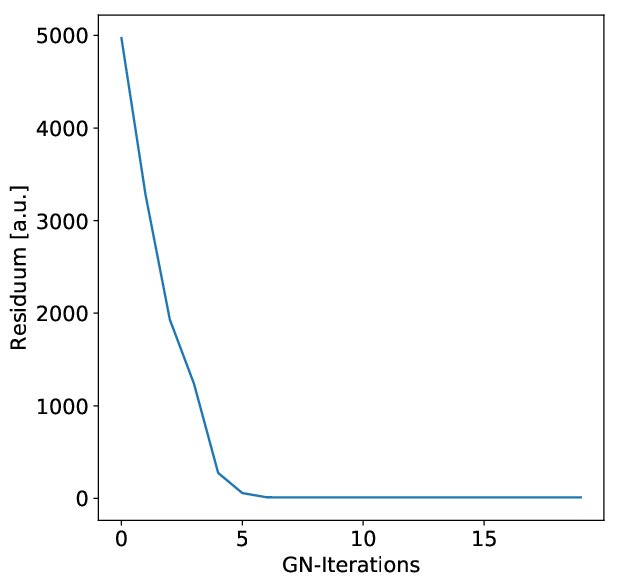}
		\caption{Visualization of the residuum during the Gauss-Newton steps of the optimization to create Figure \ref{fig::optimization_verification_Fourier_Bloch_Operator}. }
		\label{fig::optimization_verification_Fourier_Bloch_Operator_residuum}
	\end{figure}

	In the next step a Fourier operator $\mathfrak{F}$ is chained to $\mathfrak{B}$. This extends the optimization to k-space data, which avoids a previously required reconstruction of the raw data. Again, a simulated phantom is used to check whether the chaining of both operators is working. Therefore, a Fourier transformation is applied to the pixel-wise created phantom (section \ref{sec::nummerical_phantom}). The algorithm finally optimizes the non-linear inverse problem:
	\begin{align}
		\hat{x} = \underset{\bm{x_p}}{\textrm{argmin}}\|\mathfrak{F}\mathfrak{B}(\bm{x_p})-\bm{y}\|_2^2+\alpha\bm{R}(\bm{x_p})
		\label{eq::BlochModel_inverse_problem_FB}
	\end{align}
	with the acquired data $\bm{y}$ in the frequency-domain, which is an extension to equation \ref{eq::BlochModel_inverse_problem}. The optimization algorithm is based on the same settings as they are used for the pixel-wise fitting. Thus, the results are assume to be close to the previously acquired ones. They are visualized in Figure \ref{fig::optimization_verification_Fourier_Bloch_Operator}. The residuum is plotted in Figure \ref{fig::optimization_verification_Fourier_Bloch_Operator_residuum}.

	The residuum shows the same trend like the pixel-wise fitting and its approximation error for all maps is equally low. Again, the water in the background is not fitted as good, which may result from the more distant initial guess. All in all, the optimization algorithm is verified to work well also for k-space data.

	\newpage
\section{Combination with Calibrationless Parallel Imaging}
	\label{sec::NLINV_extension_optimization_algorithm}
	The model-based reconstruction developed in the previous sections works with k-space and image space data, but assumes constant and homogeneous coil profiles. As in practice, multiple receiver coils are used to increase the signal-to-noise ratio and to speed up the measurements, their sensitivities need to be determined for using the mapping technique for using raw data.\\
	The selected approach avoids preparation scans and uses the whole acquired data for coil-profile estimation \cite{Uecker_Magn.Reson.Med._2008}. This calibrationless parallel imaging is combined with the Bloch, and Fourier operator.\\
	In the following, the idea behind the concept of non-linear inversion (NLINV) is explained and how it is integrated into the reconstruction, including its verification.
	
	\subsection{Non-Linear Inversion: NLINV}
		
	To allow calibrationless parallel imaging in MRI, Uecker \textit{at al.} \cite{Uecker_Magn.Reson.Med._2008}
	introduced the reconstruction process formulated as non-linear inversion (NLINV) problem:
	\begin{align}
		\bm{F}\bm{x} = \bm{y}
		~~~\textrm{with}~~~
		\bm{x}=
		\begin{pmatrix}
		\rho\\
		c_1\\
		\vdots\\
		c_N
		\end{pmatrix},
		\label{eq::NLINV_nonlinear_problem}
	\end{align}
	with $\rho$ being the reconstructed image and $c_i$ for $i = 0,\dots,N$ describing the sensitivities for $N$ coils. Following this, the forward operator becomes
	\begin{align}
		\bm{F}: \bm{x}\mapsto
		\begin{pmatrix}
		\mathfrak{P}\mathfrak{F}(c_1\cdot\rho)\\
		\vdots\\
		\mathfrak{P}\mathfrak{F}(c_N\cdot\rho)
		\end{pmatrix},
	\end{align}
	with the Fourier operator $\mathfrak{F}$ and an additional operator for the sampling pattern $\mathfrak{P}$, which is typically a diagonal matrix of a binary sampling mask.\\
	The derivative of the operator follows
	\begin{align}
		\bm{D}\bm{F}(\bm{x})
		\begin{pmatrix}
		\textrm{d}\rho\\
		\textrm{d}c_1\\
		\vdots\\
		\textrm{d}c_N
		\end{pmatrix}
		=
		\begin{pmatrix}
		\mathfrak{P}\mathfrak{F}(\rho\cdot\textrm{d}c_1+\textrm{d}\rho\cdot c_1)\\
		\vdots\\
		\mathfrak{P}\mathfrak{F}(\rho\cdot\textrm{d}c_N+\textrm{d}\rho\cdot c_N)
		\end{pmatrix},
	\end{align}
	and the adjoint-derivative
	\begin{align}
		\bm{D}\bm{F}^H(\bm{x})
		\begin{pmatrix}
		y_1\\
		\vdots\\
		y_N
		\end{pmatrix}
		=
		\begin{pmatrix}
		\sum\limits_{j=1}^N c^*_j\cdot\mathfrak{F}^{-1}(\mathfrak{P}^H y_j)\\
		\rho^*\cdot\mathfrak{F}^{-1}(\mathfrak{P}^H y_1)\\
		\vdots\\
		\rho^*\cdot\mathfrak{F}^{-1}(\mathfrak{P}^H y_N)
		\end{pmatrix}.
	\end{align}
	It is solved with a IRGNM-algorithm with a conjugate gradient for the linearized problem. Therefore, the solution would not represent the original object because equation \ref{eq::NLINV_nonlinear_problem} is underdetermined both in the undersampled and the fully sampled case. Therefore, prior-knowledge is incorporated with a regularization of the coil profiles $\bm{c}$. To exploit their smoothness the Sobolev-norm,
	\begin{align}
		\|f\|_l=\|(I-\Delta)^{\frac{l}{2}}f\|,
	\end{align}
	with an index $l$ and the Laplacian $\Delta$, can be used. This corresponds to a weighting with $\left(1+s\|\bm{k}\|^2\right)^{\frac{l}{2}}$ in the frequency domain for the coils with a constant $s$, which results in a weighting matrix $\bm{W}$ of
	\begin{align}
		\begin{pmatrix}
		\rho\\
		\hat{c_1}\\
		\vdots\\
		\hat{c_N}
		\end{pmatrix}
		&= 
		\begin{pmatrix}
		I & & & \\
		 & \left(1+s\|\bm{k}\|^2\right)^{\frac{l}{2}} & & \\
		 & & \ddots & \\
		 & & & \left(1+s\|\bm{k}\|^2\right)^{\frac{l}{2}}		
		\end{pmatrix} 
		\begin{pmatrix}
		\rho\\
		c_1\\
		\vdots\\
		c_N
		\end{pmatrix}\\
		&=\bm{W}\bm{x},
	\end{align}
	and a reformulated non-linear inverse problem
	\begin{align}
		\bm{F}\bm{W}\bm{x} = \tilde{\bm{F}}\bm{x}=\bm{y}.
	\end{align}
	The additional coil-regularization allows to differentiate between coils and image. The smooth signal fractions can be estimated to belong to the coils, while the others are part of the image. The optimization corresponds to 
	\begin{align}
	\hat{x} = \underset{\bm{x}}{\textrm{argmin}}\|\tilde{\bm{F}}(\bm{x})-\bm{y}\|_2^2+\alpha\bm{R}(\bm{x})+\beta\bm{Q}(\bm{x_c})
	\label{eq::optimization_NLINV}
	\end{align}
	with $\bm{R}(\bm{x})$ as Tikhonov-regularization and 
	\begin{align}
	\beta \bm{Q}(\bm{x_c}) = \beta\sum\limits_{j=1}^N\left\|\left(1+s\|\bm{k}\|^2\right)^{\frac{l}{2}}\mathfrak{F}c_j\right\|^2.
	\label{eq::coil_regularization_NLINV}
	\end{align}
	following \cite{Wang_Magn.Reson.Med._2018} and \cite{Uecker_Magn.Reson.Med._2008} with the constants $\alpha$ and $\beta$.

	\subsection{Chaining NLINV into Bloch Model Reconstruction}
	To integrate the calibrationless coil profile estimation into the Bloch model, the the  profiles $\bm{x_c}$ are determined simultaneously with the parameter maps $\bm{x_p}$. Therefore, the NLINV operator $\tilde{\bm{F}}$ are included into the chained Fourier  and Bloch operator signal model. The forward operator $\bm{A}$ becomes
	\begin{align}
		\bm{A}:\bm{x} = 		
		\begin{pmatrix}
		R_1\\
		R_2\\
		M_0\\
		c_1\\
		\vdots\\
		c_N
		\end{pmatrix}
		\longmapsto
		\begin{pmatrix}
		\mathfrak{P}\mathfrak{F}(c_1 M_{t_1})\\
		\vdots\\
		\mathfrak{P}\mathfrak{F}(c_N M_{t_1})\\
		\vdots\\
		\mathfrak{P}\mathfrak{F}(c_N M_{t_n})
		\end{pmatrix},
	\end{align}
	with $\bm{x} = (\bm{x_p},~\bm{x_c})^T$ and $\bm{x_c}=(c_1,~\dots,~c_N)^T$. Using the Jacobi matrix and the product rule, the operator for the derivative follows with
	\begin{align}
		\bm{D}\bm{A}(\bm{x})
		\begin{pmatrix}
		\textrm{d}R_1\\
		\textrm{d}R_2\\
		\textrm{d}M_0\\
		\textrm{d}c_1\\
		\vdots\\
		\textrm{d}c_N
		\end{pmatrix}
		= 
		\begin{pmatrix}
		\mathfrak{P}\mathfrak{F}\left(\textrm{d}c_1 M_{t_1} + c_1\left[
		\frac{\partial M_{t_1}}{\partial R_1}\textrm{d}R_1+
		\frac{\partial M_{t_1}}{\partial R_2}\textrm{d}R_2+
		\frac{\partial M_{t_1}}{\partial M_0}\textrm{d}M_0
		\right]\right)\\
		\vdots\\
		\mathfrak{P}\mathfrak{F}\left(\textrm{d}c_N M_{t_1} + c_N\left[
		\frac{\partial M_{t_1}}{\partial R_1}\textrm{d}R_1+
		\frac{\partial M_{t_1}}{\partial R_2}\textrm{d}R_2+
		\frac{\partial M_{t_1}}{\partial M_0}\textrm{d}M_0
		\right]\right)\\
		\vdots\\
		\mathfrak{P}\mathfrak{F}\left(\textrm{d}c_N M_{t_n} + c_N\left[
		\frac{\partial M_{t_n}}{\partial R_1}\textrm{d}R_1+
		\frac{\partial M_{t_n}}{\partial R_2}\textrm{d}R_2+
		\frac{\partial M_{t_n}}{\partial M_0}\textrm{d}M_0
		\right]\right)
		\end{pmatrix}
	\end{align}
	and the adjoint-derivative
	\begin{align}
		\bm{D}\bm{A}^H(\bm{x})
		\begin{pmatrix}
		y_{1,1}\\
		y_{2,1}\\
		\vdots\\
		y_{n,N}
		\end{pmatrix}
		=
		\begin{pmatrix}
		\textrm{d}R_1\\
		\textrm{d}R_2\\
		\textrm{d}M_0\\
		\textrm{d}c_1\\
		\vdots\\
		\textrm{d}c_N
		\end{pmatrix}
		=
		\begin{pmatrix}
		\sum\limits_{j=1}^N\sum\limits_{k=1}^n
		\overline{\left(\frac{\partial M_{t_k}}{\partial R_1}\right)}
		\cdot\overline{c_j}\cdot
		\mathfrak{F}^{-1}\left[\mathfrak{P}^H y_{k,j}\right]\\
		\sum\limits_{j=1}^N\sum\limits_{k=1}^n
		\overline{\left(\frac{\partial M_{t_k}}{\partial R_2}\right)}
		\cdot\overline{c_j}\cdot
		\mathfrak{F}^{-1}\left[\mathfrak{P}^H y_{k,j}\right]\\
		\sum\limits_{j=1}^N\sum\limits_{k=1}^n
		\overline{\left(\frac{\partial M_{t_k}}{\partial M_0}\right)}
		\cdot\overline{c_j}\cdot
		\mathfrak{F}^{-1}\left[\mathfrak{P}^H y_{k,j}\right]\\
		\sum\limits_{k=1}^n\overline{M_{t_k}}\cdot\mathfrak{F}^{-1}\left[\mathfrak{P}^H y_{k,1}\right]\\
		\vdots\\
		\sum\limits_{k=1}^n\overline{M_{t_k}}\cdot\mathfrak{F}^{-1}\left[\mathfrak{P}^H y_{k,N}\right]
		\end{pmatrix}.
	\end{align}

	\subsection{Results for Numerical Phantom}
	\label{ssec::num_phantom_result_NLINV_operator}
\begin{figure}[!h]
		\centering
		\includegraphics[width=\linewidth]{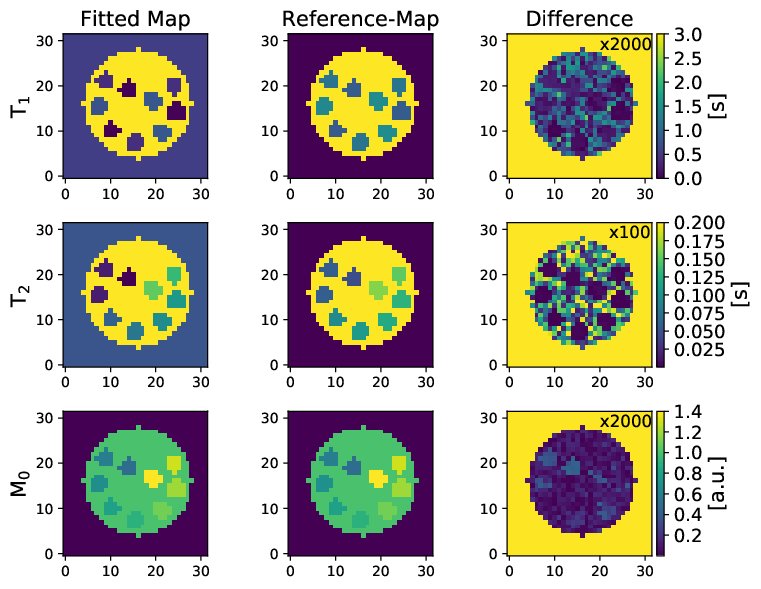}
		\caption{Visualization of the fitting results (\textbf{left}) using a simulated phantom (\textbf{center}) with the simulation parameters: inversion-prepared bSSFP sequence, 10 ms inversion pulse, $t_{RF}$= 0.90 ms, TR=4.5 ms, TE=2.25 ms, FA=45°, and $N_{\textrm{TR}}$ = 500. It is created pixel-wise and transformed to the frequency domain using a Fourier transformation to test the reconstruction. The estimated differences between simulation and approximation are plotted on the \textbf{right} and multiplied by the added factor to improve the visualization. The reconstruction model includes the chained Bloch, Fourier, and NLINV operator.}
		\label{fig::NLINV_included_Reconstruction}
	\end{figure}

\begin{figure}[!h]
		\centering
		
		\begin{subfigure}{.45\textwidth}
			\centering
			\includegraphics[width=1.1\linewidth]{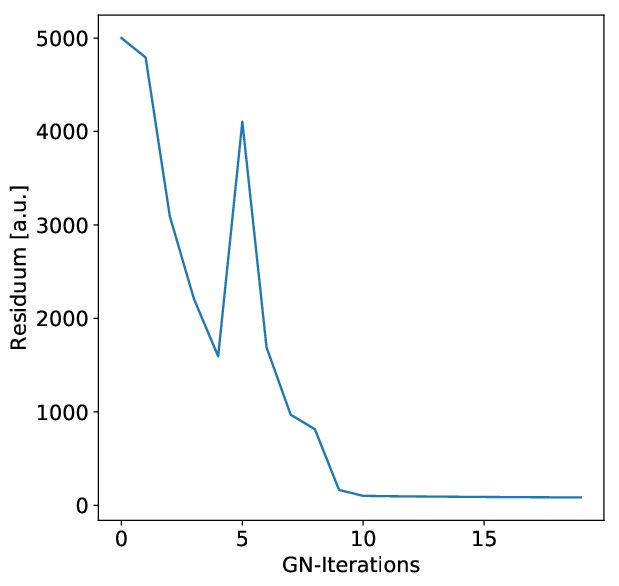}
			\caption{Visualization of the residuum during the Gauss-Newton steps of the optimization to create Figure \ref{fig::NLINV_included_Reconstruction}. }
			\label{fig::optimization_verification_NLINV_Fourier_Bloch_Operator_residuum}
		\end{subfigure}
		\hfill
		\begin{subfigure}{.45\textwidth}
			\centering
			\includegraphics[width=1.1\linewidth]{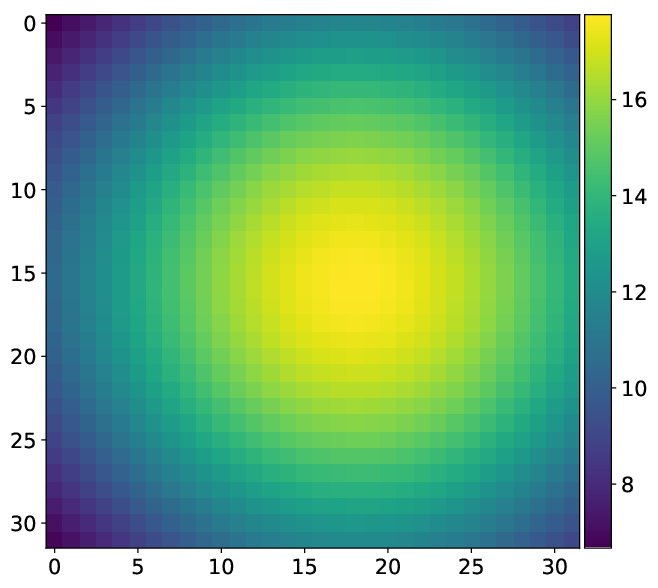}
			\caption{Visualization of the estimated sensitivity map during the reconstruction with the forward operator $\bm{A}$. }
			\label{fig::sens_NLINV_operator_test}
		\end{subfigure}
		\caption{Visualization of the residuum during the optimization with $\bm{A}$ (see Figure \ref{fig::NLINV_included_Reconstruction}) and the estimated sensitivity map.}
		\label{fig::residuum_NLINV_sens}
	\end{figure}

	In this section, the Bloch, the Fourier and the NLINV operator are combined to $\bm{A}$ and are tested using a simulated phantom, created and reconstructed analogously to section \ref{ssec::verification_optimization_simulated_phantom}. The fitting results are presented in Figure \ref{fig::NLINV_included_Reconstruction}, the residuum in \ref{fig::optimization_verification_NLINV_Fourier_Bloch_Operator_residuum}, and the sensitivities in \ref{fig::sens_NLINV_operator_test}. The reconstructed $M_0$ map is corrected by multiplication with the coil sensitivity profile $c_s$. 
	
	The error of the reconstructions visualized in Figure \ref{fig::NLINV_included_Reconstruction} is low. The intermediate water areas are not fitted as well as the tubes are but this is expected through the choice of the same initial guesses of the parameter maps like in section \ref{ssec::verification_optimization_simulated_phantom}.\\	
	Additionally, the sensitivity map in Figure \ref{fig::sens_NLINV_operator_test} looks as expected and the residuum in \ref{fig::optimization_verification_NLINV_Fourier_Bloch_Operator_residuum} converges\footnote{After 20 GN-Steps.} to 84 a.u. The peak at the 5th Gauss-Newton iteration may result from not fully optimized parameters in the IRGNM-FISTA algorithm.

	\subsection{Results for Phantom Measurement}
	\label{ssec::results_of_measurements}
	After checking the reconstruction method on simulated data in section \ref{ssec::verification_optimization_simulated_phantom} and \ref{ssec::num_phantom_result_NLINV_operator} actual phantom experiments are analyzed. The acquisition is performed as described in section \ref{ssec::measurements}. Two different approaches are chosen: The first uses a fully sampled acquisition, whereas the second is highly undersampled and more relevant for the clinical use. In the following, results for both techniques are presented and discussed.
		
	\subsubsection{Fully-Sampled Data}
	\label{ssec::analysis_fully_sampled_datasets}
\begin{figure}[!h]
		\centering
		\includegraphics[width=0.9\linewidth]{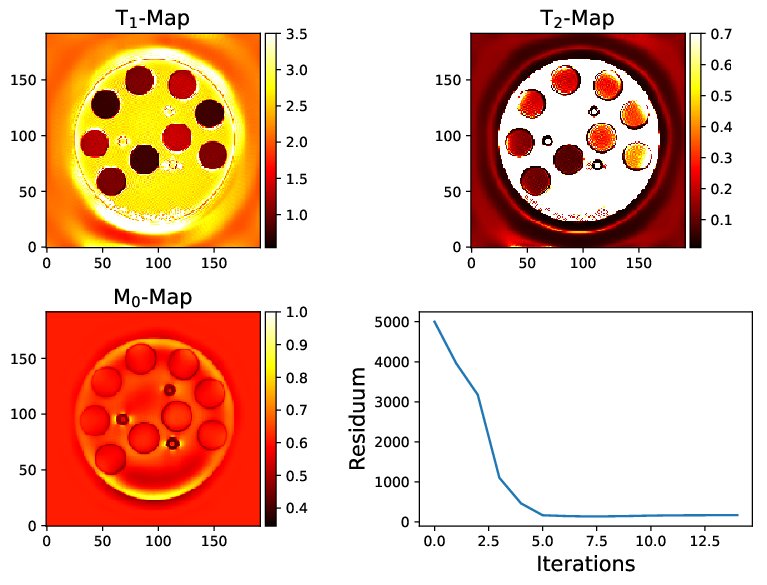}
		\caption{Visualization of the reconstructed maps in seconds of a custom-built $T_1$-$T_2$ phantom. The evolution of the residuum during the Gauss-Newton steps of the iterative reconstruction is added. All plotted values are in seconds.}
		\label{fig::RD_FS_Phantom_Maps_noWav}
	\end{figure}

	The fully sampled datasets are radially acquired, whereas the presented reconstruction works with Cartesian data. Therefore, the k-space data $\bm{y}$ is transformed into the image domain using a non-uniform FFT and is back-projected with a FFT. This gridding follows a singular value decomposition based coil compression to reduce the datasize and to speed up the reconstruction with only small losses of accuracy \cite{Zhang_Magn.Reson.Med._2013}. Afterwards, the Bloch model-based reconstruction can be applied analogously to section \ref{ssec::num_phantom_result_NLINV_operator}. Therefore, the simulation parameters added to Table \ref{tab::fitted_parameter_bloch_reco_FS} are passed to the simulation tool and an accurate balancing of the gradients is estimated: $s_{R_1}=1$, $s_{R_2}=30$, and $s_{M_0}=0.1$ for the fully sampled phantom data. Resulting parameter maps are visualized in Figure \ref{fig::RD_FS_Phantom_Maps_noWav}. The reconstruction is done using 20 Threads on a system with 40 CPUs\footnote{Intel\textregistered$~$  Xeon\textregistered$~$CPU E5-2650 v3 @ 2.30GHz} and additional other workload and takes approximately 37:30 h.
	
	\FloatBarrier

\begin{figure}[!h]
		\centering
		\begin{subfigure}{\textwidth}
			\centering
			\includegraphics[width=\linewidth]{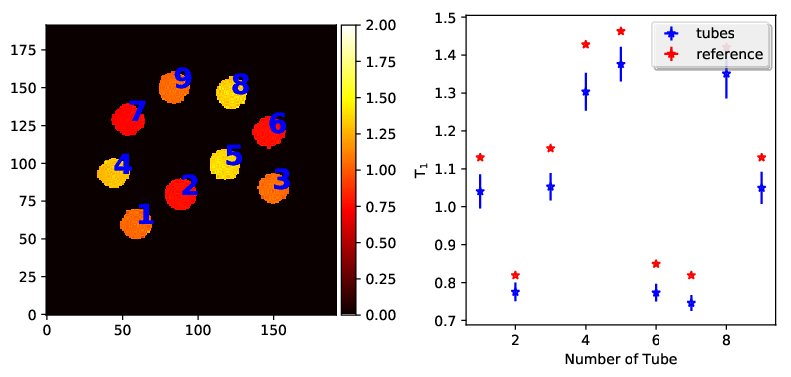}
		\end{subfigure}
		\vfill
		\begin{subfigure}{\textwidth}
			\centering
			\includegraphics[width=\linewidth]{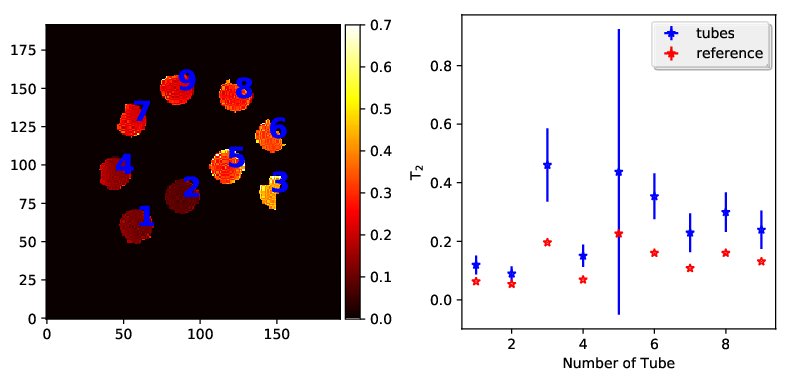}
			
		\end{subfigure}
		\caption{Visualization of the extracted parameter maps (\textbf{left}). The fully-sampled data is acquired at the 17.12.2018 and reconstructed using the calibrationless Bloch model. On the \textbf{right}, the determined parameters are plotted as well as the expected reference values determined by gold-standard measurements (see Table \ref{tab::fitted_parameter_bloch_reco_FS}). The errors are standard-derivations of the estimated regions of interests on the left. The tube numbering corresponds to the left image. All plotted values are in seconds.}
		\label{fig::tube_analysis_RD_FS_Phantom_2018-12-17_SC1_0_noWav}
	\end{figure}

	The fitting results for the tubes are shown in Figure \ref{fig::tube_analysis_RD_FS_Phantom_2018-12-17_SC1_0_noWav}. They are determined by thresholding the parameter maps of the relaxation with the values in Table \ref{tab::tresholding_para_real_data}. These values vary from the previous used ones during the $T_1$-$T_2$ phantom analysis shown in Table \ref{tab::threhLim}, because the background noise is more present. The automated thresholding method is verified by comparing it to a manual region of interest ROI selection, visualized in section \ref{sec::Comparison_Manual_Automatic_tube_analysis}.

	\begin{table}[!h]
		\centering
		\caption{Table lists the thresholding limits for the automatic segmentation of the tubes and for analyzing their relaxation parameter.}
		\begin{tabular}{c|c|c}\addlinespace[2ex]
			Parameters & Lower limit [s] & Upper limit [s]\\\hline
			$T_1$ & 0.005 & 0.5\\\hline
			$T_2$ & 0.5 & 2.0\\
		\end{tabular}
		
		\label{tab::tresholding_para_real_data}
	\end{table}

	Afterwards, two morphological transformations are applied \cite[p.21f]{Salditt__2017}: A closing (kernel size 2x2) creates uniform masks even if ringing artifacts are present and a following erosion (kernel size 3x3) removes most of the edge effects of the tubes.\\
	The whole analysis is performed on the parameter maps and is further compared to reference data. The proton density is ignored\footnote{Because it only improves the matching of the signal curves by influencing its scaling factor. There are multiple influences on the $M_0$-map. On the one hand it includes the proton density of the tissue but on the other hand it also includes data about the coil positioning. The latter might change between different experiments and therefore no reference data for the determined $M_0$-map can be provided.}.
	
\begin{figure}[!h]
		\centering
		\includegraphics[width=0.7\linewidth]{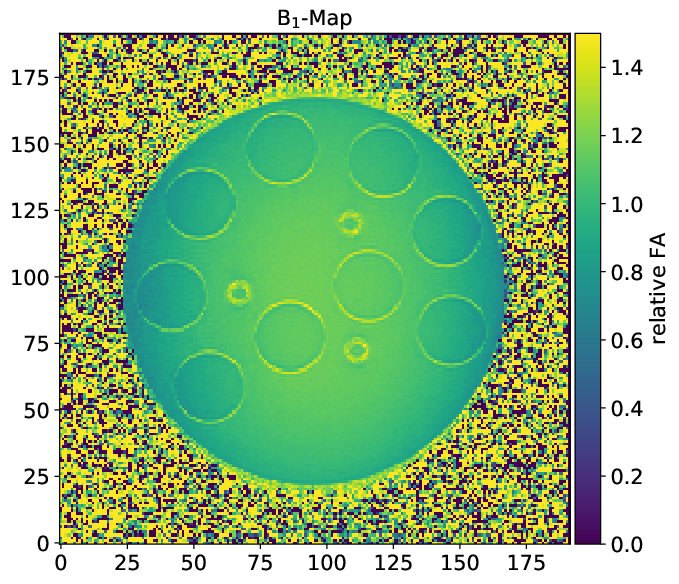}
		\caption{Visualization of an acquired $B_1$-map using a Siemens own sequence based on the Bloch-Siegert-shift. A protocol with the parameters is added to Table \ref{tab::sequence_protocol_B1_map}. }
		\label{fig::b1map_2018-12-17}
	\end{figure}

	The visualized fitting results in Figure \ref{fig::tube_analysis_RD_FS_Phantom_2018-12-17_SC1_0_noWav} differ from the measured gold-standard (also compare Table \ref{tab::fitted_parameter_bloch_reco_FS}). Especially the third and the fifth tube show high standard-derivations of the ROI and reach the limits of the thresholding for $T_2$. The large errors might result from the ringing artifacts within the tube (see Figure \ref{fig::RD_FS_Phantom_Maps_noWav}), and for the large parameter values various reasons are possible. One of them is an inhomogeneity of the applied RF-pulses in the image plane, called $B_1$-error, which can even be a reason for the increased crescent, like areas inside of the tubes. It can be reduced by acquiring a $B_1$-map of the measured slice by using the Bloch-Siegert effect \cite{Sacolick_Magn.Reson.Med._2010}. This is performed using a Siemens own sequence with the protocol added to Table \ref{tab::sequence_protocol_B1_map}. One exemplary map acquired at the 17.12.2018 is visualized in Figure \ref{fig::b1map_2018-12-17}.
	
\begin{figure}[!h]
		\centering
		\includegraphics[width=0.9\linewidth]{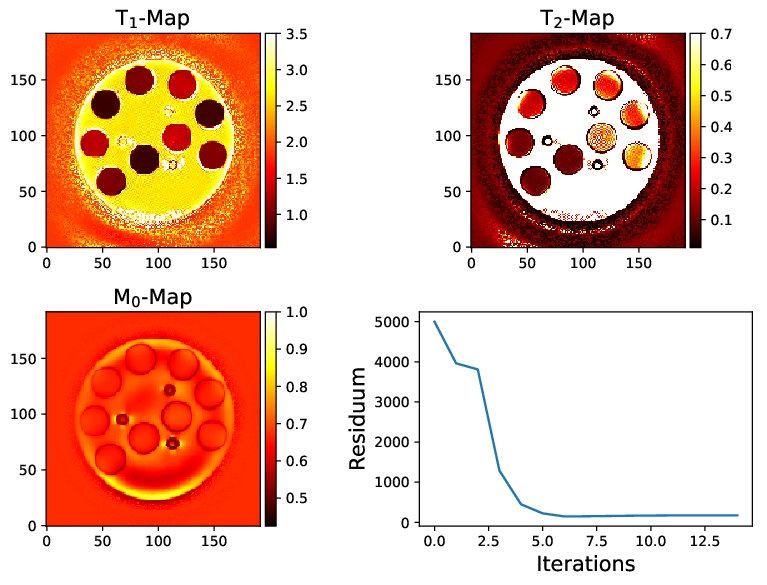}
		\caption{Visualization of the reconstructed data of the custom-built $T_1$-$T_2$ phantom. The individual maps and the evolution of the residuum during the Gauss-Newton iterations is plotted for the $B_1$-corrected matching technique. All plotted values are in seconds.}
		\label{fig::RD_FS_Phantom_Maps_B1corr_noWav}
	\end{figure}

	To include this effect in the Bloch model based reconstruction, the relative flip-angle is multiplied pixel-wise with the FA defined during the Bloch simulation in $\mathfrak{B}$. The inversion pulse is not affected because, in theory, the scanner uses an adiabatic inversion instead of a 180°-pulse. In this work an approximation with the latter is applied. The results of the $B_1$ corrected reconstructions are visualized in Figure \ref{fig::RD_FS_Phantom_Maps_B1corr_noWav}. The reconstruction time is the same like in the not corrected case.
	
\begin{figure}[!h]
		\centering
		\begin{subfigure}{\textwidth}
			\centering
			\includegraphics[width=\linewidth]{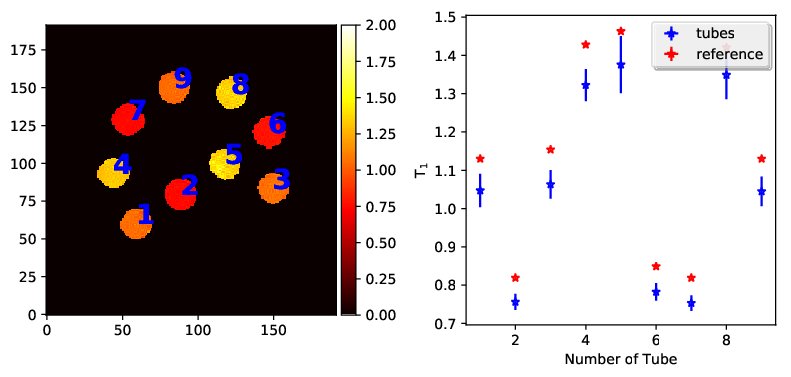}
		\end{subfigure}
		\vfill
		\begin{subfigure}{\textwidth}
			\centering
			\includegraphics[width=\linewidth]{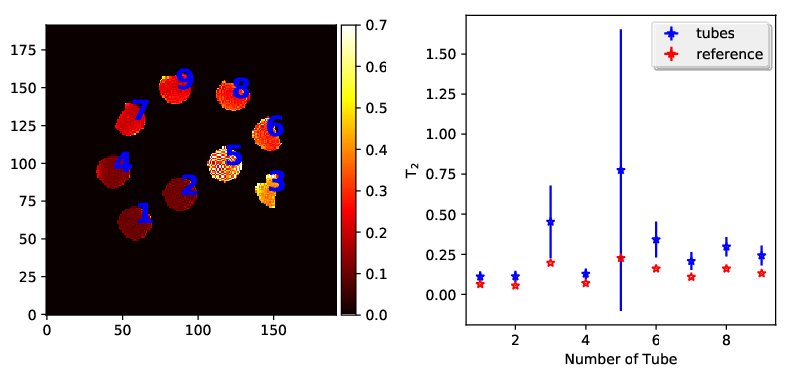}
			
		\end{subfigure}
		\caption{Visualization of the extracted parameter maps (\textbf{left}). The fully-sampled data is acquired at the 17.12.2018 and reconstructed using the calibrationless Bloch model including a $B_1$-correction. On the \textbf{right}, the determined parameters are plotted for each tube with the expected reference value (also compare Table \ref{tab::fitted_parameter_bloch_reco_FS}). The errors are standard-derivations of the estimated regions of interests on the left. The tube numbering corresponds to the left image. All plotted values are in seconds.}
		\label{fig::tube_analysis_RD_FS_Phantom_2018-12-17_B1corr_SC1_0_noWav}
	\end{figure}

	To analyze the data, the thresholding limits presented in Table \ref{tab::tresholding_para_real_data} are used and equal initialization parameters are chosen. The estimated parameter maps are visualized in Figure \ref{fig::tube_analysis_RD_FS_Phantom_2018-12-17_B1corr_SC1_0_noWav}.
	
\begin{figure}[!h]
		\centering
		\includegraphics[width=0.9\linewidth]{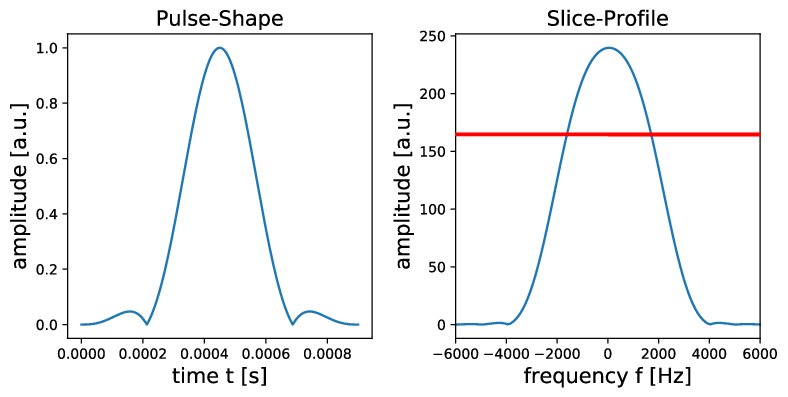}
		\caption{Visualization of the slice-profile of a typical RF-pulse with a duration of 0.9 ms and a BWTP of 3.8 on the \textbf{left}. On the \textbf{right}, its frequency-response (Fourier transformation) is plotted, which corresponds to the slice-profile. The red line corresponds to an approximated constant slice-profile of $k_{ss}=$0.7. }
		\label{fig::Slice_profile_correction_sketch}
	\end{figure}

	The result is similar to one without $B_1$-correction, which can be proven by comparing the values added to Table \ref{tab::fitted_parameter_bloch_reco_FS}. Even the errors in tube\footnote{The numbering refers to Figure \ref{fig::tube_analysis_RD_FS_Phantom_2018-12-17_B1corr_SC1_0_noWav}.} 3 and 5 and the crescent artifacts inside of the tubes are still present. The mismatching effect therefore needs to rely on something else than the $B_1$-correction in the image plane, so an imperfect slice-profile is assumed. This corresponds to a $B_1$-correction perpendicular to the imaging plane and results from the frequency response of the RF-pulse. The optimal correction would be based on simulating multiple spins along the slice profile. An estimate of the real rf-profile the scanner is using can be exported from the Siemens sequence programming platform IDEA, which simulates the scanners response. This is performed in Figure \ref{fig::Slice_profile_correction_sketch} for a used pulse with a duration of 0.9 ms and a Bandwidth-Time-Product BWTP\footnote{The BWTP is in principle the number of zero-crossings of the envelope.} of 3.8. The Figure also includes the frequency-response of the pulse, which is the same as the selected slice. Therefore, in the ideal case, it represents a rectangle but this is not the case for a BWTP of 3.8. While the need of correcting it is obvious, simulating multiple spins and averaging their signal response would significantly prolong the reconstruction in the current implementation. Therefore, the profile is approximated using a constant guess $k_{ss}$, which would be 1 for an ideal slice-profile. This parameter leads to an additional factor in the simulation of the flip-angle analogously to the $B_1$-correction.
	
\begin{figure}[!h]
		\centering
		\includegraphics[width=0.9\linewidth]{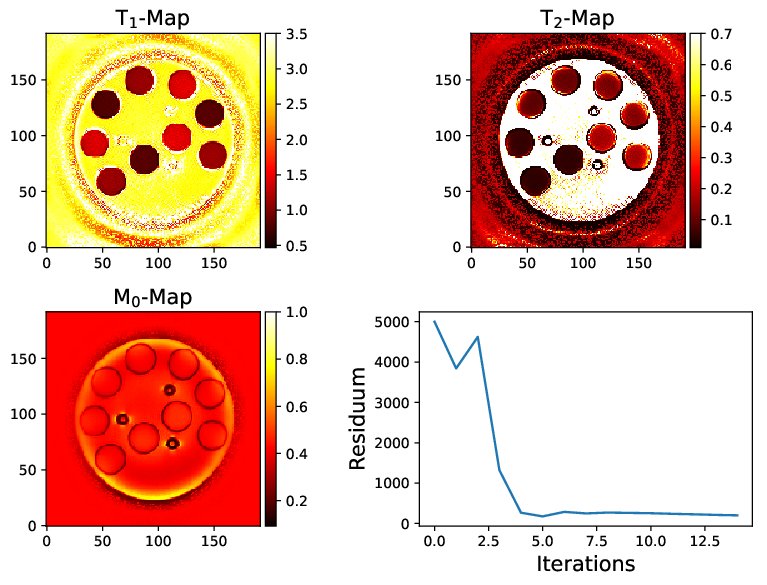}
		\caption{Visualization of the reconstructed parameter maps of a custom-built $T_1$-$T_2$ phantom. The reconstruction uses a $B_1$- and a slice-profile correction of $k_{ss}=$0.7. All plotted values are in seconds.}
		\label{fig::RD_FS_Phantom_Maps_B1Corr_SC0_7_noWav}
	\end{figure}

	After including the slice-profile correction assuming a constant factor of $k_{ss}$=0.7
	(also added to Figure \ref{fig::Slice_profile_correction_sketch}), the results of the reconstructions are presented in Figure \ref{fig::RD_FS_Phantom_Maps_B1Corr_SC0_7_noWav}.

	\FloatBarrier
	
\begin{figure}[!h]
		\centering
		\begin{subfigure}{\textwidth}
			\centering
			\includegraphics[width=\linewidth]{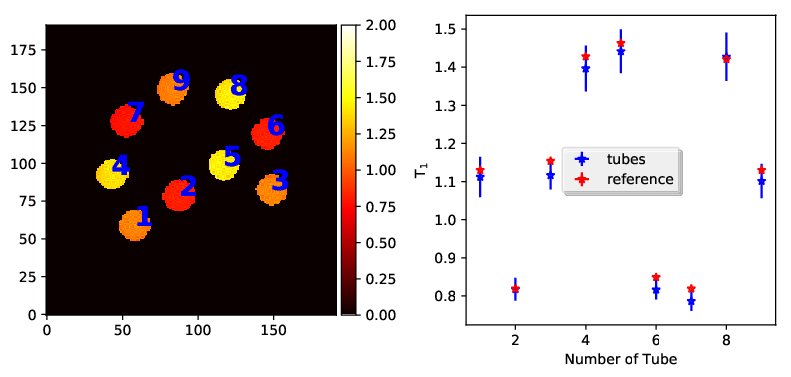}
		\end{subfigure}
		\vfill
		\begin{subfigure}{\textwidth}
			\centering
			\includegraphics[width=\linewidth]{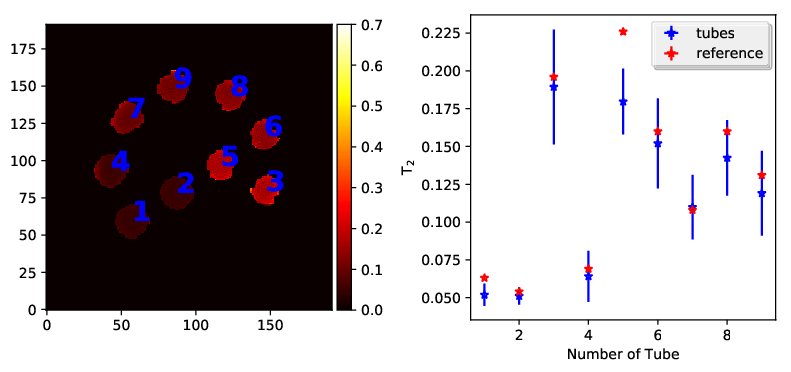}	
		\end{subfigure}
		\caption{Visualization of the extracted parameter maps (\textbf{left}). The fully-sampled data is acquired at the 17.12.2018 and reconstructed using the calibrationless Bloch model with a $B_1$-correction and a constant slice-profile correction of $k_{ss}=0.7$. On the \textbf{right} the determined parameters are plotted for each tube with the expected reference value. The errors are standard-derivations of the estimated regions of interests on the left. The tube numbering corresponds to the left image. All plotted values are in seconds.}
	\label{fig::tube_analysis_RD_FS_Phantom_2018-12-17_B1corr_SC0_7_noWav}
	\end{figure}

\begin{figure}[!h]
		\centering
		\begin{subfigure}{\textwidth}
			\centering
			\includegraphics[width=\linewidth]{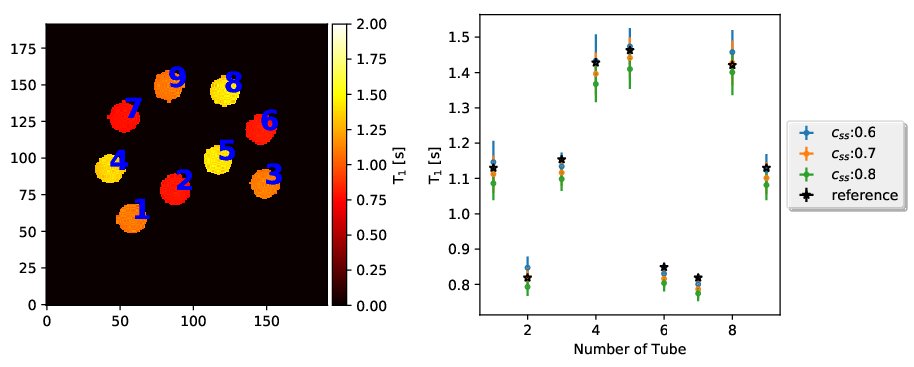}
		\end{subfigure}
		\vfill
		\begin{subfigure}{\textwidth}
			\centering
			\includegraphics[width=\linewidth]{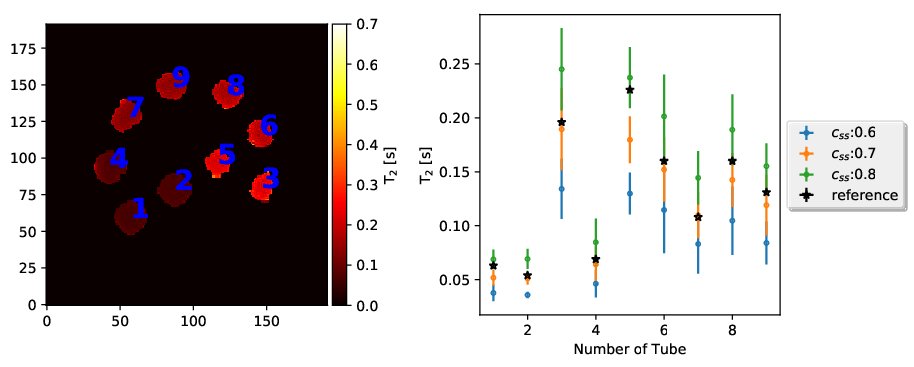}	
		\end{subfigure}
		\caption{Visualization of the extracted parameter maps (\textbf{left}) using a slice-profile correction of $k_{ss}=0.8$. The fully-sampled data is acquired at the 17.12.2018 and reconstructed using the calibrationless Bloch model including a $B_1$-correction and a slice-profile correction with $k_{ss}$. On the \textbf{right} the determined parameters are plotted for each tube and multiple correction factors together with the expected reference value. The errors are standard-derivations of the estimated regions of interests. The tube numbering corresponds to the left image.}
		\label{fig::multi_tube_2018-12-17_FS_B1corr_noWav}
	\end{figure}

	The estimated parameters are plotted in Figure \ref{fig::tube_analysis_RD_FS_Phantom_2018-12-17_B1corr_SC0_7_noWav} and listed with other values of $k_{ss}$ in Table \ref{tab::fitted_parameter_bloch_reco_FS}. The additional correction factors can be found in Figure \ref{fig::multi_tube_2018-12-17_FS_B1corr_noWav}. The reconstruction time is the same like in the not corrected case.
	
	\FloatBarrier
	
	The analysis of the maps in Figure \ref{fig::tube_analysis_RD_FS_Phantom_2018-12-17_B1corr_SC0_7_noWav} shows that they are closer to the expected reference values than previous uncorrected reconstructions. Therefore, the slice-profile correction seems to have a great effect on the listed parameter. Nevertheless, there is still a difference which might result from the constant slice approximation with the parameter $k_{ss}$. Also the large error in tube\footnote{The numbering refers to Figure \ref{fig::tube_analysis_RD_FS_Phantom_2018-12-17_B1corr_SC0_7_noWav}.} 5 and the crescent artifacts are still present. The standard-derivations of tube 6 and 9 are also increased, which may result from a not optimized segmentation.

	\FloatBarrier
	
	The effect of the slice-profile correction factor $k_{ss}$ is visualized in Figure \ref{fig::multi_tube_2018-12-17_FS_B1corr_noWav}. For $T_1$ an increased  $k_{ss}$ leads to a reduced value, whereas for $T_2$ it is the opposite. Therefore, it shows the necessity of a correct estimation of the slice-profile effect for the final result.
		

	\subsubsection{Undersampled Data}
\begin{figure}[!h]
		\centering
		\includegraphics[width=\linewidth]{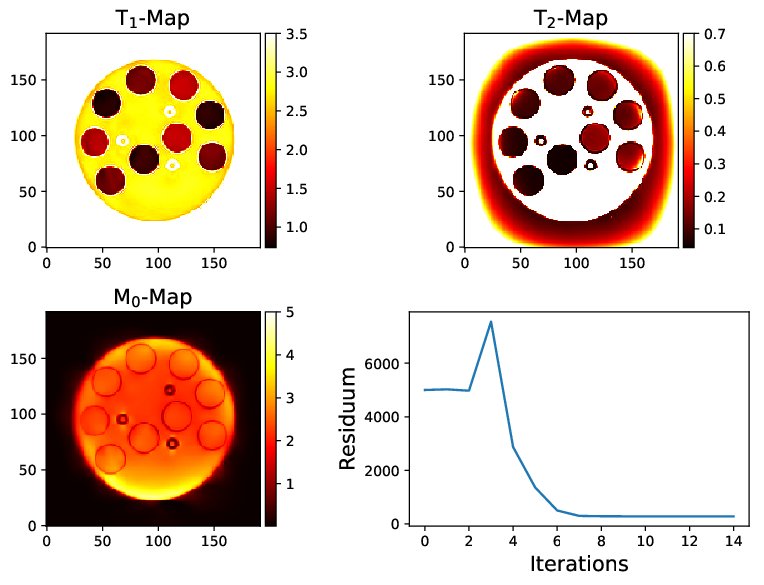}
		\caption{Visualization of the reconstructed data of the custom-built $T_1$-$T_2$ phantom acquired using the single-shot golden-angle based inversion-prepared bSSFP sequence. The reconstruction includes a $B_1$- and a slice-profile correction with $k_{ss}$=0.7. All plotted values are in seconds.}
		\label{fig::RD_SS_Phantom_Maps_B1corr_Wav}
	\end{figure}

\begin{figure}[!h]
		\centering
		\begin{subfigure}{\textwidth}
			\centering
			\includegraphics[width=\linewidth]{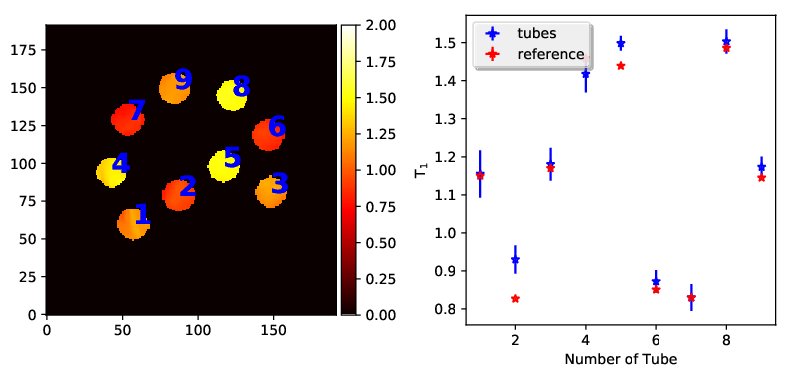}
		\end{subfigure}
		\vfill
		\begin{subfigure}{\textwidth}
			\centering
			\includegraphics[width=\linewidth]{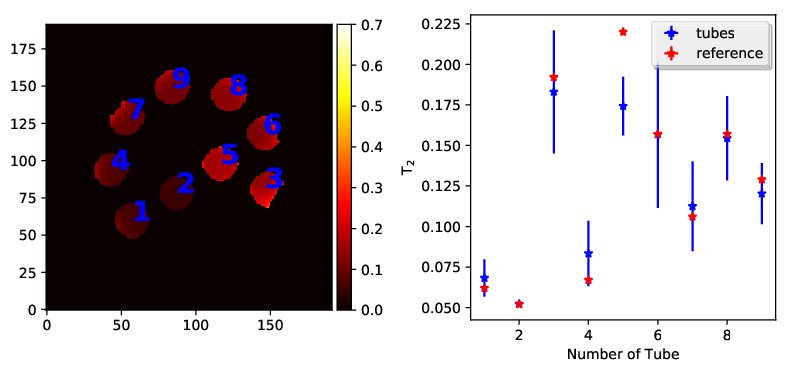}
			
		\end{subfigure}
		\caption{Visualization of extracted parameter maps (\textbf{left}). The single-shot data is acquired at the 17.12.2018 and reconstructed using the calibrationless Bloch model including a wavelet-denoising, a $B_1$-correction and a constant slice-profile correction of $k_{ss}=0.7$. On the \textbf{right} the determined parameters are plotted for each tube together with the expected reference value determined by gold-standard measurements. The errors are standard-derivations of the estimated regions of interests (see left). The tube numbering corresponds to the left image. The maps are analyzed after the 8th Gauss-Newton step to avoid over-regularizations. All plotted values are in seconds.}
		\label{fig::tube_analysis_RD_SS_Phantom_2018-12-17_B1corr_SC0.7_Wav}
	\end{figure}

	To get an overview about the possibilities of this method, it is also validated performing single-shot experiments, like explained in section \ref{ssec::measurements}. This results in a greatly reduced acquisition time, which is practically realizable in the clinical setting. The data-density is reduced by acquisitions with high undersampling factors. Typically, 10 to 20 spokes per frame compared to fully-sampling of 191 spokes are aquired, which leads to strong artifacts using conventional reconstructions. To minimize the artifacts, the method of \cite{Wang_Magn.Reson.Med._2018} is selected. Therefore, the optimization becomes similar to equation \ref{eq::optimization_NLINV} based on the Sobolev norm for exploiting the smoothness of the coils (equation \ref{eq::coil_regularization_NLINV}) but with $\bm{A}$ exchanged with operator $\tilde{\bm{F}}$. Additionally, the Tikhonov regularoization in $\bm{R}$ is replaced with a joint sparsity model introduced by \cite{Vasanawala__2011}:	
	\begin{align}
		\alpha^k\bm{R}(\bm{x_p})=\alpha_k\sum\limits_r\sqrt{\sum\limits_p |w_{rp}|^2},
	\end{align}
	including the $r$th wavelet parameter $w_{rp}$ of the parameter map $p$. This ensures that individual parameters are protected from large ones on other maps. Typically those influence smaller ones while using a non-linear optimization. \\
	For the extension with wavelet-denoising, the optimization using FISTA becomes a soft-thresholding \cite{Beck_SIAMJ.Img.Sci._2009}.\\
	For the Bloch operator $\mathfrak{B}$, the simulation is adjusted to cover the spokes averaged to one k-space similar to the golden-angle sampling scheme in the sequence illustrated in figure \ref{fig::sequence_sampling_schemes_ga}. Therefore, the signals of multiple TR are averaged to one single time point. This is done using an arithmetic mean.\\
	To reconstruct the single-shot datasets, the same preparation steps like gridding are done, as described in section \ref{ssec::analysis_fully_sampled_datasets}. The settings for the optimization algorithm are also the same. Only the gradient balancing is adjusted to $s_{R_1}=1$, $s_{R_2}=15$ and $s_{M_0}=1$.\\
	The results including a $B_1$-correction, wavelet-denoising, and a slice-profile correction of $k_{ss}$=0.7 are presented in Figure \ref{fig::RD_SS_Phantom_Maps_B1corr_Wav}, \ref{fig::tube_analysis_RD_SS_Phantom_2018-12-17_B1corr_SC0.7_Wav} for the 8th Gauss-Newton-iteration\footnote{Is determined to be the most accurate iteration before over-regularization occurs.} and the extracted relaxation parameter for the tubes are listed in Table \ref{tab::fitted_parameter_bloch_reco_SS}. The whole reconstruction takes about 9 h.

	\FloatBarrier

\begin{figure}[!h]
		\centering
		\begin{subfigure}{\textwidth}
			\centering
			\includegraphics[width=\linewidth]{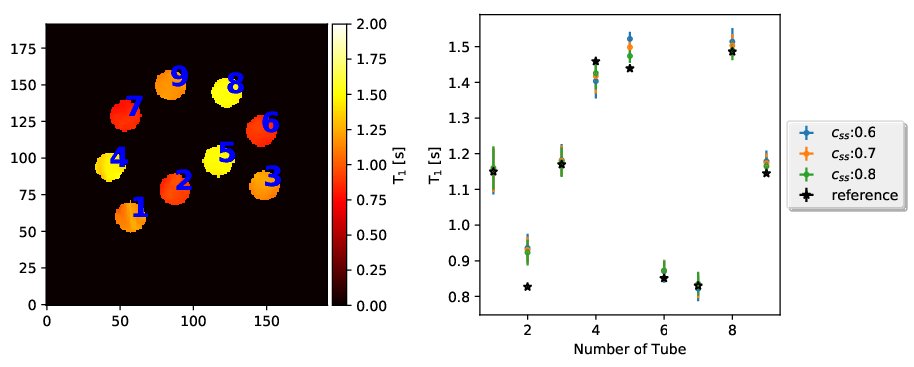}
		\end{subfigure}
		\vfill
		\begin{subfigure}{\textwidth}
			\centering
			\includegraphics[width=\linewidth]{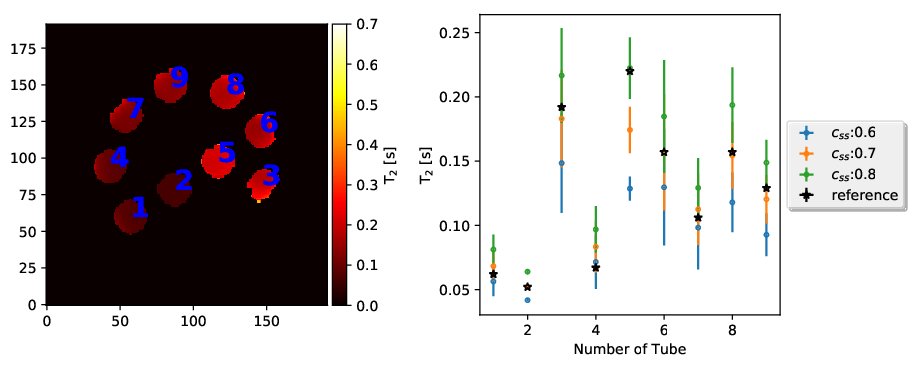}	
		\end{subfigure}
		\caption{Visualization of the extracted parameter maps (\textbf{left}) for a slice-profile correction of $k_{ss}$=0.8. The single-shot data is acquired at the 17.12.2018 and reconstructed using the calibrationless Bloch model including a wavelet-denoising, a $B_1$-correction, and a slice-profile correction with $k_{ss}$. On the \textbf{right}, the determined parameters are plotted for each tube and multiple correction factors together with their expected reference values. The errors are standard-derivations of the estimated regions of interests (see left). The tube numbering corresponds to the left image.}
		\label{fig::multi_tube_2018-12-17_SS_B1corr_Wav}
	\end{figure}

%
	The reconstructed maps visualized in Figure \ref{fig::RD_SS_Phantom_Maps_B1corr_Wav} are visually smoother than the fully-sampled. This is an effect of the wavelet-denoising. The estimated $T_1$ values have a similar precision compared to the fully-sampled images, while $T_2$ still varies a lot with no observable trend. $k_{ss}$= 0.7 is chosen as best approximation after testing different parameters, shown in Figure \ref{fig::multi_tube_2018-12-17_SS_B1corr_Wav}.

	\FloatBarrier

	\section{Discussion of Bloch Model Reconstruction}
	The Bloch model based reconstruction is validated using simulations and measured phantom data. Figure \ref{fig::optimization_verification_just_Bloch_Operator}-\ref{fig::NLINV_included_Reconstruction} visualize that the optimization algorithm works. The quantitative analysis of the method is presented in Figure \ref{fig::tube_analysis_RD_FS_Phantom_2018-12-17_B1corr_SC0_7_noWav} and \ref{fig::tube_analysis_RD_SS_Phantom_2018-12-17_B1corr_SC0.7_Wav}. Even for assumptions like a constant slice profile and a non-adiabatic inversion pulse, the determined relaxation coefficients for the tubes are close to the estimated gold-standard reference values. This shows the potential for further improvements of the technique to increases in the accuracy of the relaxation parameter maps.\\
	Nevertheless, there are still limitations, which need to be corrected in future work. The manually choice of gradient balancing factors is clinically not realizable. Additionally, the slice-profile correction has to be more accurate and the reconstruction time of multiple hours need to be improved.\\
	Nevertheless, the results presented are very accurate for the grade of assumptions and the acquisition time of 7 s is fast. Its theoretical variability for any type of sensitive sequence needs to be tested in future experiments.

\chapter{Discussion}

In this work a new model-based reconstruction technique based on the full Bloch equations and including a calibrationless estimation of the coil profiles has been developed. For its validation a phantom was created and characterized using gold-standard spin-echo methods. The forward operator based on a Dopri54 algorithm was developed and combined with a IRGNM-FISTA optimization. Afterwards, the technique has been applied to simulated and measured data using an inversion-prepared bSSFP sequence. Finally, it was validated using radial single-shot data that is acquired within 7 s.\\
The custom-built phantom was necessary to validate the accuracy of the new reconstruction technique because the gold-standard reference measurements can take up to multiple hours and are too long for volunteers to stay motionless. Additionally, the literature based reference values of human tissue have a wide range\footnote{The $T_1$ value of gray matter varies from 1331-1820 ms at 3 T \cite{Hattori_Med.Phys._2013}.} and are therefore not recommendable for precision tests. The phantoms ingredients were chosen to result in individual tubes that are similar in $T_1$ and $T_2$ to human tissue. The phantoms spatial homogeneity was found to be high but especially the $T_2$ values are only stable within a few days.\\
The forward operator includes a simulation of the full Bloch equations. Compared to other methods that are based on approximated signal models, this allows to model spin dynamics of arbitrary sequences and to estimate multiple parameters at once. It was realized by using a Dopri54 ODE-solver. Other methods like discrete simulations based on rotational matrices are limited to only one sample-size and are not exploiting that the Bloch equations are ODEs. Their sample-density needs to be high during RF-pulses but can not be decreased in periods of relaxation. This makes the ODE-solver superior because an adaptive step size can be added and no simplifications of the ODEs are necessary. The sequence blocks of the simulation were validated using the theoretical expectations and both were shown to be in good agreement. The sequence simulation was compared to an inversion-prepared bSSFP signal model including its approximations and reproduced the signal development accurately. Afterwards, the validation was repeated with a more realistic behavior and the relaxation during the RF-pulses was included. Figure \ref{fig::simultation_sequence_test_ODE} visualizes the influence of the hard-pulse approximations in the theoretical model on the temporal evolution of the signal compared to the ODE-solver. While the fitting parameters in Table \ref{tab::fitting_results_sequence_verification} indicate the strong influence of hard-pulse approximations on the final relaxation characteristics, there is still literature that actually use these simplified models and produces accurate results \cite{Schmitt_Magn.Reson.Med._2004}. This requires a further comparison between the Bloch model-based method developed in this work and the "reconstruction followed by pixel-wise fitting" mapping techniques.\\
In the next step the IRGNM-FISTA optimization algorithm, based on an adapted version of previous work by \cite{Wang_Magn.Reson.Med._2018} in the BART reconstruction toolbox, was set up. The derivative and adjoint-derivative of the Bloch operator were verified by exploiting the adjoint definition with the scalar product. Afterwards, the reconstruction was validated on simulated datasets (Figure \ref{fig::optimization_verification_just_Bloch_Operator}). Because raw data of the scanner system is acquired in the k-space and a previously needed reconstruction would limit the simple applicability of this method and might lead to additional error sources, the forward model was extended to project into the frequency domain. This was realized by chaining the Bloch operator to a Fourier operator.\\
To speed up measurements and to increase the signal-to-noise ratio of the images, modern MRI scanners use parallel imaging with multiple receiver coils. Those typically cover a small fraction of the measured object and their sensitivity profiles need to be estimated for the reconstruction. To avoid preparation scans and to use all of the acquired data for the coil profile estimation, the forward model was extended by the NLINV operator. This included the coil profile estimation directly into the reconstruction. Afterwards, the optimization algorithm was verified on simulated data, which proved its functionality. Nevertheless, the accuracy of the reconstruction had to be checked with measured phantom data because the numerical phantom was created by using the same simulation the Bloch operator is based on. Therefore, accuracy tests with the numerical phantom ended up in an inverse crime.\\
The first reconstructions of measured data were carried out with a fully-sampled dataset. This represented the gold-standard technique of testing new reconstruction methods because no missing information had to be compensated. In the resulting maps $T_1$ was constantly underestimated while $T_2$ was even stronger overestimated. Additionally, some crescent artifacts on the outer tubes edges occurred. A possible reason is an imperfect $B_1$ field in the image plane. Therefore, not all spins in the image plane experience the same alternating flip-angle during the inversion-prepared bSSFP sequence. To correct this effect, the $B_1$ map can be estimated within the presented Bloch model-based mapping technique but would lead to an additional gradient that follows from the Bloch equations but would need to be balanced, too. Because of the missing automatic balancing, the current state of the Bloch model-based reconstruction does not support the estimation of $B_1$ inaccuracies. Therefore, the $B_1$ field was estimated in a preparation scan using a Bloch-Siegert-shift based sequence. After passing the determined relative FA correction of the $B_1$ map to the simulation inside of the forward operator, the results were presented in Figure \ref{fig::RD_FS_Phantom_Maps_B1corr_noWav} and \ref{fig::tube_analysis_RD_FS_Phantom_2018-12-17_B1corr_SC1_0_noWav}. The correction had almost no influence on the relaxation parameters even though the relative flip-angle range was estimated to be high. Additionally, the edge effects in the tubes were still present.\\
Because of the weak effect of the $B_1$ correction in image domain, an other reason for the inaccuracies in the relaxation parameter maps was found to rely on the imperfect slice-profile of the used RF-pulses. Their BWTP of about 3.5 was too low to allow a rectangular frequency response and therefore resulted in a lower experienced flip-angle for the whole measured slice. The ideal simulation would be realized by exporting a good estimate of the scanners slice-profile from IDEA, simulating multiple spins for its characteristic spatial evolution, and average all spins afterwards. As this increases the reconstruction time from hours to multiple days\footnote{depending on the number of simulated spins}, the slice-profile was approximated by a constant factor $k_{ss}$ and corrected by multiplying it with the flip-angle used in the simulation for every pixel equally. This led to less crescent artifacts in the parameter maps and to more accurate results for the relaxation constants. It is not clear yet why the slice-profile had a large influence on the resulting parameter maps whereas the $B_1$-effect in the image plane had none even though both effects on the simulated sequence are similar. These differences need further investigation.\\ 
Finally, the Bloch model-based algorithm was verified on single-shot datasets. They reduced the acquisition time of a single slice from 39 min, in the fully-sampled case, to 7 s and brought the method closer to clinical applications. The increased undersampling factor reduced the amount of acquired data and therefore led to artifacts within the reconstructed images. To avoid them, a wavelet denoising, analogue to \cite{Wang_Magn.Reson.Med._2018}, is introduced. Because the denoising increased the smoothness of the images, the segmented tubes had smaller standard-derivations. The $T_1$-map had a high accuracy, which is similar to the fully-sampled measurements. The three outliers in tube\footnote{The numbering refers to Figure \ref{fig::tube_analysis_RD_SS_Phantom_2018-12-17_B1corr_SC0.7_Wav}.} 2, 4, and 5 could result from an inaccurate $B_1$-correction in the image plane. This might also led to the only partially good approximation of $T_2$.\\
In summary, the model-based reconstruction developed in this master thesis does not rely on simplified signal models, as comparable methods do, but uses the full Bloch equations as a physical model for the spin dynamics. It also does not use pre-computed dictionaries of signal curves, like MR fingerprinting, which reduces storage and avoids discretization errors. For the specific application, a single-shot radial inversion-prepared bSSFP sequence is developed, which allows a first proof-of-principle application of the acquisition of $T_1$ and $T_2$ within a single shot. This leads to an acquisition time of only 7 s, which brings quantitative mapping closer to clinical applications where acquisition time is a limiting
factor.

\chapter{Outlook}
In this work, the developed Bloch model-based method is applied to an inversion-prepared bSSFP sequence but readily extends to other sequences. Interesting extensions which should be explored are frequency-modulated bSSFP or variable flip-angle schemes, as proposed in \cite{Foxall_Magn.Reson.Med._2002}\cite{Homer_J.Magn.Reson._1985}.\\
Additionally, the reconstruction algorithm can be improved. On the simulation side, the RF-pulse might be implemented more effectively by adding the sinc-pulse directly into the Bloch equations, which would allow the use of the adaptive step-size control even during the RF-pulses and improve the speed of the simulation.\\
On the reconstruction side, the $B_1$-profile can be added to the optimization as additional parameter map to avoid the need for a preparation-scan. The additional derivative which is then required, follows directly from the sensitivity analysis of the Bloch-equations.\\
Additionally, the gradient scaling needs to be done automatically and readjusted after each iterative Gauss-Newton step.\\
These modifications will help to increase the accuracy, to reduce the reconstruction time, and to improve the robustness of the model-based reconstruction using the Bloch equations.

\appendix
\chapter{Appendix}

\section{Comparison between Manual and Automatic Tube Analysis}
\label{sec::Comparison_Manual_Automatic_tube_analysis}
\begin{figure}[!h]
	\centering
	\begin{subfigure}{0.45\textwidth}
		\centering
		\includegraphics[width=1.2\linewidth]{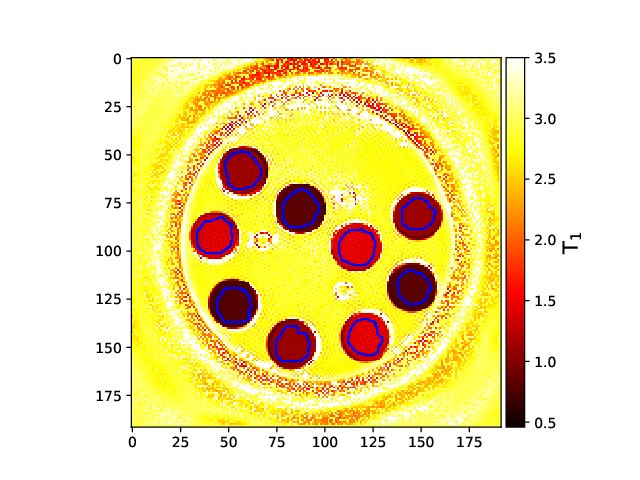}
		\label{fig::manual_roi_analysis_T1_RD_FS_Phantom_2018-12-17_B1Corr_SC0_8_noWav}
	\end{subfigure}
	\hfill
	\begin{subfigure}{0.45\textwidth}
		\centering
		\includegraphics[width=1.2\linewidth]{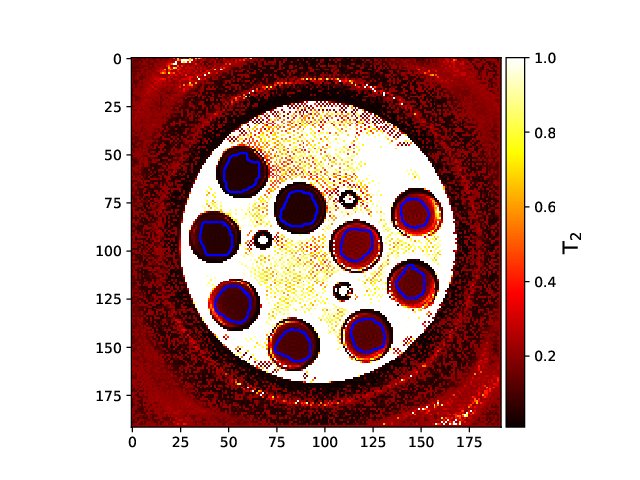}
		\label{fig::manual_roi_analysis_T2_RD_FS_Phantom_2018-12-17_B1Corr_SC0_8_noWav}
	\end{subfigure}
	\caption{Visualization of the manually determined ROIs in blue. The underlying images are fully-sampled datasets from the 17.12.2018 and the reconstruction includes a $B_1$- and a slice-profile correction of $k_{ss}=0.7$. The colorbars represent seconds. }
	\label{fig::manual_roi_analysis_RD_FS_Phantom_2018-12-17_B1Corr_SC0_8_noWav}
\end{figure}

This section evaluates the accuracy of automatic tube analysis by comparing it a manual ROI selection for an exemplary dataset. It is acquired on the 17.12.2018 and is reconstructed with $B_1$- as well as slice-profile-correction $k_{ss}=0.7$. Its thresholded automatic analysis is presented in Figure \ref{fig::tube_analysis_RD_FS_Phantom_2018-12-17_B1corr_SC0_7_noWav}. The manual analysis and the belonging ROIs are visualized in Figure \ref{fig::manual_roi_analysis_RD_FS_Phantom_2018-12-17_B1Corr_SC0_8_noWav}. A comparison of the determined relaxation and density parameter is added in Table \ref{tab::Comparison_manual_automatic_tube_analysis}.

	\begin{table}
	\footnotesize
	\centering
	\caption{Table with analysis results of the manual and the automatic thresholding method. The data is acquired at the 17.12.2018 and the reconstruction uses a $B_1$ correction and a slice-profile estimate of $k_{ss}=0.7$. The tube numbering corresponds to Figure \ref{fig::t1t2phantom_tubeOrdering}.}
	
	\begin{adjustbox}{center}
	\begin{tabular}{c||c|c|c|c|c|c|c|c|c}\addlinespace[2ex]
		Tube & 2 & 3 & 4 & 5 & 6 & 7 & 8 & 9 & 10 \\\hline\hline
		$T_{1,\textrm{man}}$ [ms] & 820(31) & 1114(55) & 1400(66) & 789(27) & 1108(43) & 1436(60) & 820(24) & 1124(37) & 1445(60) \\\hline
		$T_{1,\textrm{thresh}}$ [ms] &818(30)	&1112(53)	&1396(61)	&787(29)	&1101(46)	&1427(64)	&816(26) &1117(38)	&1442(58) \\\hline
		$T_{1,\textrm{ref}}$ [ms] &819(3)	&1131(4)	&1428(5)	&820(3)	&1130(4)	&1421(6)	&849(3) &1154(4)	& 1463(5) \\\hline\hline
		$T_{2,\textrm{man}}$ [ms] & 50(6) & 50(6) & 57(6) & 112(31) & 112(13) & 135(18) & 141(20) &173(18) & 174(18) \\\hline
		$T_{2,\textrm{thresh}}$ [ms] &51(8)	&52(8)	&64(17)	&110(22)	&119(29)	&143(25)	&152(30) &189(38)	&180(22) \\\hline
		$T_{2,\textrm{ref}}$ [ms] &54(1)	&63(1)	&69(1)	&108(1)	&131(1)	&160(1)	&160(2)	&196(1)	&226(2)
	\end{tabular}
	\end{adjustbox}

	\label{tab::Comparison_manual_automatic_tube_analysis}
\end{table}

Both methods seem to produce the same results for $T_1$ and $T_2$. Nevertheless, small changes in the way the ROI is chosen will lead to much larger differences in the result. This is a disadvantage of the manual approach. The automatic is easier to reproduce. Additionally, the previously chosen threshold limits may influence the results by removing pixels from the analysis. This is critical, if the removed pixels may belong to the tubes itself. Therefore, the range needs to be chosen high enough to cover all pixels without edge effects.\\
In the following, the thresholding is applied when the image contrast is good enough to differentiate the tubes easily from the background. In more complex cases a manual approach is preferred.

\newpage
\section{Measurement Protocols}
	
	\begin{table}[!h]
		\centering
		\caption{Protocol for quantification of $T_1$ using the gold-standard method.}
		\begin{tabular}{r|l}\addlinespace[2ex]\hline
			\multicolumn{2}{|c|}{Constant Parameter}	\\\hline
						&					\\
			Sequence	&	Spin-Echo 		\\ 
			Coils		&	HE 1-4, NE 1-2	\\
			Trajectory	&	Cartesian		\\
			FoV			&	200 mm			\\
			Slices		&	3				\\
			Slice Distance Factor	&	500 \%		\\
			TR			&	8000 ms			\\
			TE			&	15 ms			\\
			Slice-Thickness	&	5 mm		\\
			Base-Resolution	&	256			\\
			Phase-Resolution	&	100 \%	\\
			Shimming	&	Standard		\\
						&					\\\hline
			
			\multicolumn{2}{|c|}{Changed Parameter}	\\\hline
						&					\\
			Inversion-Time &	30 $\rightarrow$ 2530 ms in 250 ms steps \\\hline\addlinespace[2ex]
			Acquisition Time	&	11x 34:16 min	\\
		\end{tabular}
		
		\label{tab::sequence_protocol_T1_gold_standard}
	\end{table}
	
	\begin{table}[!h]
		\centering
		\caption{Protocol for quantification of $T_2$ using the gold-standard method.}
		\begin{tabular}{r|l}\addlinespace[2ex]\hline
			\multicolumn{2}{|c|}{Constant Parameter}	\\\hline
			&					\\
			Sequence	&	Spin-Echo 		\\ 
			Coils		&	HE 1-4, NE 1-2	\\
			Trajectory	&	Cartesian		\\
			FoV			&	200 mm			\\
			Slices		&	3				\\
			Slice Distance Factor	&	500 \%		\\
			TR			&	8000 ms			\\
			Slice-Thickness	&	5 mm		\\
			Base-Resolution	&	256			\\
			Phase-Resolution	&	100 \%	\\
			Shimming	&	Standard		\\
			&					\\\hline
			
			\multicolumn{2}{|c|}{Changed Parameter}	\\\hline
			&					\\
			TE			&	15 $\rightarrow$ 455 ms in 40 ms steps			\\\hline\addlinespace[2ex]
			Acquisition Time	&	11x 34:16 min	\\
		\end{tabular}
		
		\label{tab::sequence_protocol_T2_gold_standard}
	\end{table}

	\begin{table}[!h]
		\centering
		\caption{Protocol for acquisition of a $B_1$-map.}
		\begin{tabular}{r|l}\addlinespace[2ex]\hline
			\multicolumn{2}{|c|}{Constant Parameter}	\\\hline
			&					\\
			Sequence	&	Gradient Echo 	\\ 
			Coils		&	HE 1-4, NE 1-2	\\
			Trajectory	&	Cartesian		\\
			FoV			&	200 mm			\\
			TR			&	2000 ms			\\
			TE			&	2.14 ms			\\
			FA			&	8°				\\
			Slice-Thickness	&	8 mm		\\
			Base-Resolution	&	192			\\
			Phase-Resolution	&	100 \%	\\
			Shimming	&	Standard		\\\hline
			Acquisition Time	&	0:04 min	\\

		\end{tabular}
		
		\label{tab::sequence_protocol_B1_map}
	\end{table}

	\begin{table}[!h]
		\centering
		\caption{Protocol of a fully-sampled inversion-prepared bSSFP sequence.}
		\begin{tabular}{r|l}\addlinespace[2ex]\hline
			\multicolumn{2}{|c|}{Constant Parameter}	\\\hline
			&					\\
			Sequence	&	bSSFP	 		\\ 
			Coils		&	HE 1-4, NE 1-2	\\
			Trajectory	&	Radial			\\
			Contrast	&	Balanced		\\
			FoV			&	200 mm			\\
			Slice		&	1				\\
			TR			&	4.5 ms			\\
			TE			&	2.25 ms			\\
			FA			&	45 °			\\
			Slice-Thickness	&	5 mm		\\
			Base-Resolution	&	192			\\
			Inversion Mode	&	non slice selective \\
			Inversion-Time & 0 s			\\
			Sampling-Mode	&	MI:Al|Al	\\
			No. Inv. Exp. &		191	\\
			Measurements	&	500			\\
			Spokes		&	1				\\
			Bandwidth	&	740 Hz/pix		\\
			Shimming	&	Advanced		\\
			BWTP			&	3.5				\\
			rf-Pulse-length	&	0.9 ms		\\\hline
			Acquisition Time	&	39:07 min	\\
		\end{tabular}
		
		\label{tab::sequence_protocol_FS_ibSSFP}
	\end{table}

	\begin{table}[!h]
		\centering
		\caption{Protocol of a single-shot inversion-prepared bSSFP sequence.}
		\begin{tabular}{r|l}\addlinespace[2ex]\hline
			\multicolumn{2}{|c|}{Constant Parameter}	\\\hline
			&					\\
			Sequence	&	bSSFP	 		\\ 
			Coils		&	HE 1-4, NE 1-2	\\
			Trajectory	&	Radial			\\
			FoV			&	200 mm			\\
			Slice		&	1				\\
			TR			&	4.5 ms			\\
			TE			&	2.25 ms			\\
			FA			&	45 °			\\
			Slice-Thickness	&	5 mm		\\
			Base-Resolution	&	192			\\
			Inversion Mode	&	non slice selective \\
			Inversion-Time & 0 s			\\
			Sampling-Mode	&	GA			\\
			No. Tiny GA	&	13				\\
			No. Inv. Exp. &		1			\\
			Measurements	&	66			\\
			Spokes		&	17				\\
			Bandwidth	&	740 Hz/pix		\\
			Shimming	&	Advanced		\\
			BWTP			&	3.5				\\
			rf-Pulse-length	&	0.9 ms		\\\hline
			Acquisition Time	&	0:07 min	\\
		\end{tabular}
		
		\label{tab::sequence_protocol_SS_ibSSFP}
	\end{table}
	\FloatBarrier

\newpage

\begin{sidewaystable}[!h]
	\section{Fitted Parameter Comparison for Fully-Sampled Data}
	\centering
	\caption{Table with averaged parameters over all slices for each tube. The tube numbering is consistent with Figure \ref{fig::t1t2phantom_tubeOrdering}. The data is acquired with a fully-sampled sequence at the 17.12.2018.}
	\begin{tabular}{r|c|c|c|c|c|c|c|c|c}
		\addlinespace[4ex]
		Tube &	2&	3&	4&	5&	6&	7&	8&	9&	10\\\addlinespace[2ex]\hline
		\multicolumn{10}{|c|}{\textbf{Parameters for $T_1$}}	\\\hline\addlinespace[1ex]\hline
		\multicolumn{1}{|c|}{$T_{1,\textrm{ref}}$ [ms]}  &819(3)	&1131(4)	&1428(5)	&820(3)	&1130(4)	&1421(6)	&849(3) &1154(4)	& \multicolumn{1}{|c|}{1463(5)}	\\\hline\addlinespace[1ex]
		$T_{1,\textrm{FS}}$ [ms] &775(24)	&1041(45)	&1304(50)	&746(21)	&1050(43)	&1351(66)	&773(24) &1053(37)	&1376(46)	\\
		$T_{1,B_1,\textrm{FS}}$ [ms] &756(22)	&1147(44)	&1323(42)	&753(21)	&1045(39)	&1349(64)	&882(23) &1063(38)	&1375(75)	\\
		$T_{1,B_1,(k_{ss}=0.6),\textrm{FS}}$ [ms] &848(32)	&1146(61)	&1433(75)	&801(29)	&1024(46)	&1458(63)	&830(32) &1034(40)	&1474(52)	\\
		$T_{1,B_1,(k_{ss}=0.7),\textrm{FS}}$ [ms] &818(30)	&1112(53)	&1396(61)	&787(29)	&1101(46)	&1427(64)	&816(26) &1117(38)	&1442(58)	\\
		$T_{1,B_1,(k_{ss}=0.8),\textrm{FS}}$ [ms] &791(28)	&1085(52)	&1370(54)	&775(25)	&1082(49)	&1399(67)	&803(25) &1097(39)	&1405(57)	\\

		\addlinespace[6ex]\hline
		\multicolumn{10}{|c|}{\textbf{Parameters for $T_2$}}	\\\hline\addlinespace[1ex]\hline
		\multicolumn{1}{|c|}{$T_{2,\textrm{ref}}$ [ms]} &54(1)	&63(1)	&69(1)	&108(1)	&131(1)	&160(1)	&160(2)	&196(1)	&\multicolumn{1}{|c|}{226(2)}	\\ \hline\addlinespace[1ex]
		$T_{2,\textrm{FS}}$ [ms] &89(25)	&119(32)	&150(39)	&229(67)	&239(66)	&300(68)	&354(78) &460(125)	&437(488)	\\
		$T_{2,B_1,\textrm{FS}}$ [ms] &112(35)	&110(34)	&128(33)	&207(56)	&242(62)	&297(61)	&342(113) &457(227)	&776(879)	\\
		$T_{2,B_1,(k_{ss}=0.6),\textrm{FS}}$ [ms] &36(3)	&38(8)	&46(13)	&83(28)	&84(20)	&105(32)	&115(41) &134(30)	&130(20)	\\
		$T_{2,B_1,(k_{ss}=0.7),\textrm{FS}}$ [ms] &51(8)	&52(8)	&64(17)	&110(22)	&119(29)	&143(25)	&152(30) &189(38)	&180(22)	\\
		$T_{2,B_1,(k_{ss}=0.8),\textrm{FS}}$ [ms] &72(18)	&71(19)	&80(19)	&144(50)	&178(18)	&200(73)	&294(38) &290(90)	&353(47)	\\
	\end{tabular}
	
	\label{tab::fitted_parameter_bloch_reco_FS}
\end{sidewaystable}
\FloatBarrier

\begin{sidewaystable}[!h]
	\section{Fitted Parameter Comparison for Single-Shot Data}
	\centering
	\caption{Table with averaged parameters over all slices for each tube. The tube numbering is consistent with Figure \ref{fig::t1t2phantom_tubeOrdering}. The data is acquired with a single-shot sequence at the 17.12.2018.}
	\begin{tabular}{r|c|c|c|c|c|c|c|c|c}
		\addlinespace[4ex]
		Tube &	2&	3&	4&	5&	6&	7&	8&	9&	10\\\addlinespace[2ex]\hline
		\multicolumn{10}{|c|}{\textbf{Parameters for $T_1$}}	\\\hline\addlinespace[1ex]\hline
		\multicolumn{1}{|c|}{$T_{1,\textrm{ref}}$ [ms]}  &819(3)	&1131(4)	&1428(5)	&820(3)	&1130(4)	&1421(6)	&849(3) &1154(4)	& \multicolumn{1}{|c|}{1463(5)}	\\\hline\addlinespace[1ex]
		$T_{1,B_1,(k_{ss}=0.6),\textrm{SS}}$ [ms] &936(40)	&1150(64)	&1404(49)	&822(36)	&1179(31)	&1514(38)	&871(32) &1182(45)	&1522(22)	\\
		$T_{1,B_1,(k_{ss}=0.7),\textrm{SS}}$ [ms] &930(38)	&1155(63)	&1418(49)	&830(36)	&1173(28)	&1503(32)	&873(30) &1181(43)	&1498(20)	\\
		$T_{1,B_1,(k_{ss}=0.8),\textrm{SS}}$ [ms] &923(36)	&1060(61)	&1426(46)	&835(34)	&1165(26)	&1490(28)	&873(27) &1176(42)	&1474(20)	\\

		\addlinespace[6ex]\hline
		\multicolumn{10}{|c|}{\textbf{Parameters for $T_2$}}	\\\hline\addlinespace[1ex]\hline
		\multicolumn{1}{|c|}{$T_{2,\textrm{ref}}$ [ms]} &54(1)	&63(1)	&69(1)	&108(1)	&131(1)	&160(1)	&160(2)	&196(1)	&\multicolumn{1}{|c|}{226(2)}	\\ \hline\addlinespace[1ex]
		$T_{2,B_1,(k_{ss}=0.6),\textrm{SS}}$ [ms] &42(1)	&56(12)	&72(21)	&98(33)	&93(17)	&118(24)	&130(46) &149(39)	&130(46)	\\
		$T_{2,B_1,(k_{ss}=0.7),\textrm{SS}}$ [ms] &52(1)	&68(12)	&83(21)	&113(28)	&120(19)	&154(27)	&157(46) &183(38)	&174(19)	\\
		$T_{2,B_1,(k_{ss}=0.8),\textrm{SS}}$ [ms] &64(2)	&81(12)	&97(19)	&129(24)	&149(18)	&194(30)	&185(44) &217(37)	&222(24)	\\
	\end{tabular}
	
	\label{tab::fitted_parameter_bloch_reco_SS}
\end{sidewaystable}
\FloatBarrier

\singlespacing
\cleardoublepage
\addcontentsline{toc}{chapter}{Bibliography}
\printbibliography 

\onehalfspacing
\chapter*{Acknowledgment}
I would like to express my deep gratitude to Professor Martin Uecker for the possibility to work in his group. Thank you for showing me how much fun scientific research is and for the motivation to stay in the field of MRI. Your advice has always been a great help and I am glad to find you as my supervisor. My gratitude is also extended to Professor Ulrich Parlitz for reviewing this thesis as second supervisor.\\
I would like to thank various other people: Xiaoqing Wang for his help with the optimization algorithm and the model-based reconstruction techniques, Sebastian Rosenzweig for our very productive and fun discussions, the introduction to the scanner as well as for his help with BART, and Christian Holme for his advice on optimization algorithms, the local IT and for using BART, too. Additionally, I wish to thank Robin Wilke for introducing me to 3D-printers and for the creation of the tube holder of the $T_1$-$T_2$ phantom, Volkert Roeloffs for the discussions about my project and together with Jost Kollmeier for the fun during the joint creation of the $T_1$-$T_2$ phantom as well as Dirk Voit for his advice on sequence programming.\\
I wish to acknowledge the help provided by Tanja Otto and Ulrike Köchermann for introducing me to the scanner system.\\
Finally, I would like to extend my thanks to my parents and my girlfriend for their unconditional support and encouragement throughout my studies: especially, during the hard times... It would has not been possible without your help. Thank you so much!\\

\noindent \textit{Zum Schluss möchte ich mich nochmal bei meinen Eltern bedanken. Ihr habt mir das Studium ermöglicht und mir immer mit Rat und Tat zur Seite gestanden. Gerade in schwierigen Zeiten konnte ich mich so jedes Mal an euch wenden und Halt suchen. Ohne euch wäre mein Studium und damit auch diese Arbeit kaum möglich gewesen. Vielen vielen Dank dafür!}
\Declaration
\end{document}